\documentclass[twocolumn,showpacs,pre,floatfix]{revtex4}
\usepackage{graphicx}
\usepackage{dcolumn}
\usepackage{amsmath}
\usepackage{latexsym}
\usepackage{amssymb}
\usepackage{subfigure}

\def\bea{\begin{eqnarray}} 
\def\eea{\end{eqnarray}} 
\def\nn{\nonumber}

\def\f{\frac}
\def\la{\langle}
\def\ra{\rangle}

\def\b{\beta}

\def\om{\omega}

\def\bn{{\bf{n}}} 
\def\be{{\bf{e}}} 
\def\bk{{\bf{l}}} 
\def\b0{{\bf{0}}} 
\def\etal{{\emph{et~al.}}}

\def\mG{\mathcal{G}}
\def\mY{\mathcal{Y}}

\def\mW{\mathcal{W}}
\def\mU{\mathcal{U}}
\def\ni{\noindent}
\def\ie{\emph{i.e.}}

\begin{document}
\title{Heat transport and phonon localization in mass-disordered harmonic
  crystals}   
\author{Abhishek Chaudhuri{$^{1}$}, Anupam Kundu$^1$, 
Dibyendu  Roy$^{1}$, Abhishek   Dhar$^{1}$, 
 Joel L. Lebowitz{$^{4}$}, Herbert Spohn{$^{5}$}}
\affiliation{$^1$Raman Research Institute, C.V. Raman Avenue, Bangalore 560080, 
India}
\affiliation{$^4$Departments of Mathematics and Physics, Rutgers University, Piscataway, NJ 08854}
\affiliation{$^5$Zentrum Mathematik, Technische Universit\"{a}t M\"{u}nchen,
D-85747 Garching, Germany}
\date{\today}
\begin{abstract}
{ We investigate the steady state heat current in
 two and three  dimensional disordered harmonic crystals in a slab geometry,
 connected at  the boundaries  to
 stochastic white noise heat baths at different temperatures. 
The disorder causes short wavelength phonon modes to be localized
so the heat current in this system is carried by the extended phonon
 modes which can be either diffusive or ballistic. 
Using ideas both from localization theory and from kinetic theory 
we estimate the contribution of various modes to the  heat
 current and from this we obtain the asymptotic system size dependence
 of the current. These estimates are  compared with  results
 obtained from  a numerical
 evaluation of an exact formula  for the current, given in terms of
a frequency transmission function, as well as from direct nonequilibrium
 simulations. These yield a strong dependence of the heat flux on
 boundary conditions. 
Our analytical arguments show that for realistic boundary conditions 
 the conductivity is finite in three dimensions but we are not able to
 verify this numerically,   
except in the case where the system is subjected to an external
 pinning potential. This case is  closely related to the
 problem of localization of electrons in a random potential and here
 we  numerically verify that the pinned three dimensional system  
 satisfies Fourier's law while the two
 dimensional system is a heat insulator. We also investigate the
 inverse participation ratio of different normal modes.
}  
\end{abstract} 

\maketitle

\section{Introduction}
\label{sec:intro}

Energy transport in dielectric crystals at low temperatures is limited by
isotope mass disorder and by anharmonicities. For a crystal coupled to thermal
heat baths, in a slab geometry, the total energy transported may in addition
depend on how the bath is coupled to the boundaries of the bulk. In this paper
we will ignore anharmonicities but study the two other effects in considerable
depth.  

Based on empirical evidence, one generically  expects the validity of
Fourier's law, \ie~, for a slab in contact with heat reservoirs at different
temperatures  the average energy current, $J$, should be 
proportional to $1/N$, with $N$ being the slab length. 
There have been many attempts to derive Fourier's law from  microscopic
dynamics. Using very reasonable 
physical assumptions of local equilibrium and the notion of the mean
free path traveled by phonons between collisions yields a heuristic
derivation of Fourier's law.
There is however no fully convincing derivation.
There have also been many computer simulations of the heat flux in
the nonequilibrium stationary states of anharmonic crystals 
kept in contact with thermal reservoirs at different temperatures, and
theoretical analysis based on the Green-Kubo formalism 
\cite{dharrev09,LLP03,BLR00}. 
These  studies suggest (but there is no proof despite some claims)   
that Fourier' law is  not valid for one and two dimensional
systems, even in the presence of anharmonic interactions, unless the
system is also subjected to an  external substrate pinning potential. 
Generically it is
found that, for a system in contact with heat reservoirs at different
fixed temperatures, the heat current density $J$ scales anomalously with
system length $N$ as 
\bea
J \sim \f{1}{N^{\mu}}~,
\eea
with $\mu \neq 1$. 
The effective thermal conductivity behaves then as $\kappa \sim
N^\alpha$ where $\alpha=1-\mu$. For two dimensional systems there are some
analytic studies which suggest $\kappa \sim \ln (N)$. 
Recent experiments on heat conduction  in nanotubes and graphene flakes have
reported 
observations which indicate such divergence of $\kappa$ with system size
\cite{chang08,nika09}.

For heat conduction in the ordered harmonic crystal there are exact results
from which  one has $\mu=0$ in all dimensions \cite{rieder67,nakazawa70}.  
Heat conduction in  a disordered harmonic crystal will be affected by
Anderson localization  
\cite{anderson58} and by phonon scattering.

In this paper we report results of heat conduction studies in
$2D$ and $3D$ disordered harmonic lattices with scalar displacements,
connected to heat baths modeled by Langevin equations with white
noise. 
We pay particular attention to the interplay between localization
effects, boundary effects, and the role of long wavelength modes. 
The steady  state heat current is given exactly as an integral
over all frequencies of a phonon transmission coefficient. 
Using this formula and heuristic arguments, based on localization
theory and kinetic theory results, we estimate the system size
dependence of the current. The main idea behind our
arguments is that the phonon states can be classified as  ballistic 
modes, diffusive modes and localized modes. The classification refers
both to the character of the eigenfunctions  as well as to their
transmission properties. Ballistic modes are 
spatially extended and approximately periodic; their transmission is
independent of system size. Diffusive modes are extended but non-periodic and
their transmission decays as $1/N$. For localized modes  transmission
decays exponentially with $N$.
In  the context of kinetic theory calculations, the
ballistic modes are the low frequency modes with phonon mean free path
$\ell_K(\om) \stackrel{>}{\sim} N$, and their contribution to the current
leads to divergence of the thermal conductivity. 
Here we will carefully examine the effect of boundary conditions on
these modes. 

Numerically we use two different approaches to study the
nonequilibrium stationary state. 
The first is a numerical one which relies on  the result that the
current can be expressed in terms of a transmission coefficient. 
This transmission coefficient can be
written in terms of phonon Green's functions and we implement
efficient numerical schemes to evaluate this.  
The second approach is through direct nonequilibrium
simulations of the Langevin equations of motion and finding the 
steady state current and
temperature profiles. 
We have also studied properties of the isolated system, \ie, of the disordered
lattice without coupling to heat baths and looked at the normal mode frequency
spectrum and the wavefunctions.  
One measure of the degree of localization of the normal modes of the
isolated system is the so-called inverse participation
ratio [IPR, defined in Eq.~(\ref{ipr}) below].
We have carried out studies of the IPR
and linked these with the results from the transmission study. 

{\bf{ Phonon localization}:} This is closely related to the electron
localization problem. 
The effect of localization on linear waves in disordered
media has been most extensively studied in the context of the
Schr\"{o}dinger equation for 
non-interacting electrons moving in a disordered potential. 
Looking at the eigenstates and eigenfunctions of the
isolated system of a single electron in a disordered potential one
finds that, in contrast to the spatially extended Bloch states in periodic
potentials, there are now many eigenfunctions which are exponentially
localized  in space. 
It was argued by Mott and Twose \cite{mott61} and
by Borland  \cite{borland63}, and  proven rigorously by Goldsheid
\etal ~\cite{goldsheid77},  
that in one dimension ($1D$) all states are exponentially localized.
 In two dimensions ($2D$) there is no
proof but it is believed that again all states
are  localized. In three dimensions ($3D$)   there
is expected to be a transition from extended to localized states as the energy
is moved towards the band edges  \cite{lee85}.  The transition from extended
to localized states, which occurs when the disorder is increased, changes the 
system from a conductor to an insulator. 
 The connection between localization and heat
transport in a  crystal is complicated by the fact that
phonons of all frequencies can contribute to energy transmission
across the system. 
In particular  account has to be
taken of the fact that
low frequency phonon modes are only weakly affected by
disorder and always remain  extended. 
The heat current carried by  a mode which is 
localized on a  length scale $\ell$, decays with system length $N$ as
$e^{-N/\ell}$. This $\ell$ depends on the phonon frequency and low
frequency modes for which $\ell \sim N$ will therefore be  
carriers of  the heat current. The net
current then depends on the  nature of these low
frequency modes and their scattering due to boundary conditions (BCs). 

A renormalization group study of
phonon localization in a continuum 
vector displacement model was carried out by John \etal~
\cite{john83}. They found that much of 
the predictions of the scaling theory of localization for electrons
carry over to the phonon case.  Specifically they showed that in one and
two dimensions all non-zero frequency phonons are localized with the low
frequency localization length diverging as $\ell \sim \om^{-2}$ and
$\sim e^{c/ \om^{2}}$, respectively (where $c >0$ is some constant). This means that 
in $1D$ all modes with $\om \stackrel{>}{\sim} \om_c^L= N^{-1/2}$ are
localized while in $2D$   
all modes with $\om \stackrel{>}{\sim} \om_c^L= [\log(N)]^{-1/2}$ are
localized.   
In $3D$ the prediction is that there is an $\om_c^L$ independent of $N$ above
which all modes are localized. 
However this study does not make any 
statements on the system size dependence of the conductivity.

{\bf{Kinetic theory}:} If one considers the low frequency extended
phonons, then  the effect of disorder  is weak and in dimensions $d >1$ one
expects that localization effects can be neglected and  kinetic theory should
be able to provide an accurate  description.   
In this case one can think of Rayleigh
scattering of phonons. This  gives an
effective mean free path $\ell_K(\omega) \sim \omega^{-(d+1)}$ [see
  appendix~\ref{sec:appa}] , for dimensions $d >1$, and 
a diffusion constant $D(\omega)=v \ell_K (\omega)$ where $v$, the
sound velocity, can be taken to be a constant. For a finite system of
linear dimension $N$ we have $D(\om)=v N$ for $\om \stackrel{<}{\sim} N^{-1/(d+1)}$. 
Kinetic theory then predicts 
\bea
 \kappa = \int_{N^{-1}}^{\om_{\rm max}} d \om \rho (\om) D(\om)~, \label{kinetick}
\eea 
where $\rho(\om)\sim \om^{d-1}$ is the density of states 
 and we get $\kappa \sim N^{1/(d+1)}$ implying $\mu=d/(d+1)$. 
The divergence of the phonon mean free path at low frequencies and the
resulting divergence of the thermal conductivity of a disordered harmonic
crystal has been discussed in the literature and it has been argued
that anharmonicity is necessary to make $\kappa$ finite
\cite{callaway59,ziman72}. 

{\bf{Simulation results}:} There have been  only few
simulation studies of heat conduction in  three dimensional disordered
systems and none have been definitive concerning  the validity of
Fourier's law \cite{shimada00,shiba08}. In two dimension a diverging
thermal conductivity was reported in \cite{lee05}.  
Some other studies have also looked at heat conduction in glassy systems at
low temperatures where the harmonic approximation was used
\cite{allen89,xu09} but these 
did not address the  questions of
$N$-dependence of $\kappa$ and the validity of Fourier's law.

In $1D$ it is well known from rigorous results and numerical studies
that $\alpha \neq 0$ and its value is strongly dependent on boundary
conditions
\cite{matsuda70,rubin71,casher71,oconnor74,dhar01,roydhar08}. 
For fixed BCs one has $\alpha=-1/2$ while for free BCs,
$\alpha=1/2$. 
The precise definitions of the different BCs will be given later. Here we
explain the different physical situations they 
correspond to.  If we model the heat reservoirs  themselves  by  infinite
ordered harmonic crystals then  Langevin type equations
for the system \cite{dharroy06} are obtained on
eliminating the bath degrees of freedom.  
The  two different BCs then emerge naturally.
Fixed BCs  correspond to reservoirs   with  properties different from 
the system ({\emph{e.g.}}   different spring constants) and in this
case one finds that effectively the particles at the boundaries (those
coupled to reservoirs) experience an additional harmonic pinning
potential. Free BCs   correspond to   the case where  the
reservoir is simply an  extension of the system (without   disorder) 
and in this case the end particles are unpinned. 
Free BCs have been studied in the literature in the context of heat conduction
in one dimensional chains \cite{rubin71} and in studies on nanotubes
\cite{savic08,stoltz08}.   
In this paper we study lattices
with both fixed and free BCs although we think that fixed BCs are  more  realistic.

In the presence of an external pinning potential low frequency modes
are suppressed, hence one expects qualitative differences 
in transport properties. 
The pinned system has often been used as a model
system to study the validity of Fourier's law. It has no translational invaraince and is thus more
closely related to the problem of electrons moving in a random
potential. 
Here we consider systems  with  and without external pinning potentials.

The rest of the paper is organized as follows. 
In Sec.~(\ref{sec:model}) we define the specific model studied by us and
present some general results for heat conduction in harmonic Hamiltonian
systems connected to Langevin baths. 
We also give some details of the numerical and simulation methods
used in the paper.
The transfer matrix approach used in evaluating the phonon
transmission function is explained in Appendix~\ref{sec:appb}. 
In Sec.~(\ref{sec:general}) a brief review of results for the one
dimensional case and the 
heuristic arguments for the higher dimensional cases are given.
In Sec.~(\ref{sec:results}) we present results from both the numerical
approach and from nonequilibrium simulations. 
The main results presented are for transmission functions, IPRs of
normal modes and the system size dependence of the 
current in two and three dimensional disordered harmonic lattices.  Along the
way we also present results for the density of states $\rho(\om)$. 
Finally we conclude with a discussion in Sec.~(\ref{sec:discussion}). 

\section{Models and methods}
\label{sec:model}

For simplicity we consider only the  case where longitudinal and
transverse vibration modes are decoupled and hence we can describe the
displacement at each site by a scalar variable. Also we restrict our
study to $d$-dimensional hypercubic lattices. Let us denote the
lattice points by the vector $\bn=\{n_1,n_2,...,n_d\}$ with
$n_\nu=1,2,...,N$. The displacement of
a particle at the  lattice site $\bn$ is given by
$x_\bn$.
In the harmonic approximation the system Hamiltonian  is given by
\begin{eqnarray}
H &=& \sum_{\bn}\frac{1}{2}{ m_\bn \dot{x}_{\bn}^2} + \sum_{\bn,\hat{\be}}
\frac{k}{2}(x_{\bn}-x_{\bn+\hat{\be}})^2+\f{k_o}{2} \sum_\bn {x_{\bn}^2}~,  \label{ham}
\end{eqnarray}
where $\hat{\be}$ refers to the $2d$ nearest neighbors of any site and
we impose boundary conditions which will be specified later. We have
also included an external pinning harmonic potential with spring 
constant $k_o$, which we will sometimes set equal to zero. 
We consider a binary mass disordered crystal. Specifically we set the
masses of exactly half the particles at randomly chosen sites to be
$\bar{m}-\Delta$ and the rest to be $\bar{m}+\Delta$. Thus $\Delta$ gives a measure 
of the disorder.

We couple all the particles at $n_1=1$ and $n_1=N$ 
to heat reservoirs, at temperatures $T_L$ and $T_R$ respectively, and
use periodic boundary conditions in the other $(d-1)$ directions. 
The heat conduction takes place along the $1$ direction. 
Each layer with constant $n_1$ consists of $N'=N^{d-1}$ particles. 
The heat baths are modeled by white noise Langevin equations
of motion for the particles coupled to the baths.  
Using the notation $\bn=(n_1,\bn')$, the equations of motion are 
given by: 
\bea
&& {m}_{\bn} \ddot{x}_{\bn} = -\sum_{\hat{\be}}k (x_\bn-x_{\bn+\hat{\be}}) -k_o x_{\bn} +
\delta_{n_1,1} (-\gamma \dot{x}_{\bn} \nn \\ && ~~~~~~~~~+\eta^L_{\bn'}-k' x_{\bn})
  +\delta_{n_1,N} (-\gamma \dot{x}_{\bn}+\eta^R_{\bn'}-k' x_{\bn})~,
~~~~~\label{eqm}
\eea
where the dissipative and noise terms are related by the usual fluctuation
dissipation relations 
\bea
\la \eta^L_{\bn'}(t) \eta^L_{\bk'}(t') \ra &=& 2 \gamma k_B T_L \delta
(t-t') \delta_{\bn' \bk'}~,~~~\nn \\
  \la \eta^R_{\bn'}(t) \eta^R_{\bk'}(t') \ra &=& 2 \gamma k_B T_R \delta
(t-t') \delta_{\bn' \bk'}~. \label{nncorr}
\eea
The particles at the surfaces $n_1=1,N$ experience additional harmonic
pinning potentials with  spring constants $k'$ arising from coupling
to the heat reservoirs.   
We consider two kinds of boundary conditions at the surfaces connected
to reservoirs:  
(i) fixed BCs  $k' > 0$ and  (ii) free BCs $k' = 0$.
As discussed in the introduction fixed BCs correspond to reservoirs
with properties different from the system while free reservoirs
correspond to the case where the reservoir is really an extension of
the system but without disorder. 
For the pinned case we only consider fixed BC. 
A schematic of the models and the different boundary
conditions   that we have studied is given in Fig.~(\ref{scheme}). 

Henceforth we will use dimensionless variables: 
force-constants are measured in units of $k$,  masses in
units of the average mass $\bar{m}$, time  in units of the inverse
frequency $\Omega^{-1} = (\bar{m}/k)^{1/2}$, displacements are in units of the
lattice spacing $a$, friction constant $\gamma$ is in units of $\bar{m}
\Omega$,   and finally temperature  is measured in units of $\bar{m} a^2
\Omega^2/k_B$.

Driven by the reservoirs at two different temperatures $T_L$ and
$T_R$  the system
reaches a nonequilibrium steady state. Our main interest will be in the steady
state heat current in the system. 
Given the Langevin equations of motion Eq~(\ref{eqm}), one can 
find a formal general expression for the current.
Let us
denote by ${X}$ a column vector with 
$N^d$ elements consisting of the displacements at all lattice
sites. Similarly let ${\dot{X}}$ represent the vector for velocities at all
sites. Then we can write the  Hamiltonian in Eq.~(\ref{ham}) in the
compact form $ H=\f{1}{2} {\dot{X}}^T \mathcal{M} {\dot{X}} +\f{1}{2}
{X}^T \mathcal{V} {X}~$,   
which defines the diagonal mass matrix $\mathcal{M}$ and the force constant matrix $\mathcal{V}$.  
With this notation we have the following form for the steady state current
per bond 
from the left to the right reservoir \cite{dharroy06,casher71} :
\bea
\mathcal{J} &=& \f{\Delta T } {4\pi N'} \int_{-\infty}^\infty  d \om {\mathcal{T}}_N
(\om)~, \label{jeq}   
\eea
 where  
\bea
{\mathcal{T}}_N(\om) &=& 4~Tr [ \mathcal{I}_L (\om) \mathcal{G}^+ (\om )
  \mathcal{I}_R (\om ) \mathcal{G}^- (\om) ]~,  \\
 \mathcal{G}^+(\om)&=&[-\om^2 \mathcal{M} + \mathcal{V}
  -\mathcal{S}^+_L -\mathcal{S}^+_R]^{-1}~,~~\mathcal{G}^-=[\mathcal{G}^+]^* \nn  
\eea
and $\Delta T=T_L-T_R$. The ${\mathcal{S}}^+_{L}$, ${\mathcal{S}}^+_{R}$
represent terms arising from the coupling to  
the left and right baths respectively, and 
${\mathcal{I}}_{L,R}=Im[{\mathcal{S}}^+_{L,R}]$.  
The specific form of $\mathcal{S}^+_{L,R}$ for our system described by
Eqs.~(\ref{eqm}) is given in Appendix~\ref{sec:appb}.   
The matrix $\mathcal{G}^+(\om)$ can be identified as the phonon
Green's function  of the system with self-energy corrections due to
the baths \cite{dharroy06}. The integrand in Eq.~(\ref{jeq}) 
$\mathcal{T}_N(\omega)$ can be thought of as  the transmission coefficient of
phonons at 
frequency $\om$ from the left to the right reservoir. 
It will vanish, when $N \to \infty$ , at values of $\om$ for which the
disorder averaged  density of states is zero.  
Note that due to the
harmonic nature of the forces the dependence of the heat flux on the reservoir
temperatures enters only through the term $\Delta T$ in Eq.~(\ref{jeq}).  
The above expression for the current is of the Landauer form and
has been derived  using various other approaches such as
scattering theory \cite{rego98,blencowe99} and  the nonequilibrium
Green's function formalism \cite{yamamoto06,wang06}.

 {\bf Numerical approach}: 
In Appendix~\ref{sec:appb} we describe how ${\mathcal{T}}_N$ can be
expressed in a form amenable to accurate numerical computation.
The system sizes we study are sufficiently large so that $\mathcal{T}_N(\om)$
has appreciable values only within the range of frequencies of normal
modes of the isolated system,  {\emph{i.e.}}, corresponding to $\gamma=0$
in Eq.~(\ref{eqm}).  Outside this range we find that the transmission
rapidly goes to zero.
By performing a discrete sum over the transmitting range of frequencies we
do the integration in 
Eq.~(\ref{jeq})  to obtain  the heat current density $\mathcal{J}$. In
evaluating the 
discrete sum over $\om$, step sizes of $\delta \om =0.01-0.0001$ are
used and we verified convergence in most cases. With our choice of
units we have $k=1, \bar{m}=1$ and we fixed $\Delta T=1$. Different values of the mass
variance $\Delta$ and the on-site spring constant $k_o$ were studied
for two and three dimensional lattices of different sizes. It is expected that the value
of the exponent $\mu$ will not depend on $\gamma$ and in our
calculations we mostly set $\gamma=1$, except  when  otherwise
specified.

{\bf Simulation approach}: The  simulations of Eq.~(\ref{eqm}) are 
performed using a velocity-Verlet scheme as given in
\cite{allen}. The current and temperature profiles in the
system are obtained from the following time averages in the
nonequilibrium steady state:
\bea
\mathcal{J}_1&=& \f{1}{N'}\sum_{\bn'} \f{\gamma}{m_{(1,\bn')}} \left[ 
T_L- m_{(1,\bn')}  \la \dot{x}^2_{(1,\bn')} \ra \right], \nn \\
\mathcal{J}_{n}&=&-\f{1}{N'}\sum_{\bn'}  ~\la~
[x_{(n,\bn')}-x_{(n-1,\bn')}]~\dot{x}_{(n,\bn')}~\ra~,\nn \\&&~~~~~~~~~~~~~~~~~~~~~~~~~n=2,3,...,N~, \nn \\
\mathcal{J}_{N+1}&=& -\f{1}{N'}\sum_{\bn'} \f{\gamma}{m_{(N,\bn')}} \left[ 
 T_R- m_{(N,\bn')}  \la \dot{x}^2_{(N,\bn')} \ra \right]~, \nn \\ 
 T_n&=& \f{1}{N'}\sum_{\bn'} m_{(n,\bn')} \la~ \dot{x}^2_{(n,\bn')}~\ra~,~~~n=1,2,...,N~. \nn
\eea
We then obtained the average current $\mathcal{J}=(\sum_{n=1}^{N+1}
\mathcal{J}_N)/(N+1)$. 
In the steady state one has $\mathcal{J}_n=\mathcal{J}$ for all $n$ and 
stationarity can be tested by checking how accurately
this is satisfied. 
We chose a  
step size of $\Delta t = 0.005$ and  equilibrated the system for over
$10^8$ time steps. Current and temperature profiles were
obtained by averaging  over another 
$10^8$ time steps. The parameters  $T_L = 2.0, T_R = 1.0$ are kept
fixed and different values of the mass 
variance $\Delta$ and the on-site spring constant $k_o$ are simulated. 

The value of $\mathcal{T}_N (\omega)$, $\mathcal{J}$ and $T_n$
depend, of course, on the
particular disorder realization. Mostly we will here be interested in
disorder averages of these quantities which we will denote by
$[\mathcal{T}], J=[\mathcal{J}]$ and $[T_n]$. We also define the disorder averaged
transmission per bond with the notation
\bea
T(\omega)=\f{1}{N'} [\mathcal{T}(\omega)]~. \nn
\eea

{\bf Numerical analysis of eigenmodes and eigenfunctions}:~ 
We have  studied the properties of the normal modes of the
 disordered harmonic lattices in the absence
of coupling to reservoirs, again with both free and fixed boundary conditions. 
The $d$-dimensional lattice has $p=1,2,...,N^d$ normal modes and  we denote the
displacement field corresponding to the $p^{\rm th}$ mode by
$a_{\bn}(p)$ and the corresponding eigenvalue by
$\omega_p^2$.  The normal mode equation  corresponding to
the Hamiltonian in Eq.~(\ref{ham}) is given by:
\bea
m_{\bn} \om_p^2 a_{\bn}=  (2d +k_o) a_{\bn}- \sum_{\hat{\be} } a_{\bn+\hat{\be}}~,
\eea
where the $a_\bn$ satisfy appropriate boundary conditions.
Introducing variables $\psi_{\bn}(p)=m_{\bn}^{1/2} a_{\bn}(p)$,
$v_\bn=(2d  +k_o)/m_\bn$ and $t_{\bn,\bk}= 1/(m_\bn m_\bk)^{1/2}$ for
nearest neighbour sites $\bn,~\bk$ the
above equation transforms to the following form:
\bea
\omega_p^2 \psi_{\bn}(p)=v_\bn \psi_{\bn} (p) - \sum_\bk t_{\bn,\bk}
\psi_\bk (p)~.  
\eea
This has the usual structure of an  eigenvalue equation for a single
electron moving in a $d$-dimensional lattice corresponding
to a  tight-binding Hamiltonian with nearest neighbour hopping
$t_{\bn,\bk}$ and on-site energies $v_\bn$. Note that  $t_{\bn,\bk}$
and $v_\bn$ are correlated random variables, hence the disorder-energy
diagram might differ considerably from a single band Anderson
tight-binding model.  

We have numerically evaluated all eigenvalues and eigenstates of the above
equation for finite cubic lattices of size upto $N=64$ in 
$2D$ and $N=16$ in $3D$. One measure of the degree of localization
of a given mode is the inverse participation ratio (IPR) defined as
follows:
\bea
P^{-1}= \f{ \sum_\bn a_\bn^4 }{(\sum_\bn a_\bn^2)^2}~. \label{ipr}
\eea
 For a completely localized state, {\emph{i.e.}}
 $a_\bn=\delta_{\bn,\bn_0}$,  $P^{-1}$ takes the value $1$. On the
 other hand for a
 completely delocalized state, for which $a_\bn=N^{-d/2}e^{i
   \bn. {\bf{q}}}$ where ${\bf{q}}$ is a wave vector, $P^{-1}$ takes    the value $N^{-d}$.
We will present numerical results for the IPR calculated
for all eigenstates of given disorder realizations, in both $2D$ and $3D$. 
Finally we will show some results for the density of states, $\rho(\om)$, of the
disordered system defined by:
\bea
\rho(\om)= \sum_p \delta ( \om_p - \om ) ~.
\eea
The density of states of disordered binary mass harmonic crystals was 
studied  numerically by Payton  and Visscher in 1967 \cite{payton67} and  
reviewed by Dean in 1972 \cite{dean72}.

\section{ Heat conduction in disordered
  harmonic crystals: General considerations}
\label{sec:general}
Let us first briefly consider 
heat conduction in the  one dimensional disordered harmonic chain. This has
been extensively studied and is well understood
\cite{matsuda70,rubin71,casher71,oconnor74,dhar01,roydhar08}. The matrix 
formulation explained in the last section leads  to  a
clear analytic understanding of the main results. The current is
given by the general expression Eq.~(\ref{jeq}). From Eq.~(\ref{transmeq}) the
transmission is   given by 
$\mathcal{T}_N(\om)=4 \gamma^2 \omega^2 |G^+_{N}(\om)|^2$ where $G^+_{N}(\om)$
is now just a complex number. The disorder averaged transmission is
given by $T_N(\om)=[\mathcal{T}_N(\om)]$. There are three observations
that enable one to determine the asymptotic system size dependence of the 
current. These are:

(i)$P^{(1,N)}=[G^+_N]^{-1}$ given
by Eqs.~(\ref{peqn}),~(\ref{qeqn}) is a complex number which can be
expressed in terms of the product of $N$  random $2 \times 2$
matrices. Using Furstenberg's theorem it
can be shown that for almost all disorder realization, the large $N$
behaviour of $P^{(1,N)}$ for fixed $\om >0$ is $|P^{(1,N)}|
\sim e^{b N \omega^2}$, where $b>0$ is a constant. This is to be understood in
the sense that $\lim_{N \to \infty} (1/N) \log |P^{1,N)}| \sim b \om^2$ for
$\om \to 0$.    
Since $\mathcal{T}_N(\omega) \sim |P^{(1,N)}|^{-2} \sim e^{-2 b N \om^2}$, this implies that
transmission is significant only for low frequencies $\omega
\stackrel{<}{\sim} \om_c(N) \sim 1/N^{1/2}$. 
The current is therefore dominated by the small $\om$ behaviour of ${T}_N(\om)$.

(ii) The second observation made in \cite{dhar01} is
that the transmission for $\om < \om_c(N)$ is ballistic in the sense that
${T}_N(\om)$ is insensitive to the disorder. 

(iii) The final important observation 
is that the form of the prefactors of $e^{-b N \om^2}$ in ${T}_N(\omega)$  for
$\om < \om_c(N)$  
depends strongly on boundary conditions and bath properties \cite{dhar01,roydhar08}. For the white
noise Langevin baths one finds  ${T}_N(\omega) \sim 
\omega^2 e^{-b N \om^2} $ for fixed BC and ${T}_N(\omega)\sim 
\omega^0 e^{-b N \om^2} $ for free BC \cite{roydhar08}. This
difference arises because 
of  the scattering of long wavelength modes by the boundary pinning
potentials.

In Fig.~(\ref{tw1d0.4}) we plot  numerical results showing ${T}_N(\om)$ for
the $1D$ binary mass-disordered lattice  with both fixed and free
boundary conditions.  One can clearly see the two features discussed above namely (i) dependence of
frequency cut-off  on system size and (ii) dependence of form of  
${T}_N(\om)$  on boundary conditions.
Using the three observations made above  it is easy to arrive at the
conclusion that  $J \sim N^{-3/2}$  for fixed BC and $J \sim
N^{-1/2}$ for free BC. In the presence of a pinning potential
the low-frequency modes are suppressed and  one obtains 
a heat insulator with $J \sim e^{-c N}$, with $c$ a constant \cite{oconnor74} (see also
\cite{dharleb08} and references there).

{\bf{Higher dimensions}}. Let us try to extend the analysis of the $1D$
case to higher dimensions. For this we will  use inputs from
both kinetic theory and the theory of phonon  localization. The main
point of our arguments involves the assumption that normal modes can
be classified 
as ballistic, diffusive or localized. Using localization theory we determine
the frequency region where states are localized. 
The lowest frequency states with $\om \to 0$ will be ballistic 
and we use kinetic theory to determine the fraction of extended states which
are ballistic. We assume that at sufficiently low frequencies the effective
disorder is always weak (even when the mass variance $\Delta$ is large) and one
can still use kinetic theory.   
Corresponding to the three observations made above for the $1D$ case
we now make the following arguments:

(i) From localization theory one expects all
fixed non-zero frequency states in a $2D$ disordered system to be
localized when the size of the system goes to infinity. 
As discussed in Sec.~(\ref{sec:intro}) localization theory gives us a
frequency cut-off $\om_c^L=(\ln N)^{-1/2}$ in $2D$ above which states are
localized.  
In $3D$ one obtains a  finite frequency cut-off $\om_c^L$ independent
of system size above which states are 
localized. 

(ii) For the unpinned case with finite $N$ there will exist low frequency
states below $\om_c^L$, in 
both $2D$ and $3D$, which are extended states. These states are either
diffusive or ballistic. 
Ballistic modes are insensitive to the disorder and their
transmission coefficient are almost the same as for
the ordered case. To find the frequency cut-off below which states are
ballistic we use kinetic theory results (see Appendix~\ref{sec:appa}).     
For the low-frequency extended states we expect kinetic
theory to be reliable and this gives us a mean free
path for phonons $\ell_K \sim 
\om^{-(d+1)}$. This means that for low frequencies  $\om
\stackrel{<}{\sim} \om_c^K = N^{-1/(d+1)}$ we have 
$\ell_K(\om)  > N $ and  phonons transmit ballistically.
 We now proceed  to calculate  the contribution of these ballistic
 modes to the total  current. This can be obtained by looking at the
 small $\om$ form of $T_N(\om)$ for the ordered lattice. 

(iii) For the ordered lattice $T_N(\om)$
 is typically a highly oscillatory function with the oscillations
 increasing with system size. An effective transmission coefficient in
 the $N \to \infty$ limit can be obtained by considering the
 integrated transmission.
This asymptotic {\emph{ effective}} low-frequency form of
 $T_N(\om)$, for the ordered lattice can be calculated using methods described
 in \cite{roydhar08} and is given by:
\bea
T(\om) &\sim& \om^{d+1}~,~~~~~{\rm  fixed~ BC} \nn \\
T(\om) &\sim& \om^{d-1}~, ~~~~~~{\rm  free~ BC}~, \label{Tasym}
\eea
the result being valid for $d=1,2,3$  \cite{anupam09}.

Using the above arguments we then get the ballistic 
contribution  to the total current density (for the unpinned case) as:
\bea
J_{\rm ball} &\sim & \int_0^{\om_c^K} d \om~ \om^{d+1} \sim \f{1}{N^{(d+2)/(d+1)}}~,~~~~{\rm
  fixed~~BC,} \nn \\
&\sim & \int_0^{\om_c^K} d \om~ \om^{d-1} \sim
\f{1}{ N^{d/(d+1)}}~,~~~~{\rm free~~BC}~.~~~ 
\label{jball}
\eea 
We can now make predictions for the asymptotic system size dependence
of total current density in two and three dimensions. 

{\emph{ Two dimensions}}: From localization theory one expects that all 
 finite frequency modes $\om \stackrel{>}{\sim} \om_c^L=(\ln N)^{-1/2}$  are localized and
 their contribution to the total
 current falls exponentially with system size. Our kinetic theory arguments 
show that  the low frequency extended states with  
 $\om_c^K \stackrel{<}{\sim} \om \stackrel{<}{\sim} \om_c^L$ are diffusive
(where $\om_c^K =N^{-1/3}$) while the 
remaining modes with     $\om \stackrel{<}{\sim} \om_c^K$ are ballistic. The diffusive contribution to total current
 will then scale as $J_{\rm diff} \sim (\ln N)^{-1/2} N^{-1}$. The
 ballistic contribution depends on BCs and 
 is given by Eq.~(\ref{jball}). This gives 
 $J_{\rm ball} \sim N^{-4/3}$ for fixed BC and $J_{\rm ball} \sim N^{-2/3}$ for free BC. Hence,
 adding all the different contributions, we conclude that
 asymptotically:
\bea
J &\sim &   \f{1}{(\ln N)^{1/2} N} ~,~~~~{\rm
  fixed~~BC,}~~d=2~, \nn \\
&\sim &   \f{1}{N^{2/3}}~,~~~~~~~{\rm free~~BC,}~~d=2. 
\label{analy2d}
\eea 
In the presence of an onsite pinning potential at all sites the low
frequency modes get cut off and  all the remaining states are
localized, hence we expect:
\bea
J \sim e^{-b N},~~~~~~~~~~~{\rm pinned }~,~~d=2~,
\eea
where $b$ is some positive constant. 

{\emph{Three dimensions}}: In this case localization theory tells us that 
modes with  $\om \stackrel {>}{\sim} \om_c^L$ are localized and $\om_c^L$ is
independent of $N$. From kinetic theory we find that 
the extended states with $\om_c^K \stackrel{<}{\sim} \om \stackrel{<}{\sim}
\om_c^L$ are diffusive (with 
$\om_c^K = N^{-1/4}$) and those with $\om \stackrel{<}{\sim} \om_c^K$ are
ballistic.   
The contribution to current from diffusive modes scales as $J_{\rm diff} \sim N^{-1}$. 
The ballistic contribution (from states with $\om \stackrel{<}{\sim}
N^{-1/4}$) is obtained from Eq.~(\ref{jball}) and 
gives $J_{\rm ball} \sim N^{-5/4}$ for fixed BC and $J_{\rm ball} \sim
N^{-3/4}$ for free BC. 
Hence,  adding all contributions, we conclude that  asymptotically:
\bea
J &\sim &   \f{1}{N}~,~~~~~~{\rm
  fixed~~BC~,}~~d=3~, \nn \\
&\sim & \f{1}{ N^{3/4}}~,~~~~~~{\rm free~~BC~,}~~d=3~. 
\label{analy2d}
\eea 
In the presence of an onsite pinning potential at all sites the low
frequency modes get cut off and, since in this case the remaining
states form bands of diffusive and localized states, hence we expect:
\bea
J \sim \f{1}{N}~,~~~~~~~{\rm pinned~, }~~d=3.
\eea
Thus in $3D$ both the unpinned lattice with fixed boundary conditions and the
pinned lattice are expected to show Fourier type of behaviour as far
as the system size dependence of the current is considered.

Note  that for free BC,  the prediction for the current contribution from the
ballistic part $J_{\rm ball } \sim N^{-d/(d+1)}$ is identical to that from kinetic
theory discussed in Sec.~(\ref{sec:intro}). This agreement can be traced
to the small $\om$ form of $T(\om) \sim \om^{d-1}$ for free BC [see
  Eq.~(\ref{Tasym})] which is identical  
to the form of the density of states $\rho(\om)$ used in kinetic
theory. The typical form of density of states for ordered and disordered
lattices in different dimensions 
is shown in Fig.~(\ref{dosdis}) and we can see that the
low frequency form is similar in both cases and has the expected
$\om^{d-1}$ behaviour. However it seems reasonable to expect that, since the transport
current phonons are  injected at the boundaries, 
in kinetic theory one needs to use the {\emph{local density of
    states}} evaluated at the boundaries. For fixed BC this will then give rise to an
extra factor of $\om^2$ (from the squared wavefunction)  and then the kinetic theory prediction matches   
with those given above.

We note that the density of states in Fig.~(\ref{dosdis}) show
apparent  gaps in the middle ranges of $\om$ for $d=2,3$. These might be
expected to disappear when the size of the system goes to infinity when there
should be large regions containing only masses of one type
\cite{casher71,oconnor74}. These regions will however be rare.   
In Fig.~(\ref{dosdispin}) we show plots of the density of states for the 
ordered and disordered harmonic lattices in the presence of pinning. 
In this case the gaps in the spectrum are more pronounced and, for large
enough values of $k_o$ and $\Delta$, may be present even in the thermodynamic
limit.

\section{Results from Numerics and Simulations}
\label{sec:results}
We now present the numerical and simulation results for transmission
coefficients, heat current density, temperature profiles and IPRs for the disordered
harmonic lattice in various dimensions. The numerical scheme for
calculating $J$ is
both faster and more accurate than  nonequilibrium simulations. Especially, for strong disorder,
equilibration times in nonequilibrium simulations become very large and in such
cases only the numerical method can be used.  However we  also show some
nonequilibrium simulation results. Their almost perfect agreement with the
numerical results provides  additional confidence in the accuracy of our
results. 
In Sec.~(\ref{twod}) we give the results for the $2D$ lattice for the
unpinned case with both fixed and free boundary conditions and then
for the pinned case. In Sec.~(\ref{threed}) we present the results for
the three dimensional case with and without substrate pinning potentials.

\ni \subsection{Results in two dimensions}
\label{twod}
In this section we consider $N\times N$ square lattices 
with periodic BCs in the $\nu=2$ direction and
either fixed or free BCs in the conducting direction ($\nu=1$). 
One of the interesting questions here is as to how the three
properties for the $1D$ case discussed in Sec.~(\ref{sec:general}) get
modified for the $2D$ case. 

\ni \subsubsection{Disordered $2D$ lattice without pinning}

{\emph {Fixed BC}}:  we have computed the transmission
coefficients and the corresponding heat currents for different
values of $\Delta$ and  for system sizes from $N=16-1024$. The number of averages
varied from over $100$ samples for $N=16$ to about two samples for
$N=1024$.  In
Figs.~(\ref{tw2dfix0.95},\ref{tw2dfix0.8},\ref{tw2dfix0.2}) we plot the disorder
averaged transmission coefficient for three different disorder strengths,
$\Delta=0.95$, $\Delta =0.8$ and $\Delta=0.2$, 
for different system sizes. The corresponding plots of IPRs as a
function of normal mode frequency $\om_p$, for 
single disorder realizations, are also given. From the IPR plots we get
an idea of the typical range of allowed normal mode frequencies and
their degree of localization. Low IPR values which scale as $N^{-2}$
imply extended states while large IPR values which do not change much
with system size denote localized states. In Fig.~(\ref{tw2dfix0.8})
we also show typical plots of small IPR and large IPR
wavefunctions. From
Figs.~(\ref{tw2dfix0.95},\ref{tw2dfix0.8},\ref{tw2dfix0.2}) we make
the following observations:

(i) As expected we see significant transmission only over the range of
 frequencies with extended states. Thus in Fig.~(\ref{tw2dfix0.95})
 for $\Delta =0.95$ we see that, 
 while there are normal modes in the range $\omega \approx (0-12)$, 
 transmission is appreciable only 
 in the range $\om \approx (0-1.5)$ and this is also roughly the range
 where the  IPR data shows a $N^{-2}$ scaling behaviour. 
This can also be seen in Fig.~(\ref{tw2dfix0.8}) where the inset shows
 the decay of $T(\om)$ in the localized region. 
 Unlike the  $1D$ case we  see a very weak dependence on
 system size of the upper frequency cut-off $\om_c^L$ beyond which
 states are localized and transmission is  negligible. As discussed
 earlier,  localization theory   predicts $\om_c^L \sim (\ln N)^{-1/2}$
but this may be difficult to observe numerically.
The overall transmission function ${T}_N (\om)$ decreases
with increasing system size, with $T(\om) \sim 1/N$ at higher
 freqencies  and $T(\om) \sim N^0 $ at the lowest frequencies.

(ii) In Fig.~(\ref{tw2dfix0.2}) we have also plotted $T(\om)$ for
the ordered binary mass case and we note that over a range of small
frequencies, 
$T(\om)$ for the disordered case is very close to
the  curve for the ordered case, which means that these modes are
ballistic. As expected from the arguments in Sec.~(\ref{sec:general}) we
roughly find $T(\om) \sim \om^3$ at small frequencies. The 
remaining transmitting states are either diffusive (with a $1/N$
scaling) or are in the cross-over
regime between diffusive and ballistic and so do not have a simple
scaling. 

We next look at the the integrated transmission which gives the net
heat current. 
The system size dependence of the disorder averaged current $J$ for different values of
$\Delta $ is shown in Fig.~(\ref{jvsn2dfix}). For the case $\Delta
=0.2$, we also show  simulation results and one can see that there is
excellent agreement with the numerical results.
For $\Delta=0.2$ we get
an exponent $\mu \approx 0.6$ which is close to the value obtained
earlier in \cite{lee05} for a similar disorder
strength. However with increasing disorder  
we see  that this value changes and seems to settle to around 
$\mu \approx 0.75$. 
It seems reasonable to expect (though we have no rigorous arguments)
that there is only one asymptotic exponent and for small disorder one
just needs to go to very large system sizes to see the true value.   
In Fig.~(\ref{temp2dfix}) we show  temperature profiles obtained
from simulations for lattices  of different sizes  with
$\Delta=0.2$. The jumps at the boundaries indicate that the asymptotic
size limit has not yet been reached. This is consistent with our result
that the exponent $\mu$ obtained at $\Delta=0.2$ is different from
what we believe is the correct asymptotic value (obtained at larger values of $\Delta$).   
We do not have temperature plots at strong disorder where simulations
are difficult. 

Thus contrary to the arguments in Sec.~(\ref{sec:general}) which predicted $J
\sim (\ln N)^{-1/2} N^{-1}$ we find a much larger current scaling as
$J \sim N^{-0.75}$. It is possible that one needs to go to larger
system sizes to see the correct scaling.

{\emph{Free BC}}: In this case from the arguments in
Sec.~(\ref{sec:general}) we expect ballistic states to contribute
most significantly  to the current density giving $J \sim
N^{-2/3}$. 

In Figs.~(\ref{tw2dfree0.8},\ref{tw2dfree0.2}) we plot the disorder
averaged transmission
coefficient for $\Delta=0.8$ and $\Delta =0.2$ for different system
sizes. Qualitatively these results look very similar to those for
fixed boundaries. However transmission is now significantly larger in
the region of extended states. The behaviour at  frequencies $\om \to
0$  is also 
different and we now find  $T(\om) \sim \om$ in contrast to $T(\om)
\sim \om^3$ for fixed boundaries.
From the plots of IPRs in Fig.~(\ref{tw2dfree0.8}) we note that there
is not much qualitative difference with the fixed boundary plots
except in the low frequency region (see below).

The system size dependence of the disorder averaged current $J$ for two different values of
$\Delta $ is shown in Fig.~(\ref{jvsn2dfree}). For $\Delta=0.2$ we get
an exponent $\mu \approx 0.5$ 
while for the stronger disorder case $\Delta=0.8$  we see a 
 different exponent $\mu \approx 0.6$. Again we
believe that the strong disorder value of $\mu=0.6$ is closer to
the value of the true asymptotic exponent. This value  is
close to the expected $\mu = 2/3$ for free BC and significantly
different from the value obtained for fixed 
BC ($\mu \approx 0.75$). 
Thus the dependence of the value of $\sigma$ on boundary
conditions  exists even in the $2D$ case.

For the case of free BCs, we find that the values of  $T(\om)$ in the diffusive
regime matches with those for fixed BCs but are completely
different in the ballistic regime. This is seen in
Fig.~(\ref{leff2d}) where we plot the
effective mean free path $l_{\rm eff}(\om)=NT(\om)/w^{d-1}$ in the
low-frequency region [this is obtained by comparing Eq.~(\ref{jeq}) with the
  kinetic theory expression for conductivity Eq.~(\ref{kinetick})]. 
For free BC,  $l_{\rm eff}$ is roughly consistent with 
the kinetic theory prediction $l^{-1}_{\rm eff} \sim
N^{-1}+{\ell_K}^{-1}(\om)$ but the behaviour for fixed BC is very
different. The inset of Fig.~(\ref{leff2d})  plots  
$l_{\rm eff}$ for the equal mass ordered case and we find that in the
ballistic regime it  is very close to the disordered case, an
input that we used in the heuristic derivation. 
The numerical data also confirms  
that for small $\om$, $T(\om) \sim \om$ for free BCs and as $\om^3$
for fixed BCs.  
The transmission for fixed BC shows rapid oscillations which 
increase with system size, and arise from scattering and interference
of waves at the interfaces.

\ni \subsubsection{Disordered $2D$ lattice with pinning}

We now study the effect of introducing a harmonic pinning potential
at all sites of the lattice. It is expected that this will cut off low
frequency modes and hence one should see strong localization effects. The
localization length $\ell$ will decrease both with increasing $\Delta$ and
increasing $k_o$ (in $1D$ heuristic arguments give $\ell \sim
1/(\Delta^2 k_o)$ \cite{dharleb08}).   
In Figs.~(\ref{tw2dpin10},\ref{tw2dpin02}) we plot the transmission
coefficients for two cases with on-site potentials $k_o=10.0$ and
$k_o=2.0$ respectively, and $\Delta=0.4$. We also plot the IPR in
Fig.~(\ref{tw2dpin10}). Unlike in the unpinned case
we now find that the transmission  coefficients are much smaller and
fall more rapidly with system size. 

From the plot of $P^{-1}$
we find that for all the modes, the value 
of $P^{-1}$ does not change much with system size which implies 
that all modes are localized. The allowed frequency
bands  correspond to the  transmission bands.
The two wavefunctions plotted in Fig.~(\ref{tw2dpin10}) correspond
to one relatively small and one large $P^{-1}$ value  
and clearly show that both states are localized.

The system size dependence of the
integrated current is shown in Fig.~(\ref{jvsn2donpin}) for the two
parameter sets. The values of $\mu \approx 1.6,~ 3.65$ for the two
sets indicate that at large enough length scales one will get a
current falling exponentially with system size and hence we have an insulating
phase. In Fig.~(\ref{temp2donpin}) we plot the temperature profiles
for the set with $\Delta = 0.4, k_o=10.0$ . In this case it is
difficult to obtain steady state temperature profiles from simulations
for larger system sizes. The reason is that the temperature (unlike
current) gets contributions from all modes (both localized and
extended) and equilibrating the localized modes takes a long time.

\ni \subsection{Results in three dimensions}
\label{threed}
In this section we  mostly consider  $N\times N \times N$  lattices 
with periodic boundary conditions in the $\nu=2,3$ directions.
Some results
for $N\times N_2 \times N_3$ lattices with $N_2=N_3 < N$ will also be
described.
Preliminary results for the case of free BCs are given and indicate
that there is no dependence of the exponent $\mu$ on BCs. 
It is not clear to us whether this is related to the boundedness of the
fluctuations in $x_\bn$ and the decay of the correlations between $x_\bn$ and
$x_\bk$ (like $|\bn-\bk|^{-1}$) in $d=3$ and their growth (with $N$) in $d <
3$.

\ni \subsubsection{Disordered $3D$ lattice without pinning}
\label{3Dunpin}

{\emph{Fixed BC}}: we have used both the numerical approach
and simulations for sizes up to $32 \times 32 \times 32$ for which we
have data for $T(\om)$. For larger systems  the
matrices  become too big and we have not been able to use the
numerical approach. Hence, for larger system sizes we have only performed
simulations, including some on $N \times N_2 \times N_2$ lattices. For
these cases only the current $J$ is obtained.
The number of averages
varies from over $100$ samples for $N=16$ to two samples for
$N=64$.  In Figs.~(\ref{tw3dfixed0.8},\ref{tw3dfixed0.2}) we plot the disorder
averaged transmission coefficient for two different disorder strengths,
$\Delta=0.8$ and $\Delta =0.2$, 
for different system sizes. The corresponding plots of IPRs as a
function of normal mode frequency $\om_p$, for 
single disorder realizations, are also given. From the IPR plots we get
an idea of the typical range of allowed normal mode frequencies and
their degree of localization. Low IPR values which scale as $N^{-3}$
imply extended states while large IPR values which do not change much
with system size denote localized states. 

From Figs.~(\ref{tw3dfixed0.8},\ref{tw3dfixed0.2}) we make
the following observations.

(i) From the  $3D$ data it is clear the effect of localization is
weaker than in $1D$ and $2D$. Both for $\Delta = 0.2$ and $\Delta
=0.8$ we find that there is transmission over almost the entire range
of frequencies of the allowed normal modes. From the IPR plots we  see
that for $\Delta = 0.2$ most states are extended except for a small
region in the high frequency band-edge. For $\Delta=0.8$ the allowed modes form two
bands and one finds significant transmission over almost the full
range. At the band edges (except the one at $\om=0$) there are again localized states. It also
appears that there are some large IPR states interspersed within the
high frequency band. 
As in the $2D$ case and unlike the $1D$ case, the frequency range over which
transmission takes place does not change with system size, only the
overall magnitude of transmission coefficient changes.

(ii) The  plot of  $N T(\omega)$ in Fig.~(\ref{tw3dfixed0.8}) shows 
the nature of the extended states. 
The high frequency band and a portion of the lower frequency band have
the scaling $T(\omega) \sim N^{-1}$  
and hence corresponds to diffusive states. In the lower-frequency band
the fraction of  diffusive states seems to be increasing with system 
size but it is difficult to verify the $\om_c^K \sim N^{-1/4}$ scaling. 
The ballistic nature of the low-frequency states is confirmed 
 in Fig.~(\ref{tw3dfixed0.2}) where we see that $T(\om)$
for the binary-mass ordered and disordered lattices match for small
$\om$ [with a $T(\om) \sim \om^4$ dependence].

In Fig.~(\ref{jvsn3dunpin}) we show the
system size dependence of the disorder averaged current density $J$
for the two cases with weak disorder strength ($\Delta =0.2$) and
strong disorder strength 
($\Delta = 0.8$). The results for cubic lattices of sizes up to  $N=32$
are from the numerical method while the results for larger sizes are
from simulations. We find an exponent $\mu \approx 0.6$ at small
disorder and $\mu \approx 0.75$ at large disorder strength. 
As in the $2D$ case here too we believe that at small disorder, the
asymptotic system size limit will be reached at much larger system
sizes and that the exponent obtained at large disorder strength is
probably close to the true asymptotic value.    
The value ($\mu = 0.75$) does not agree with the  prediction
($J \sim N^{-1}$) made from the heuristic arguments in
Sec.~(\ref{sec:general}).  A study of larger system sizes is necessary
to confirm whether or not the asymptotic size limit has been reached.

The data point at $N=128$ for the set with $\Delta=0.2$ in
Fig.~(\ref{jvsn3dunpin}) actually corresponds to a lattice 
of dimensions $128 \times 48 \times 48$ and we believe that the
current value is very close to the expected fully $3D$ value. To see 
this point, we have plotted in Fig.~(\ref{scale3d}) results from
nonequilibrium simulations
with $N\times N_2 \times N_2$ lattices with $N_2 \leq N$.

Finally, in Fig.~(\ref{temp3dfix}) we show  temperature profiles (for
single disorder realizations) obtained
from simulations for lattices  of different sizes and with
$\Delta=0.2$. The jumps at the boundaries again indicate that the asymptotic
system size limit has not been reached even at the largest size.

{\emph{Free BC}}: In this case from the arguments in
Sec.~(\ref{sec:general}) we expect ballistic states to contribute
most significantly  to the current density giving $J \sim
N^{-3/4}$. 

In Fig.~(\ref{tw3dfree0.8}) we plot the disorder
averaged transmission
coefficient for $\Delta =0.8$ for different system
sizes. The transmission function is very close to that for the fixed
boundary case except in the frequency region corresponding to non-diffusive states.
At $\om \to 0$  we now expect, though it is hard to verify from the
data, that   $T(\om) \sim \om^2$ in contrast to $T(\om)
\sim \om^4$ for fixed boundaries.
 
The system size dependence of the disorder averaged current $J$ for two different values of
$\Delta $ is shown in Fig.~(\ref{jvsn3dunpin}). 
We find that the current values are quite close to the fixed BC case
and the exponent obtained at the largest system size studied for this
case is $\mu \approx 0.71$. 
This value  is close to the expected $\mu = 3/4$ for free BC.

We now compare the transmission coefficient for free and fixed BCs in
the ballistic regime.
This is plotted in
Fig.~(\ref{leff3d}) where we show the
effective mean free path $l_{\rm eff}(\om)=NT(\om)/w^{d-1}$ in the
low-frequency region. 
As in the $2D$ case we again find that for free BCs,  $l_{\rm eff}$ is
roughly consistent with  
the kinetic theory prediction $l^{-1}_{\rm eff} \sim
N^{-1}+{\ell_K}^{-1}(\om)$ and the behaviour for fixed BCs is very
different. The inset of Fig.~(\ref{leff3d})  plots  
$l_{\rm eff}$ for the equal mass ordered case and we find that in the
ballistic regime it  is very close to the disordered case.
The numerical data  confirms the input in our theory on the form of
$T(\om)$ for small $\om$, {\emph{i.e.}}~ $T(\om) \sim \om^2$ for free
BCs and as $\om^4$ 
for fixed BCs.  
The transmission for fixed BC shows rapid oscillations which
increase with system size, and arise from scattering and interference
of waves at the interfaces.   

\ni \subsubsection{Disordered $3D$ lattice with pinning}
\label{3Dpin}

For the pinned case,  we again use both the numerical method and
simulations for sizes up to
$N=32$. For $N=64$ only nonequilibrium simulation results are reported. 

In Figs.~(\ref{tw3donpin100.2},\ref{tw3donpin100.8}) we
plot the disorder averaged transmission coefficient for
$\Delta =0.2$ and $\Delta=0.8$ with  $k_o=10.0$.  
The corresponding IPRs $P^{-1}$ and scaled IPRs $N^3 P^{-1}$ are also shown.

From the IPR plots we notice that the spectrum of the $3D$ disordered
pinned chain has a similar  interesting  structure as in the $2D$ case
with two bands and a gap which is seen at strong disorder. However
unlike the $2D$ case where all states were localized, here  the IPR
data indicates that most states except  those  at the band edges are
diffusive. We see localized states at the band edges and also there
seem to be some localized states interspersed among the extended
states within the bands. 
The insets in Figs.~(\ref{tw3donpin100.2},\ref{tw3donpin100.8}) show
that there  is a reasonable $N^{-1}$ scaling of the transmission data
in most of the transmitting region. This is clearer at the larger
system sizes. Thus, unlike 
the unpinned case where low frequency 
extended states were ballistic or super-diffusive, here we find that
there is no transmittance at small ($\om \to 0$) frequencies and that
all states are diffusive. 

From the above discussion we expect Fourier's law to be valid in the
$3D$ pinned disordered lattice. 
The system size dependence of the disorder
averaged current $J$ for different disorder strengths is plotted in
Fig.~(\ref{jvsn3donpin}). For all the parameter sets the exponent
obtained is close to $\mu =1$ corresponding to a finite
conductivity and validity of Fourier's law. 
The temperature profiles plotted in Fig.~(\ref{temp3donpin}) 
have small boundary temperature jumps and indicate
that the asymptotic size limit has already been reached.

One might expect that at very strong disorder, all states should
become localized and then one should get a heat insulator. 
The parameter set corresponding to Fig.~(\ref{tw3donpin100.8})
corresponds to  strong disorder and for this we still find a 
significant fraction of extended states. 
Thus for the binary mass case it appears that there are always extended
states. 
We have some results for the case with a continuous mass distribution
( masses are chosen from a uniform distribution between $1-\Delta$ and
$1+\Delta$).
In this case we find that the effect of disorder is stronger and the
transmission at all frequencies is much reduced compared  to the binary mass
case.  However we cannot see the exponential decrease in transmission
with system size and so it is not clear if an insulating behaviour is obtained.
Further numerical studies are
necessary to understand the asymptotic behaviour.

\section{Discussion}
\label{sec:discussion}

\begin{table}
\begin{tabular}{|c|c|c|c|c|}
\hline
\multicolumn{1}{|c|}{} &
\multicolumn{2}{|c|}{$d=2$} & 
\multicolumn{2}{|c|}{$d=3$}  \\
\cline{2-5}
 & Analytical & Numerical & Analytical & Numerical  \\ 
\hline 
Pinned & $\exp{(-b N)} $ & $N^{-3.7}$ & $N^{-1}$ & $N^{-1.0}$ 
  \\ 
\hline 
Fixed & $N^{-1}(\ln{N})^{-1/2} $ & $N^{-0.75}$ & $N^{-1}$ & $N^{-0.75}$   \\
\hline 
Free & $N^{-2/3}$ & $N^{-0.6}$ & $N^{-3/4}$ & $N^{-0.71}$   \\
\hline
\end{tabular}
\caption{The table  summarizes the main results of the paper. The
  numerical (and nonequilibrium simulation) results obtained in the paper are
  compared, in two and three dimensions, with the analytical
  predictions obtained from our heuristic arguments. The error bar for the
  numerically obtained exponent values is of the order $ \pm 0.02$. This error
  is estimated   from the errors in the last few points of the
  $J$-versus-$N$ data. NB: The system sizes used may well be far from asymptotic.}  
\label{table}
\end{table}

We have studied heat conduction in isotopically disordered harmonic
lattices with scalar displacements in two and three dimensions. 
The main question addressed is the system size dependence of the
heat current, which is computed using Green's function based numerical
methods as well as nonequilibrium simulations. We have tried to
understand the size dependence by  
looking at the phonon transmission function $T(\om)$ and examining the 
nature of the energy transport in  different  frequency regimes. 
We also described a heuristic analytical calculation based on 
localization theory and kinetic theory and compared their
predictions with our numerical and simulation results.  
This comparison is summarized in Table~(\ref{table}).

The most interesting findings of this work are: \\
(i) For the unpinned system we find that in $2D$ there are a
large number of localized modes for which phonon transmission is
negligible. In $3D$ the number of localized modes is much
smaller. The extended modes are either diffusive or ballistic.
Our analytic arguments  show that the
contribution of ballistic modes to conduction is 
dependent on BCs and is strongly suppressed for the case of fixed BCs, the more
realistic case. In $3D$ this leads to diffusive modes dominating for
large system sizes and Fourier's law is satisfied. 
Thus a finite heat conductivity is obtained for the $3D$
disordered harmonic crystal without the need of invoking
anharmonicity as is usually believed to be necessary \cite{callaway59,ziman72}. 
This is similar to what one obtains when one adds stochasticity to the time
evolution in the bulk as shown by \cite{basile06}.
Our numerical results verify the predictions for free BCs and we
believe that much larger system sizes are necesary to verify the fixed
BC results ( this is also the case in $1D$ \cite{dhar01,roydhar08}).\\
(ii) In two dimensions the pinned disordered lattice shows clear evidence of
localization and we obtain a heat insulator with exponential decay 
of current with system size. \\
(iii) Our result for the $3D$ pinned disordered lattice provides the
first microscopic verification of Fourier's law in a three dimensional
system. For the binary mass distribution we do not see a transition to
insulating behaviour with increasing disorder. For a continuous mass
distribution we find that the current is much smaller (than the binary mass
case with the same value of $\Delta$) but it is not clear whether all states
get localized and if an insulating phase exists.

{\bf Acknowledgements:} We thank G. Baskaran, Michael Aizenman, Tom Spencer and especially David Huse for 
useful discussions. We also thank Srikanth Sastry and Vishwas Vasisht for use of
computational facilities. The research of J. L. Lebowitz was supported
by NSF grant No. DMR0802120 and by AFOSR grant No. FA9550-07.

\appendix

 \section{ KINETIC THEORY}
\label{sec:appa}

Kinetic theory becomes valid in the limit of small disorder. Its basic object is the
Wigner function, $f$, which describes the phonon density in phase space and
is governed by the transport equation
\begin{equation}\label{B.1}
\frac{\partial}{\partial t} f(r,k,t) + \nabla \omega(k) \cdot
\nabla_r f(r,k,t) = \mathcal{C} f(r,k,t)\,.
\end{equation}
Here $r\in \mathbb{R}^d$ (boundary conditions could be imposed),
$k\in[-\pi,\pi]^d$ is the wave number of the first Brioullin zone,
$\omega$ is the dispersion relation of the constant mass harmonic crystal,
and $\mathcal{C}$ is the collision operator. It acts only on wave
numbers and is given by
\begin{eqnarray}\label{B.2}
\mathcal{C} f(k)=&& ( 2\pi)^{-d+1}\omega(k)^2 \Delta^2
\int_{[-\pi,\pi]^d} dk' \nn \\
&& \delta\big(\omega(k)-\omega(k')\big)\big(f(k')-f(k)\big)\,.
\end{eqnarray}
We refer to \cite{lukk07} for a derivation. In the range of validity of
(\ref{B.1}), (\ref{B.2}) we can think of phonons as classical
particles with energy $\omega$ and velocity $\nabla \omega(k)$. 
They are scattered by the impurities from $k$ to $dk'$ with the rate
\begin{equation}\label{B.3}
(2\pi)^{-d+1} \omega(k)^2 \Delta^2
\delta\big(\omega(k)-\omega(k')\big) dk'.
\end{equation}
 Collisions are
elastic. We distinguish\medskip\\
\textit{(i) no pinning potential}. Then for small $k$ one has
$\omega(k)=|k|$ and $|\nabla\omega(k)|=1$. From (\ref{B.2}) the
total scattering rate behaves as $|k|^{d+1}$. This is the basis for the
discussion in connection with Eq.~(\ref{kinetick}).\medskip\\
\textit{(ii) pinning potential}. In this case
$\omega(k)=\omega_0+k^2$ for small $k$. The prefactor in (\ref{B.2}) can be
replaced by $\omega^2_0$. The velocity is $k$ and the scattering is
isotropic with rate $|k|^{d-2}$. Thus the diffusion coefficient
results as $D(k)\cong |k|^{-d+4}$ which vanishes as $|k|\to 0$ for
$d=2,3$. Hence there is no contribution to the thermal conductivity from the
small $k$ modes.

\ni \section{Transfer matrix approach}
\label{sec:appb}
We now outline steps by which $\mathcal{T}_N$ can be expressed in
forms which are amenable to accurate numerical evaluation. We
will give results whereby we express $\mathcal{T}_N$ in terms of product
of random matrices. These are related to the Green's function and
transfer matrix methods used earlier in the calculation of localization
lengths in disordered electronic systems \cite{mackinnon83}.  
Some related discussions for the phonon case can be found in \cite{hori68}.
For heat conduction in one dimensional disordered chains, the transfer
matrix approach has  been shown to be very useful in obtaining
analytic as well as accurate numerical  results and here we  study the
extension of this to higher dimensions.  

The transmission coefficient is given by $\mathcal{T}_N(\om)= 4 Tr [
  \mathcal{I}_L(\om) \mathcal{G}^+(\om) \mathcal{I}_R(\om) \mathcal{G}^-(\om)$ where  
$\mathcal{G}^+(\om)=[-\om^2 \mathcal{M} + \mathcal{V}
  -\mathcal{S}^+_L
  -\mathcal{S}^+_R]^{-1}$, $\mathcal{G}^-=[\mathcal{G}^+]^*$,
$\mathcal{I}_{L,R}=Im[\mathcal{S}^+_{L,R}]$ and we now specify the form of
  $\mathcal{S}^+_{L,R}$ corresponding to the equations of motion in
  Eqs.~(\ref{eqm}). Note that we have transformed to
  dimensionless variables 
$\om \to \om/\Omega, \mathcal{M}\to \mathcal{M}/\bar{m},~
  \mathcal{V}\to\mathcal{V}/k,~ \gamma \to \gamma/(\bar{m} \Omega)$
  where $\Omega=(k/\bar{m})^{1/2}$. 
We are considering heat conduction in the $\nu=1$
direction of a $d$-dimensional lattice with particles on the layers 
$n_1=1$ and $n_1=N$ being connected to heat baths at temperatures $T_L$
and $T_R$ respectively. 
The matrices $\mathcal{S}^+_{L}$ and $\mathcal{S}^+_R$ represent the
  coupling of the system to the left and right reservoirs
  respectively, and can   
be written as $N\times N$ block matrices where each block is a
$N' \times N'$ matrix. The block structures are as follows:
\bea
&& \mathcal{S}^+_L=\left( \begin{array}{cccc}
{\Sigma_L^+}& 0 & ...& 0 \\
0 & 0 & .~.~. &0  \\
0 & 0 & .~.~. &0  
\end{array} \right) 
~,
\mathcal{S}^+_R=\left( \begin{array}{cccc}
0 & 0 & .~.~. & 0 \\
0 & 0 & .~.~. &0  \\
0 & 0 & ... &\Sigma_R^+  
\end{array} \right)~,~~  \label{slr1}
\eea
where
\bea
\Sigma^+_L= \Sigma^+_R= i \gamma \om I ~,\label{slr2}
\eea
$I$ is a $N' \times N'$ unit matrix, and $0$ is a $N' \times N'$
matrix with all elements equal to zero. Similarly the matrices
$\mathcal{M}$ and $\mathcal{V}$ have the following block structure:
 \bea
\mathcal{M}=\left( \begin{array}{cccc}
M_1& 0 & ...& 0 \\
0 & M_2 & .~.~. & 0 \\
0 & 0 & .~.~. &0  \\
0 & 0 & .~.~ &M_N  
\end{array} \right), 
~
\mathcal{V}=\left( \begin{array}{cccc}
\Phi & -I & .~.~.& 0 \\
-I & \Phi & .~.~. & 0 \\
0 & 0 & .~.~. &0  \\
0 & 0 & ..-I~ &\Phi  
\end{array} \right) ,~~ 
\eea
where $M_n$ denotes the diagonal mass-matrix for the $n_1=n$ layer and
$\Phi$ is a force-constant matrix whose off-diagonal terms
correspond to coupling to sites within a layer. 
Hence the matrix $\mathcal{G}^{-1}=[-\mathcal{M} \om^2 +
  \mathcal{V}-\mathcal{S}_L^+-\mathcal{S}^+_R]$ has the following structure:
 \bea
[\mathcal{G}]^{-1}=\left( \begin{array}{ccccc}
a_1& -I & 0& ...& 0 \\
-I & a_2 & -I&0~... & 0 \\
...&...&...&...&...\\
0 & ...~0 & -I& ~a_{N-1} &-I  \\
0 & ... & 0 & -I &a_N  
\end{array} \right)~, 
\eea
where $a_l=-M_l \om^2 +\Phi -\delta_{l,1} \Sigma^+_L -\delta_{l,N}
\Sigma^+_R$. 
Now defining $\Gamma_{L,R}=Im[\Sigma^+_{L,R}]$ and with   
the  form of $\mathcal{S}^+_{L,R}$ given in Eqs.~(\ref{slr1}),(\ref{slr2}), we find that the expression
for the transmission coefficient reduces to the following form:
\bea
{\mathcal{T}}_N(\om) &=& 4~Tr [ {\Gamma}_L (\om) {G}^+_N (\om )
  {\Gamma}_R (\om ) {G}^-_N(\om) ]~, \label{transmeq}
\eea
where $G^+_N$ is the $(1,N)^{\rm th}$ block element of
$\mathcal{G}$ and $G^-_N=[G^+_N]^\dagger$. We now show that $G^+_N$
satisfies  a simple recursion equation.

We first introduce some notation.
Let ${\mathcal{Y}}^{(l,l+n-1)}$ with $1 \leq n \leq N-l+1$  denote a $n \times
n$ tridiagonal block matrix whose diagonal entries are
$a_l,a_{l+1},...a_{l+n-1}$, where each $a_l$ is a $N' \times N'$ matrix.
The off-diagonal entries are given by $-I$.
For an arbitrary block matrix $\mathcal{A}^{(l,m)}$, 
 ${\mathcal{A}}^{(l,m)}_{(i,j)}$ will denote the block sub-matrix of 
$\mathcal{A}^{(l,m)}$ beginning with $i^{\rm th}$ block row 
and column and ending with the $j^{\rm th}$ block row and column, while
${A}^{(l,m)}_{i,j} $ will denote the ${(i,j)}^{th}$ block element of
$\mathcal{A}^{(l,m)}$. Also  $\mathcal{I}_n$ will denote a $n \times
n$ block-diagonal matrix with diagonal elements $I$.

The inverse of ${\mathcal{Y}}^{(1,N)}$ is denoted by 
$[\mathcal{Y}^{(1,N)}]^{-1}={\mathcal{G}}^{(1,N)}$ and satisfies the equation:
\bea
{\mathcal{Y}}^{(1,N)}~{\mathcal{G}}^{(1,N)} &=& {\mathcal{I}}_{N}
\nn~. \label{invYeq}
\eea
According to our  notation we have
${\mathcal{G}}^{(1,N)}=\mathcal{G}^+$ and $G^{(1,N)}_{1,N}=G^+_N$. 
The matrix $\mathcal{Y}^{(1,N)}$ 
has the following structure: 
\bea
{\mathcal{Y}}^{(1,N)}=
\left( \begin{array}{cc}
{\mathcal{Y}}^{(1,N-1)}&{\mathcal{W}}_{N} \\
{\mathcal{W}}_{N}^T & a_{N} 
\end{array} \right)~,
\label{YN} 
\eea
where ${\mathcal{W}}_{N}^T=(0,0,...,-I)$ is a ${1 \times N-1}$ block vector.
We then write Eq.~(\ref{invYeq}) in the form
\bea
\left( \begin{array}{cc}
{\mathcal{Y}}^{(1,N-1)}&{\mathcal{W}}_{N} \\
{\mathcal{W}}_{N}^T & a_{N} 
\end{array} \right)
\left( \begin{array}{cc}
{\mathcal G}^{(1,N)}_{(1,N-1)} & {\mathcal{U}}_{N} \\
{\mathcal{U}}_{N}^T & G^{(1,N)}_{N,N}
\end{array} \right) = 
\left( \begin{array}{cc}
{\mathcal{I}}_{N-1} & 0 \\
0 & I
\end{array} \right),~~
\label{invrseN}
\eea
where ${\mathcal{U}}_{N}^T =
[G^{(1,N)T}_{1,N},G^{(1,N)T}_{2,N}...,G^{(1,N)T}_{N-1,N}]$ is a $1
\times N-1$ block vector.   
From  Eq~({\ref{invrseN}}) we get the following four equations:
\bea
&{\mathcal{Y}}^{(1,N-1)}~{\mathcal G}^{(1,N)}_{(1,N-1)} +
    {\mathcal{W}}_{N}~{\mathcal{U}}_{N}^T  
= {\mathcal{I}}_{N-1}~, \nn \\
&{\mathcal{W}}_{N}^T~{\mathcal G}^{(1,N)}_{(1,N-1)} + a_{N}~{\mathcal{U}}_{N}^T = 0~, \nn \\
&{\mathcal{Y}}^{(1,N-1)}~{\mathcal{U}}_{N} + {\mathcal{W}}_{N}~G^{(1,N)}_{N,N} = 0~, \nn \\
&{\mathcal{W}}_{N}^T~{\mathcal{U}}_{N} + a_{N}~G^{(1,N)}_{N,N} = I~.
\label{step1}
\eea
Noting that $[\mY^{(1,N-1)}]^{-1}=\mG^{(1,N-1)}$ we get, 
using the third equation  above and the form of $\mW_{N}$:
\bea
\mU_{N}&=&-\mG^{(1,N-1)} \mW_{N} G^{(1,N)}_{N,N}~, \nn \\
{\rm or}~~
G^{(1,N)}_{i,N} &=&
{G}^{(1,N-1)}_{i,N-1}G^{(1,N)}_{N,N}~,~{\rm for}~~ i=1,2,...,N-1.~~~~ 
\label{step2}
\eea 
From the fourth equation in Eq.~(\ref{step1}) we get:
\bea
G^{(1,N)}_{N-1,N} &=& a_{N}~G^{(1,N)}_{N,N} - I~.
\label{step3} 
\eea
We will now use Eqs.~(\ref{step2}),(\ref{step3}) to obtain a recursion for
$G^{(1,N)}_{1,N}$ $=G_{N}^+$ in Eq.~(\ref{transmeq})], which is the
main object of interest. Let us define 
$P^{(l,n)}=[G^{(l,n)}_{1,n-l+1}]^{-1}$ where
$\mathcal{G}^{(l,m)}=[\mathcal{Y}^{(l,m)}]^{-1}$. Then setting $i=1$ in Eq.~(\ref{step2}) and
taking an inverse on both sides we get:
\bea
P^{(1,N)}=[G^{(1,N)}_{N,N}]^{-1}~P^{(1,N-1)}. 
\eea
Setting $i=N-1$ in Eq.~(\ref{step2}) we get
$G^{(1,N)}_{N-1,N}=G^{(1,N-1)}_{N-1,N-1}G^{(1,N)}_{N,N}$
and using this in Eq.~(\ref{step3}) we get
$[G^{(1,N)}_{N,N}]^{-1}=[a_{N}- G^{(1,N-1)}_{N-1,N-1}]$~. 
Inserting this in the above equation we finally get our required
recursion relation: 
\bea
P^{(1,N)}=a_{N} P^{(1,N-1)}- P^{(1,N-2)}~.\label{Precur}
\eea
The initial conditions for this recursion are: $P^{(1,0)}=I_M$ and
$P^{(1,1)}=a_1$. 
By proceeding similarly
as before we can also obtain the following recursion relation:
\bea
P^{(n,N)}=P^{(n+1,N)} a_1- P^{(n+2,N)} \label{Precur2}~,~~~n=1,2,...,N-1~,
\eea 
and $P^{(1,N)}$ can be recursively obtained using the initial
conditions $P^{(N+1,N)}=I_M$ and $P^{(N,N)}=a_N$. 
Given the set $\{a_i\}$, by iterating either of the above
equations one can 
numerically find $P^{(1,N)}$ and then invert it to find
$G_{1,N}^{(1,N)}$. However this scheme  runs into accuracy problems since the
numerical values of the  matrix elements of the iterates grow
rapidly. We describe now a different way of performing the recursion
which turns out to be numerically more efficient. 
We first define
\bea
r_N=P^{(1,N)} [ P^{(1,N-1)}]^{-1}  ~.
\eea
From Eq.~(\ref{Precur}) we immediately get:
\bea
r_N=a_N-\f{1}{r_{N-1}} \label{erecur}~,
\eea
with the initial condition $r_1=a_1$. 
Then $G_{1,N}^{(1,N)}$ is given by:
\bea
G_{1,N}^{(1,N)}&=&[P^{(1,N)}]^{-1}= [r_N r_{N-1}...r_1]^{-1} \nn \\
 &=& r_1^{-1} r_2^{-1}...r_N^{-1} \label{Geff}~.
\eea
This form where at each stage $r_l^{-1}$ is evaluated turns out to be
numerically more accurate.

Finally we show that one can  express $G_{1,N}^{(1,N)}$ in the form of a product of matrices.
The product form is such that the system and reservoir contributions are separated.
First we note that the form of the matrices $a_l$ for our specific
problem is:
$a_l = c_l-\delta_{l,1}\Sigma_1-\delta_{l,N} \Sigma_N$ where $c_l = -M_l \om^2 + \Phi$.
We define system-dependent matrices $Q^{(1,n)},~Q^{(n,N)}$ 
by replacing $a_1,a_N$ by $c_1,c_N$ in the recursions for $P$s'.  
Thus $Q^{(1,n)}=P^{(1,n)}(a_1\to c_1,a_N\to c_N)$ and 
$Q^{(n,N)}=P^{(n,N)} (a_1 \to c_1, a_N \to c_N )$. Clearly $Q$s'
satisfy the same recursion as the $P$s' with $a_l$ replaced by $c_l$.
Then using Eqs.~(\ref{Precur}),(\ref{Precur2}), and similar equations
for the $Q$s' we get:
\bea
&&P^{(1,N)} \nn \\
&=& Q^{(1,N)}-Q^{(2,N)}~ \Sigma_1-
\Sigma_N~Q^{(1,N-1)}+\Sigma_N~Q^{(2,N-1)}~\Sigma_1 \nn \\
&=& (1~~~ -\Sigma_N) \left( \begin{array}{cc}
Q^{(1,N)} & -Q^{(2,N)} \\
Q^{(1,N-1)} & -Q^{(2,N-1)}  
\end{array} \right)
\left( \begin{array}{c}
1 \\ \Sigma_1
\end{array} \right)~. \label{peqn}
\eea
From the recursion relations for the $Q$s' it is easy to see
that
\bea
&& \left( \begin{array}{cc}
Q^{(1,N)} & -Q^{(2,N)} \\
Q^{(1,N-1)} & -Q^{(2,N-1)}  
\end{array} \right) \nn \\
&=& \left( \begin{array}{cc}
a_N & -I \\
I & 0 
\end{array} \right) 
\left( \begin{array}{cc}
Q^{(1,N-1)} & -Q^{(2,N-1)} \\
Q^{(1,N-2)} & -Q^{(2,N-2)}  
\end{array} \right) \nn \\
&=&\hat{T}_N \hat{T}_{N-1}...\hat{T}_1~, 
\eea
where
\bea
\hat{T}_l&=&  \left( \begin{array}{cc}
a_l & -I \\
I & 0 
\end{array} \right)~. \label{qeqn}
\eea 
We  then obtain $G^+_N=[P^{(1,N)}]^{-1}$.

In our numerical calculations we use the recursion relations in
Eqs.~(\ref{erecur}),(\ref{Geff}) to evaluate the required Green's
function. Computing the trace in Eq.~(\ref{transmeq}) then gives us the
transmission coefficient as a function of frequency.


\begin{widetext}

\begin{figure*}[ht]
\centering
\subfigure[Free boundaries]{
\includegraphics[scale=0.18]{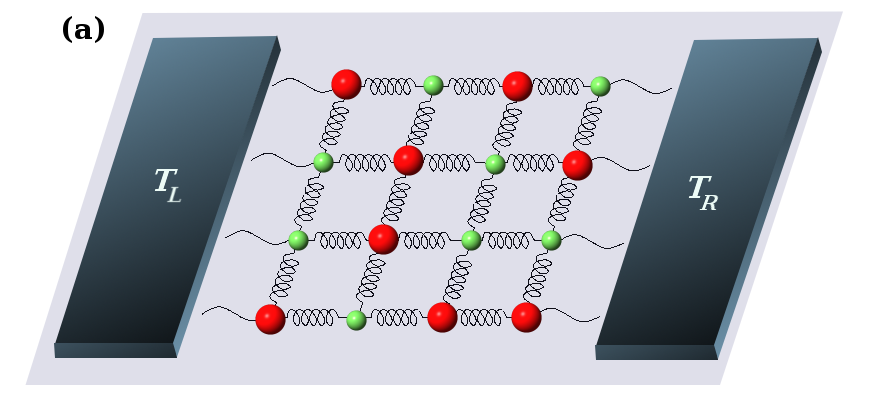}
\label{fig:subfig1}
}
\subfigure[Fixed boundaries]{
\includegraphics[scale=0.18]{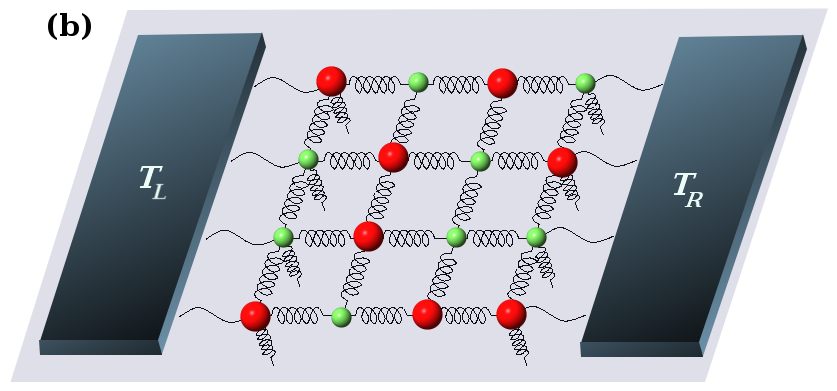}
\label{fig:subfig2}
}
\subfigure[Pinned lattice]{
\includegraphics[scale=0.18]{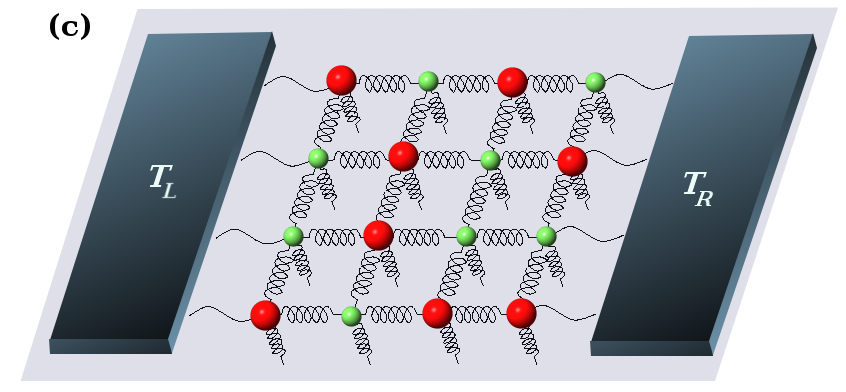}
\label{fig:subfig3}
}
\label{fig:subfigureExample}
\caption[Optional caption for list of figures]{A schematic diagram of
  a two-dimensional mass-disordered
  lattice of particles connected by harmonic springs and connected to heat
  baths at temperatures $T_L$ and $T_R$. 
Red and green colours indicate particles of different masses. 
Pinning refers to the presence of a spring attaching a particle to the
substrate.  
In (a) there is no pinning, in (b) boundary particles are pinned and in (c) 
all sites are pinned.} 
\label{scheme} 
\end{figure*}

\begin{figure}
\includegraphics[width=4in]{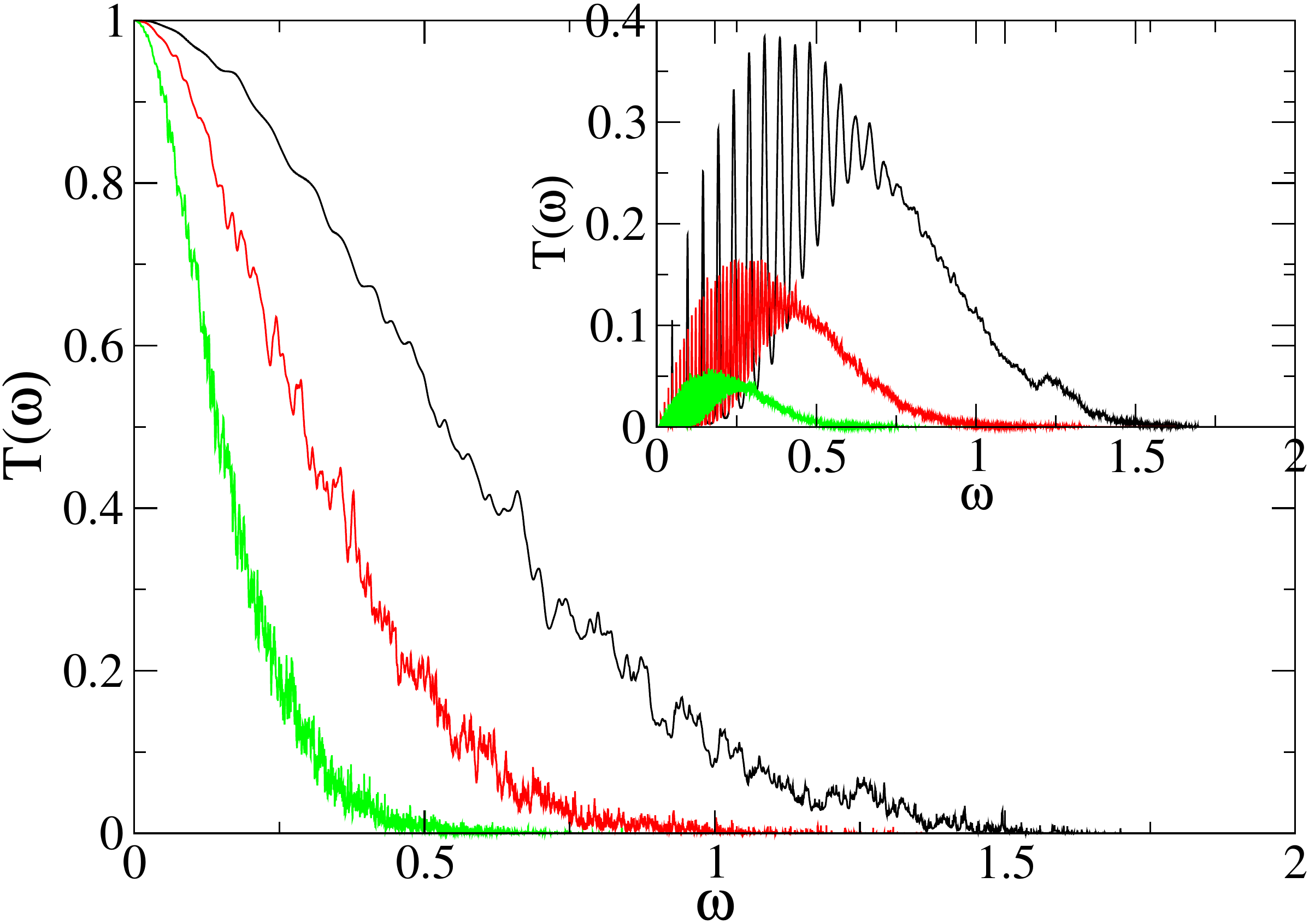}
\caption{(color online)
$1D$ unpinned case with both free and fixed (INSET)
  boundary conditions: plot of the disorder averaged transmission 
  $T(\om)$   versus $\omega$ for  $\Delta=0.4$. 
The various curves  (from top to   bottom) correspond to  lattices of sizes
$N=64,256,1024$  respectively.} 
\label{tw1d0.4}
\vspace{0.75cm}
\end{figure}

\begin{figure}
\includegraphics[width=4in]{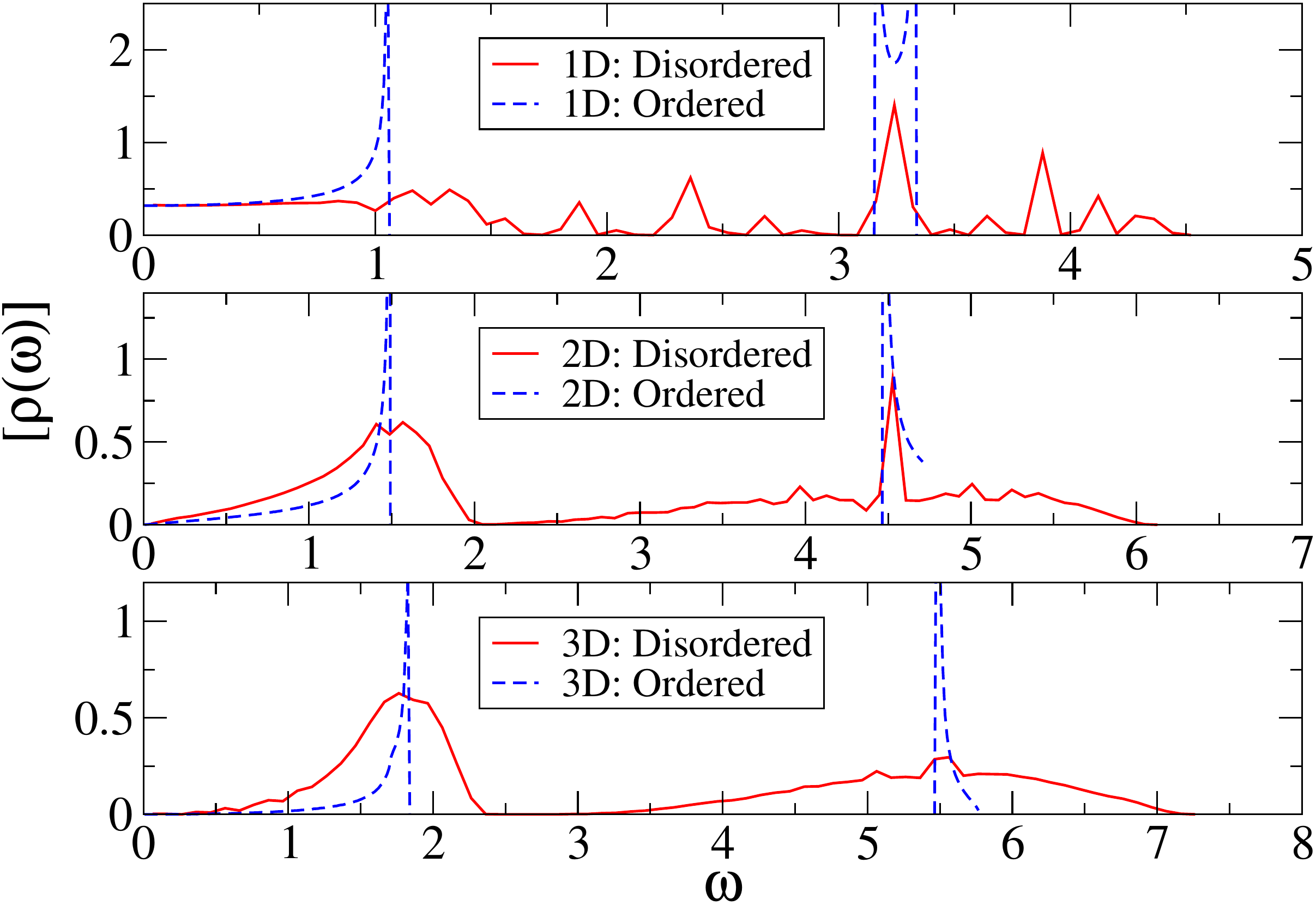}
\caption{(color online) Unpinned  lattices with fixed BC in one direction and periodic in all  others.  \\ 
 Disorder averaged density of states   obtained numerically from the eigenvalues
  of several disorder realizations in $1D, 2D$ and $3D$ for lattice
  sizes $N=4096,64,16$ respectively. Note that the low frequency
  behaviour is unaffected by disorder and one has  $\om^{d-1}$
   as $\om \to 0$. We set $\Delta=0.8, k=1$ and averaged over
 $30$ realizations in $1D$ and over $10$ realizations in $2D$ and
 $3D$. In $2D$ and $3D$ there is not much variation in $\rho(\om)$ for
 different disorder samples. Also shown are the density of states for the
  binary mass ordered lattices.}
\label{dosdis}
\vspace{0.75cm}
\end{figure} 

\begin{figure}
\includegraphics[width=4in]{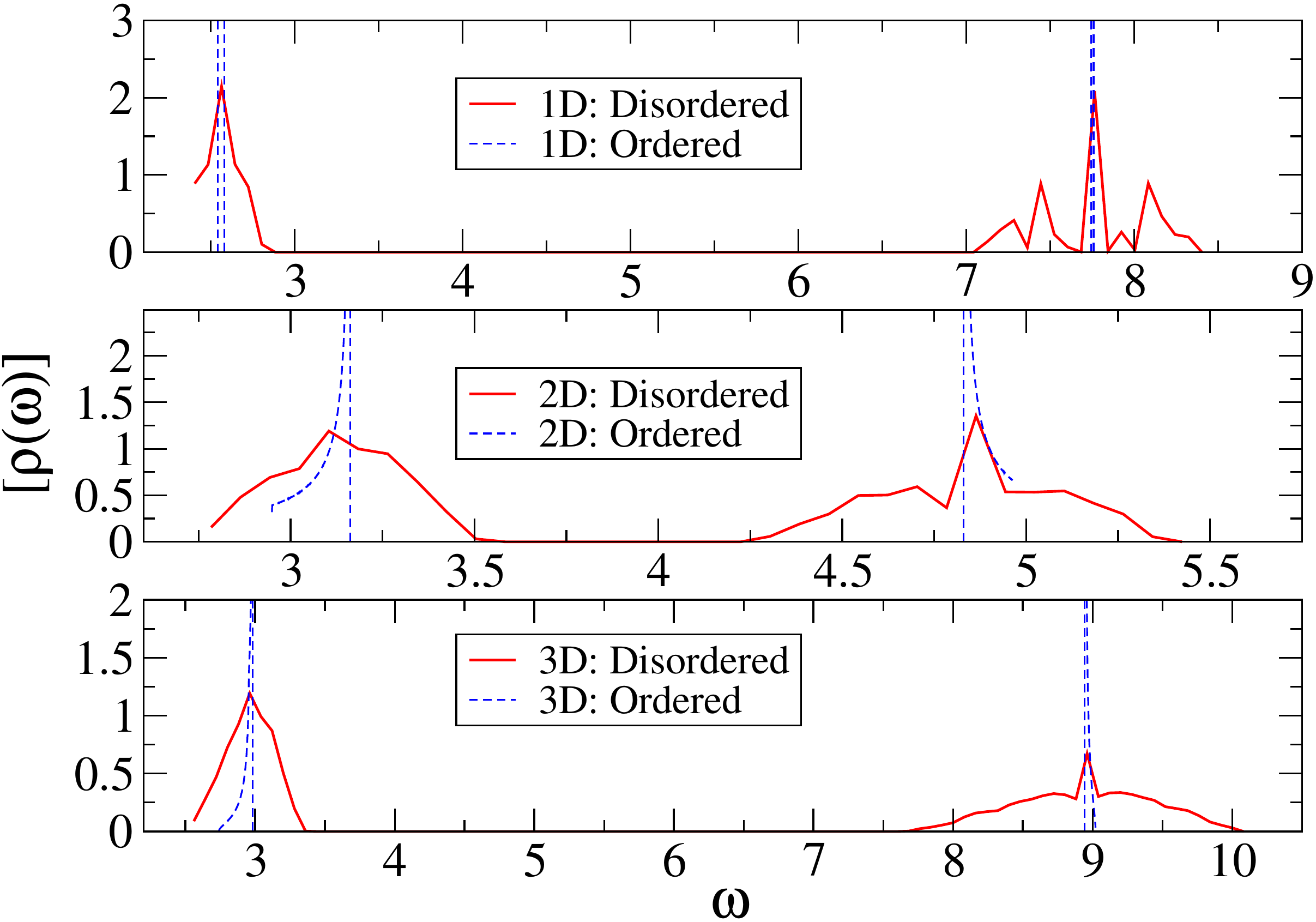}
\caption{(color online) Pinned lattices. \\  
 Disorder averaged density of states   obtained numerically from the eigenvalues
  of several disorder realizations in $1D, 2D$ and $3D$ for lattice
  sizes $N=4096,64,16$ respectively. Note that low frequency modes are absent. 
   We set  $k=1, k_o=10.0$ and $\Delta=0.4$ in $2D$ and $\Delta=0.8$ in $1D,
   3D$.  
Averages were taken over $30$ realizations in $1D$ and $10$ realizations in
$2D, 3D$. We find that in $2D$ and $3D$ there is not much
variation in $\rho(\om)$ for  different disorder samples. Also shown are the
density of states for the   binary mass ordered lattices.}
\label{dosdispin}
\vspace{0.75cm}
\end{figure}

\begin{figure}
\includegraphics[width=4in]{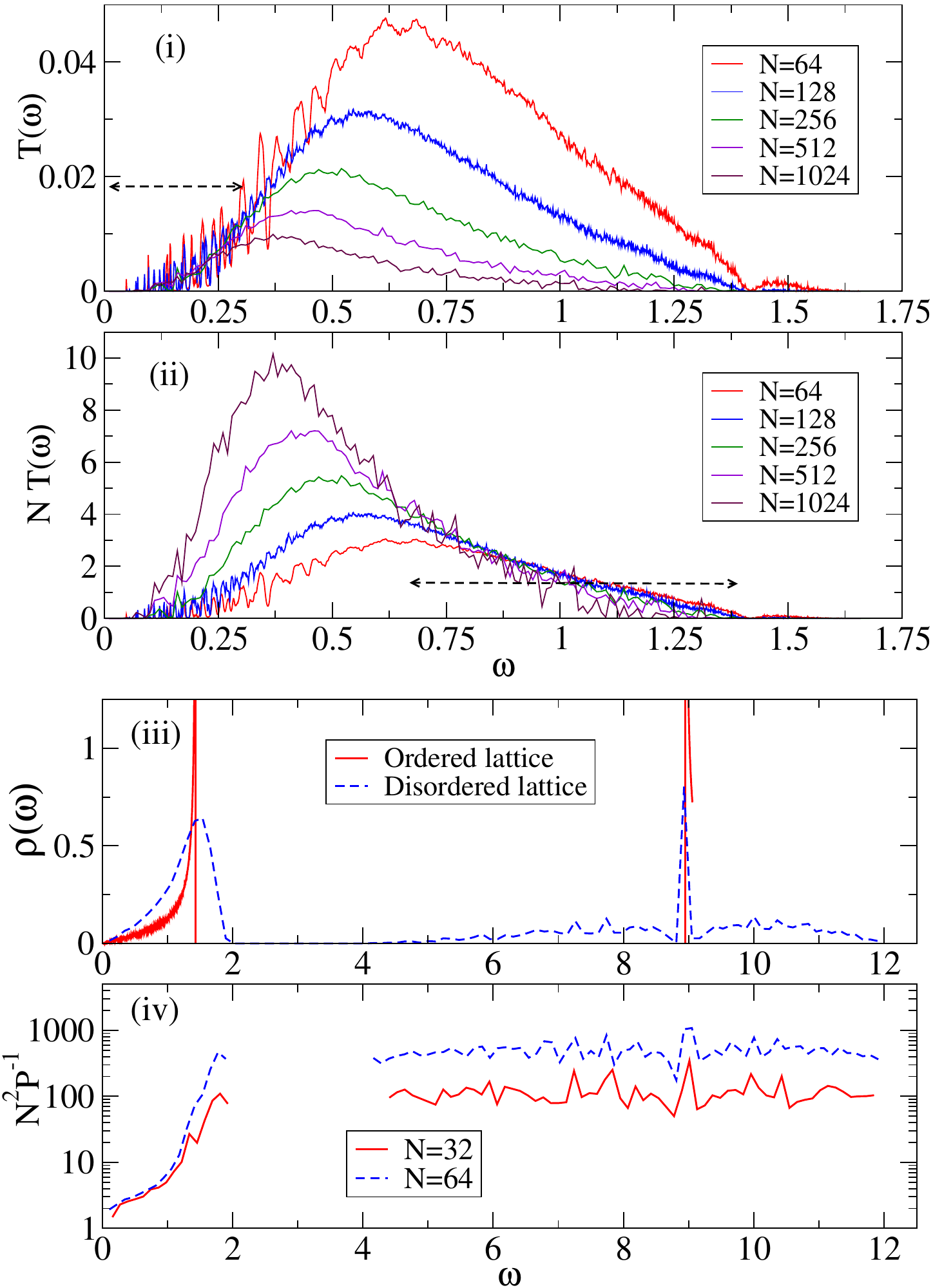}
\caption{(color online) $2D$ unpinned case with fixed BC for $\Delta=0.95$.
(i) Plot of the disorder averaged transmission 
  $T(\om)$  versus $\omega$. (ii) Plot of $N T(\om)$. The range of
    frequencies for which $T(\om) \sim 1/N$ is indicated by the dashed
    line.  (iii) Plot of $\rho(\om)$ for binary mass ordered and
    single disordered sample. (iv) Plot of $N^2 P^{-1}$ for single
    samples (smoothed data).  We see that
  even though the allowed   normal modes occur over a large frequency
  band $\approx (0-12)$, transmission takes place in a small band
  $\approx (0-1.25)$ and is negligible elsewhere. The IPR plots
  confirm that the non-transmitting states correspond to localized modes. 
In (i) we see that $\om_c^L$ is slowly decreasing with inrease of $N$.
}
\label{tw2dfix0.95}
\vspace{0.75cm}
\end{figure}

\begin{figure}
\includegraphics[width=4in]{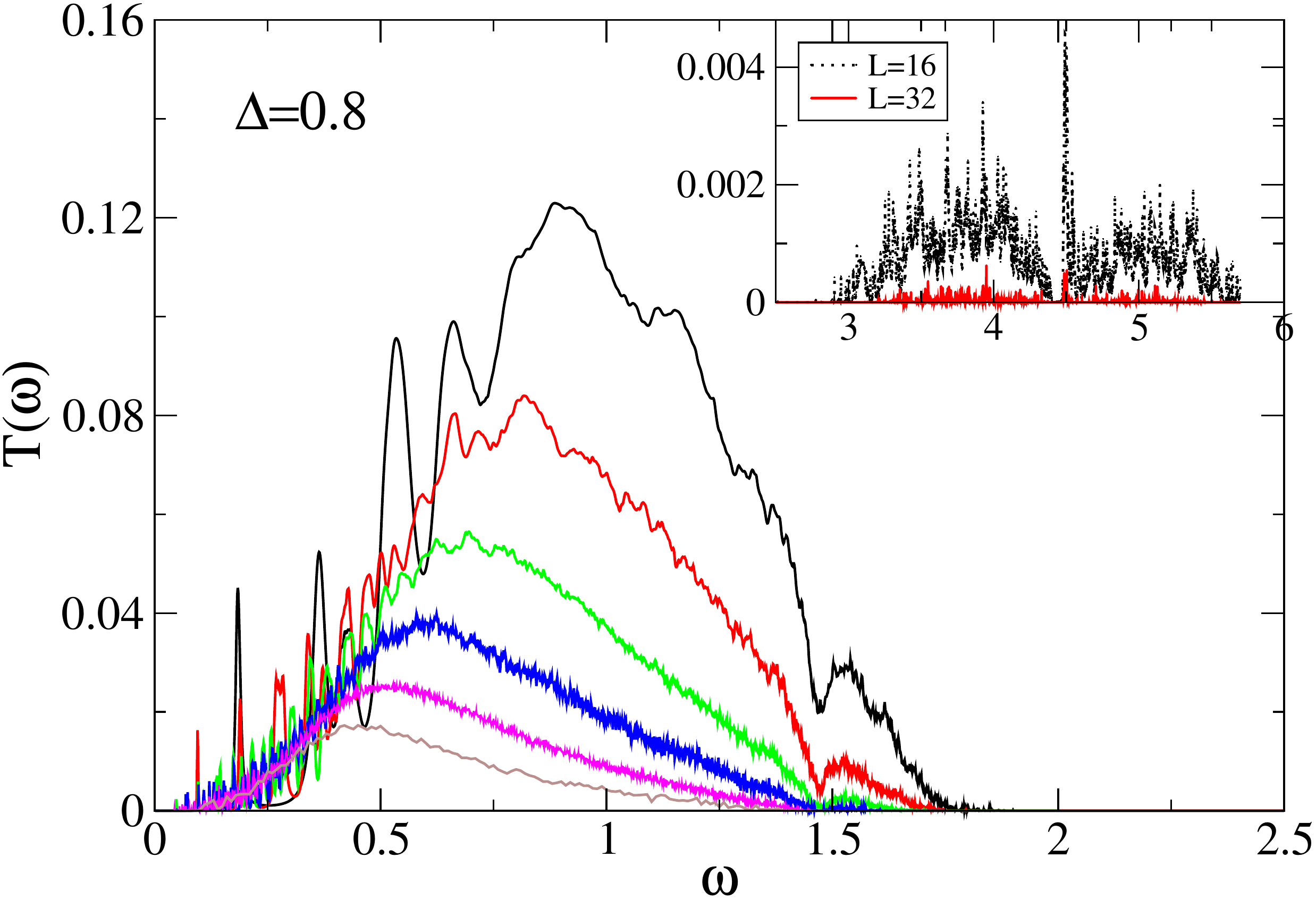}
\includegraphics[width=5in]{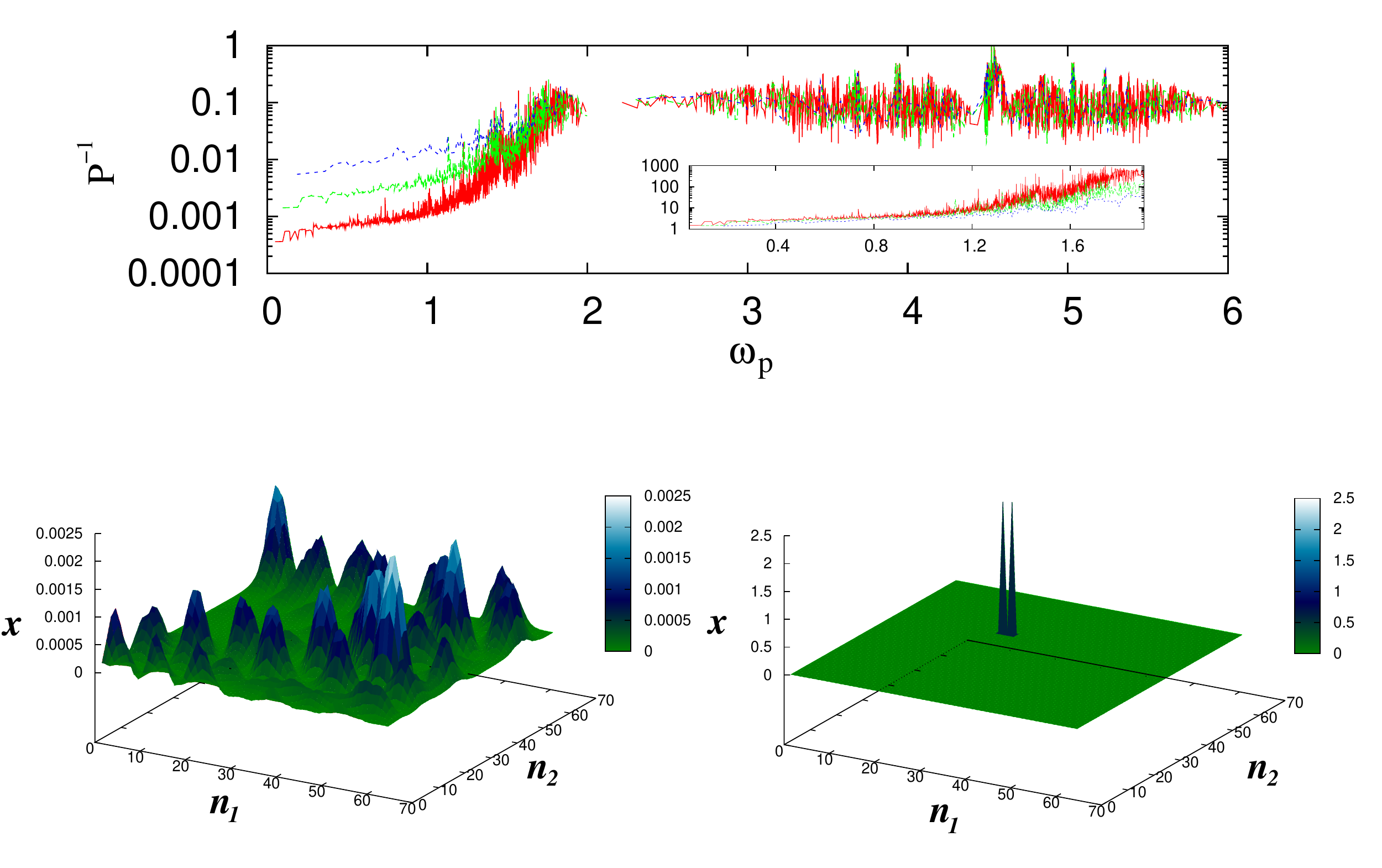}
\caption{(color online) {$2D$ unpinned case with fixed BC for $\Delta=0.8$.
\\ TOP: Plot of the disorder averaged transmission 
  $T(\om)$  versus $\omega$. 
The various curves  (from top to   bottom ) correspond to  square
  lattices with  $N=16,32,64,128,256,512$  respectively. 
We see again that most modes are localized and transmission takes
place over a small range of requencies. \\
BOTTOM: {Plot shows the IPR ($P^{-1}$) as a function of normal mode-frequency 
$\om_p$
  for the $2D$ lattice with $\Delta=0.8$. The curves 
 are for   $N=16$ (blue), $32$ (green) and $64$ (red). 
The inset plots $N^2 P^{-1}$ and the
  collapse at low frequencies shows that these modes are
  extended. Also  shown are two typical normal modes   for one small
  (left) and   one large value of $P^{-1}$  for $N=64$.}}} 
\label{tw2dfix0.8}
\vspace{0.75cm}
\end{figure}

\begin{figure}
\includegraphics[width=4in]{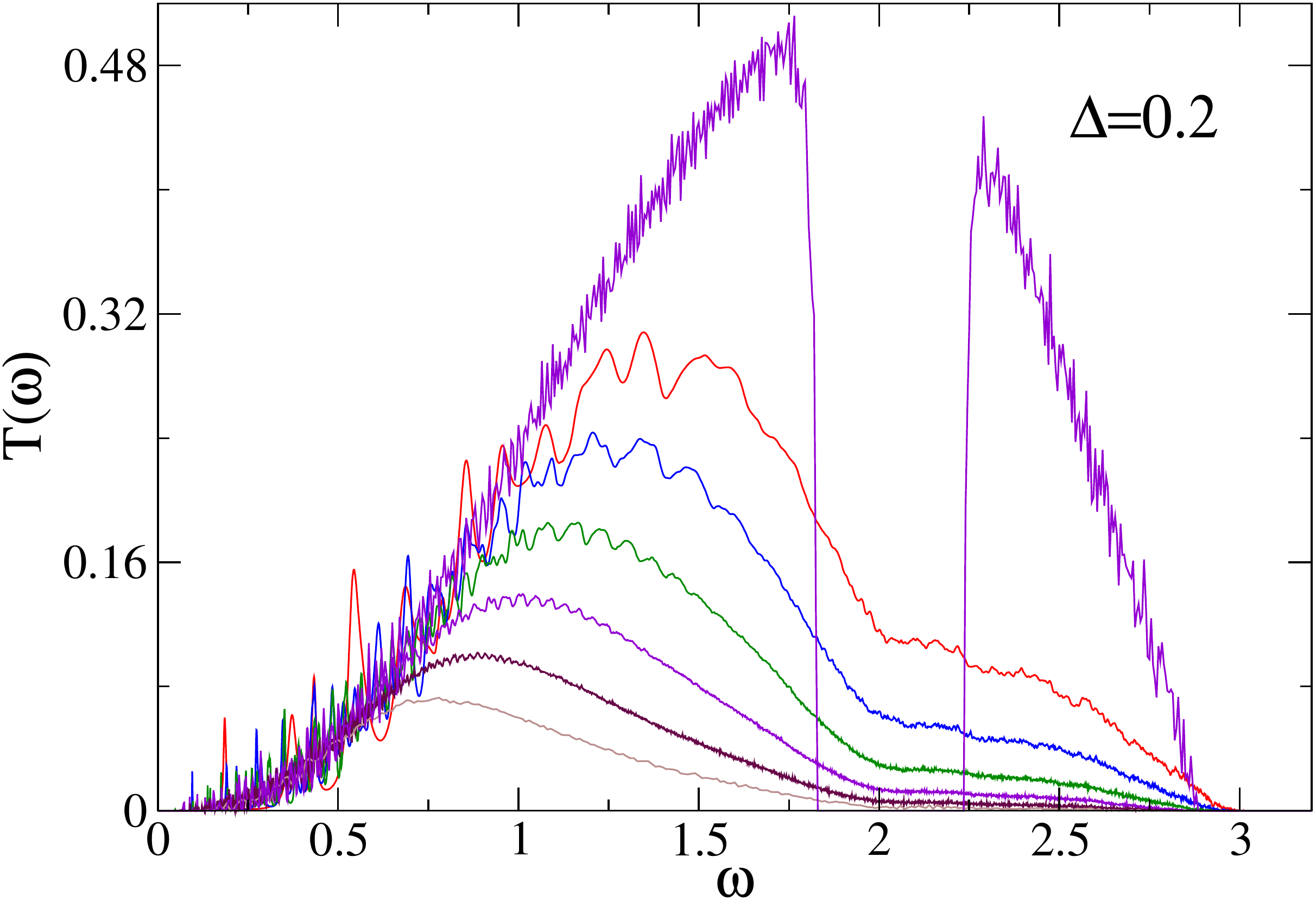}
\includegraphics[width=4in]{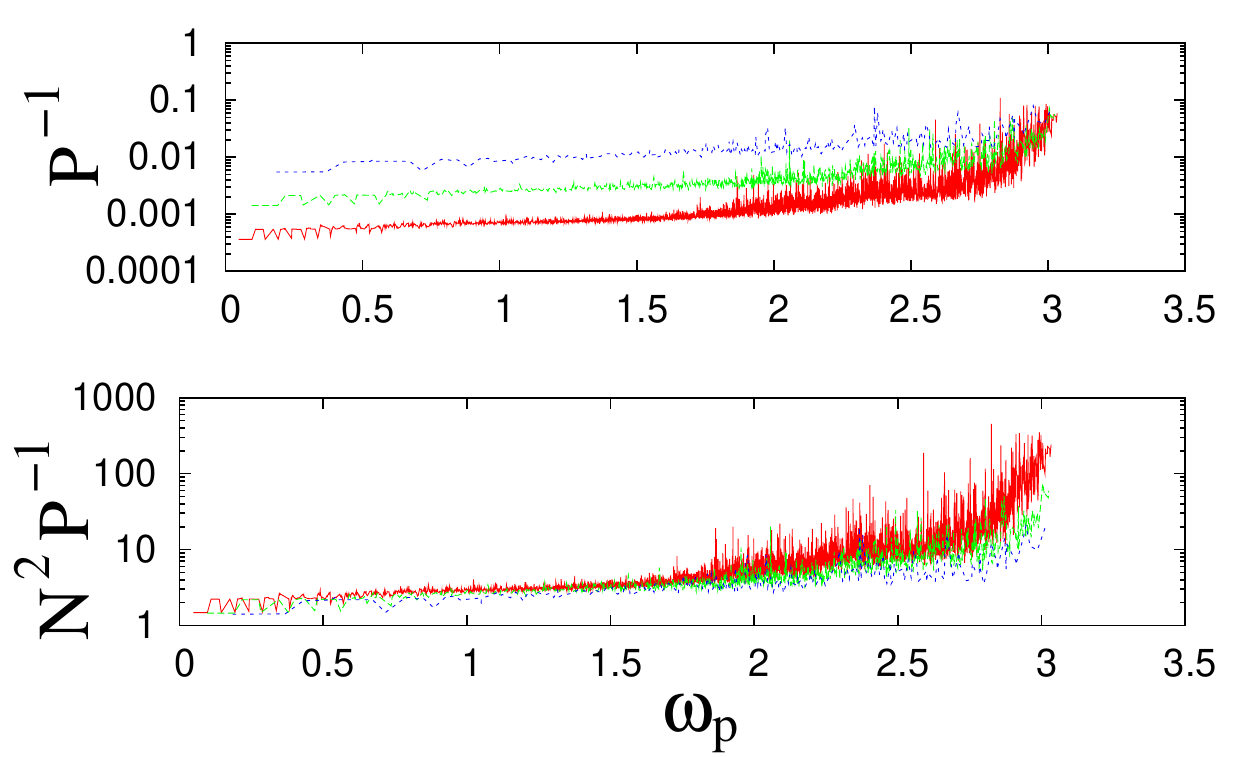} 
\caption{(color online) {$2D$ unpinned case with fixed BC  for $\Delta=0.2$.\\
TOP: Plot of the disorder averaged transmission  
  $T(\om)$  versus $\omega$.
  The upper-most 
  curve corresponds to a binary-mass  ordered lattice with $N=128$ while the
  remaining curves  (from top to bottom) correspond to  square
  lattices with 
$N=16,32,64,128,256,512$ 
  respectively. } \\ 
BOTTOM: {Plot shows the IPR ($P^{-1}$) and scaled IPR ($N^2 P^{-1}$) as a
  function of normal mode-frequency $\om_p$. 
   The curves  are for   $N=16$ (blue), $32$ (green) and $64$ (red). 
 }}
\label{tw2dfix0.2}
\vspace{0.75cm}
\end{figure}

\begin{figure}
\includegraphics[width=4in]{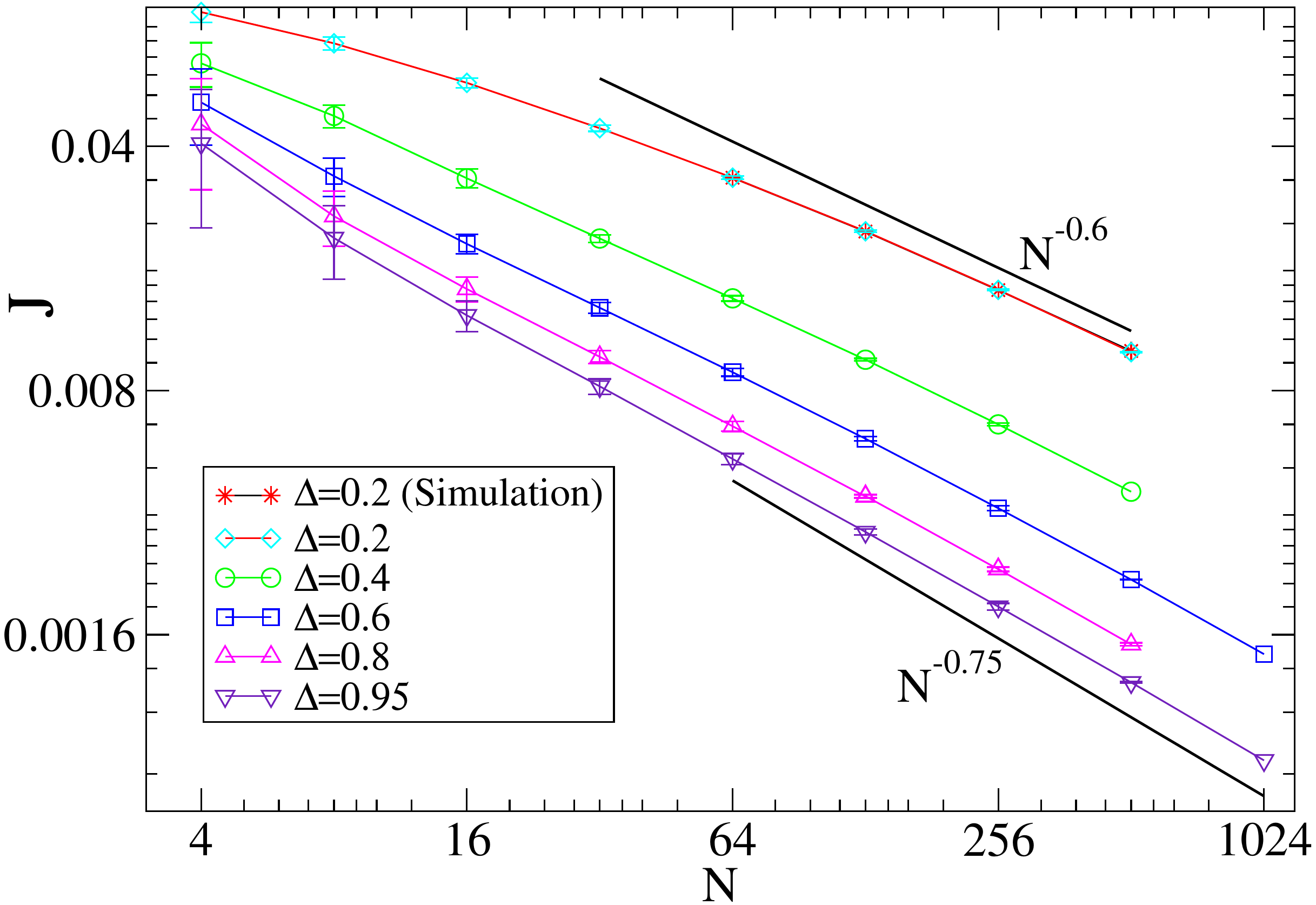}
\caption{(color online) $2D$ unpinned lattice with fixed BC. \\
 Plot of
  disorder-averaged current $J$ versus system size for different
  values of $\Delta$. 
 The error-bars show the actual standard deviations from
  sample-to-sample fluctuations. Numerical 
  errors are much smaller. For $\Delta =0.2$, simulation data is also plotted. }
\label{jvsn2dfix}
\vspace{0.75cm}
\end{figure} 

\begin{figure}
\includegraphics[width=4in]{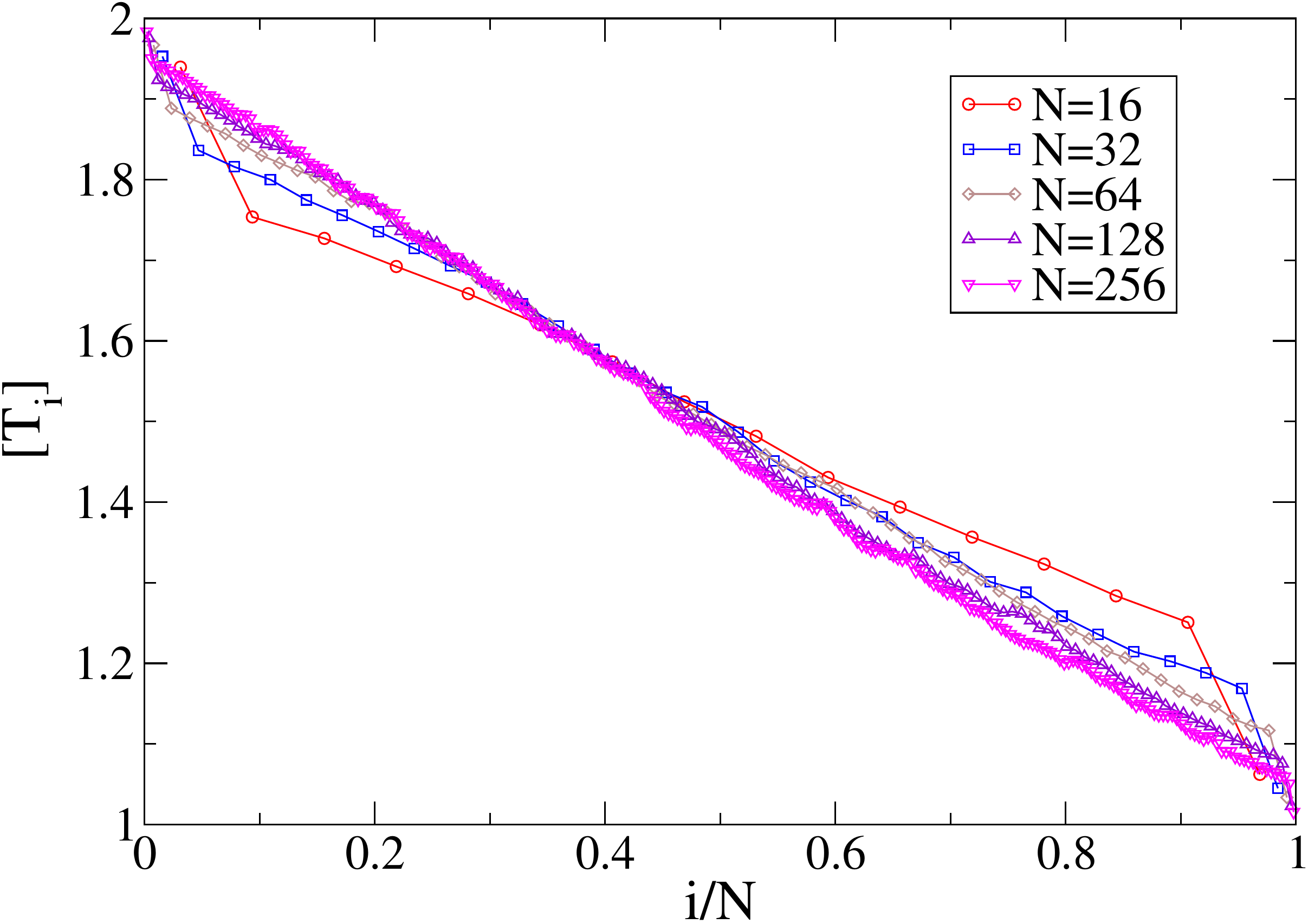}
\caption{(color online) $2D$  unpinned case with fixed BC for $ \Delta=0.2$. \\
Plot of disorder-averaged temperature
  profile   $[T_i]$  for  different system sizes obtained from
  simulations. 
}
\label{temp2dfix}
\vspace{0.75cm}
\end{figure}

\begin{figure}
\includegraphics[width=4in]{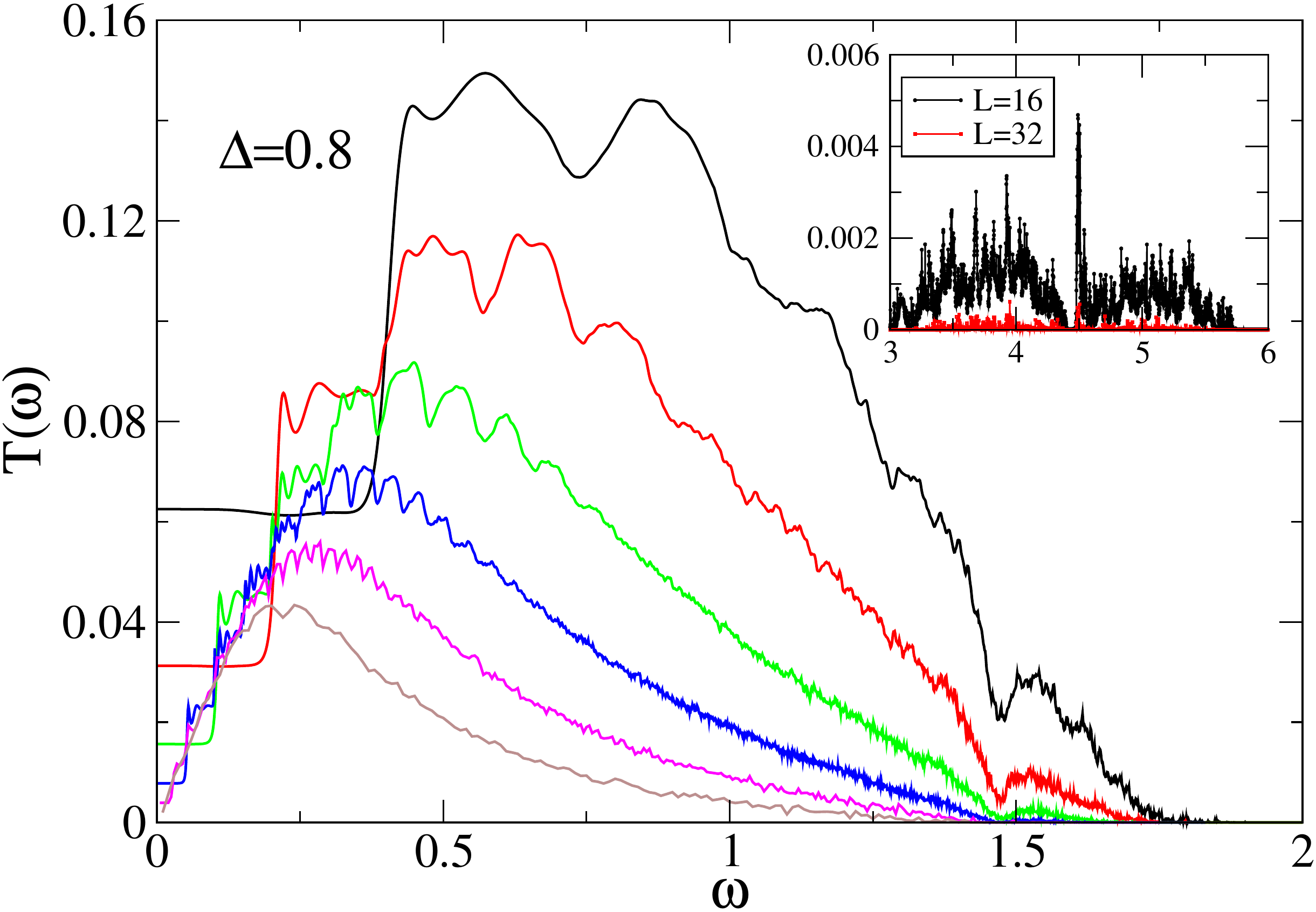}
\includegraphics[width=5in]{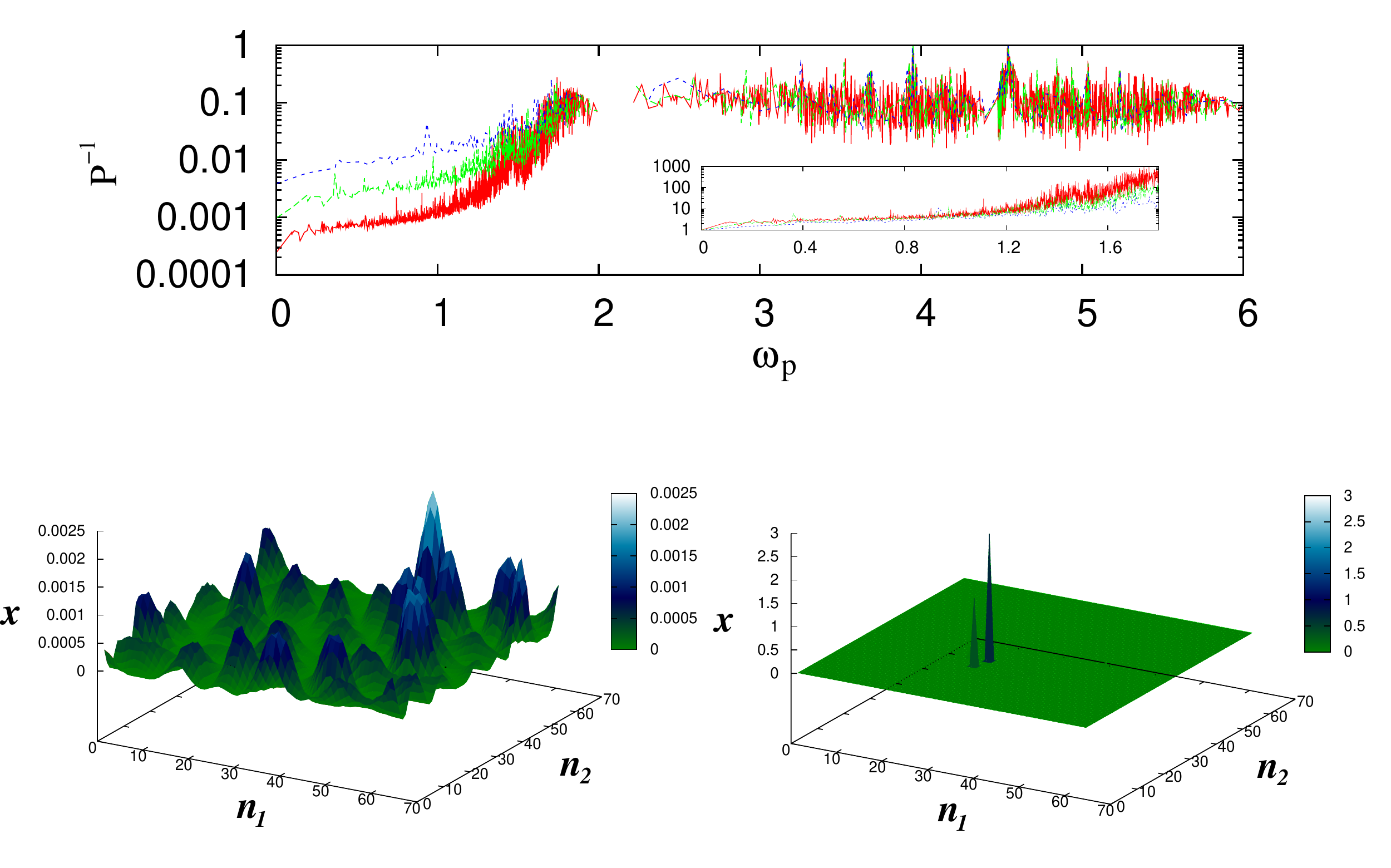}
\caption{(color online)
 {$2D$ unpinned case with free BC for $\Delta=0.8$. \\ TOP: Plot of the disorder averaged transmission 
  $T(\om)$  versus $\omega$. 
The various curves  (from top to bottom) correspond to  square lattices with
 $N=16,32,64,128,256,512$  respectively. We see that
  transmission takes  place in a small band $\approx (0-2)$ of the full range 
$\approx (0-6)$ of normal modes
   and as can be seen in the inset is
  negligible elsewhere.} \\
BOTTOM: {Plot shows
  the IPR ($P^{-1}$) as a function of normal
  mode-frequency $\om_p$. The curves
  are for   $N=16$ (blue), $32$ (green) and $64$ (red). 
In the inset we plot $N^2 P^{-1}$ and the 
  collapse at low frequencies shows that low frequency modes are
  extended. Also shown are two typical normal modes   for one small
  (left) and   one large value of $P^{-1}$  for $N=64$.} }
\label{tw2dfree0.8}
\vspace{0.75cm}
\end{figure} 

\begin{figure}
\includegraphics[width=4in]{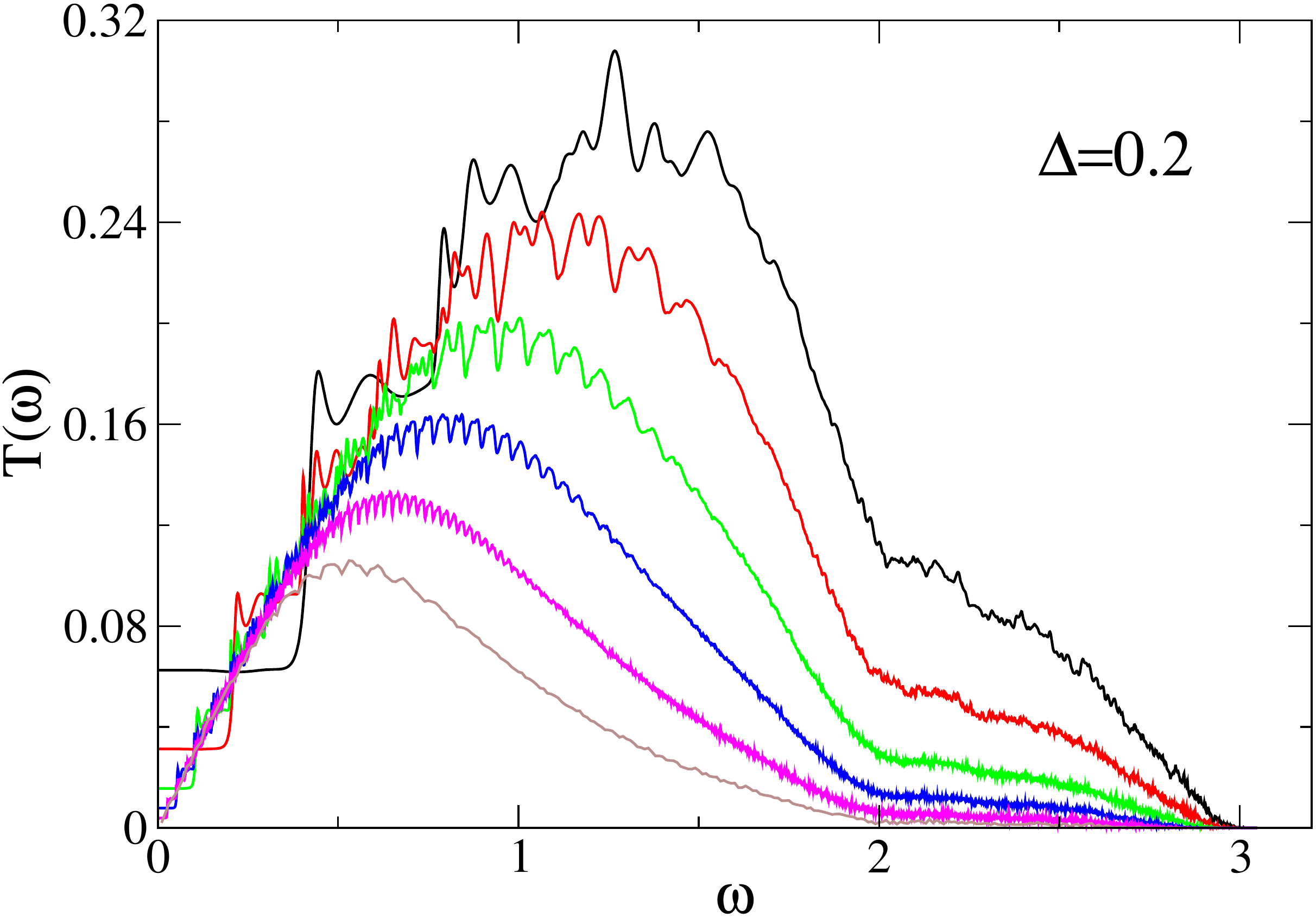}
\caption{(color online) $2D$ unpinned case with free BC for $\Delta=0.2$. \\
 Plot of the disorder
  averaged transmission   $T(\om)$   versus $\omega$ for. The curves 
(from top to bottom) are for $N=16,32,64,128,256,512$  respectively. Note the
  linear form at small $\om$.}
\label{tw2dfree0.2}
\vspace{0.75cm}
\end{figure} 

\clearpage

\begin{figure}
\includegraphics[width=4in]{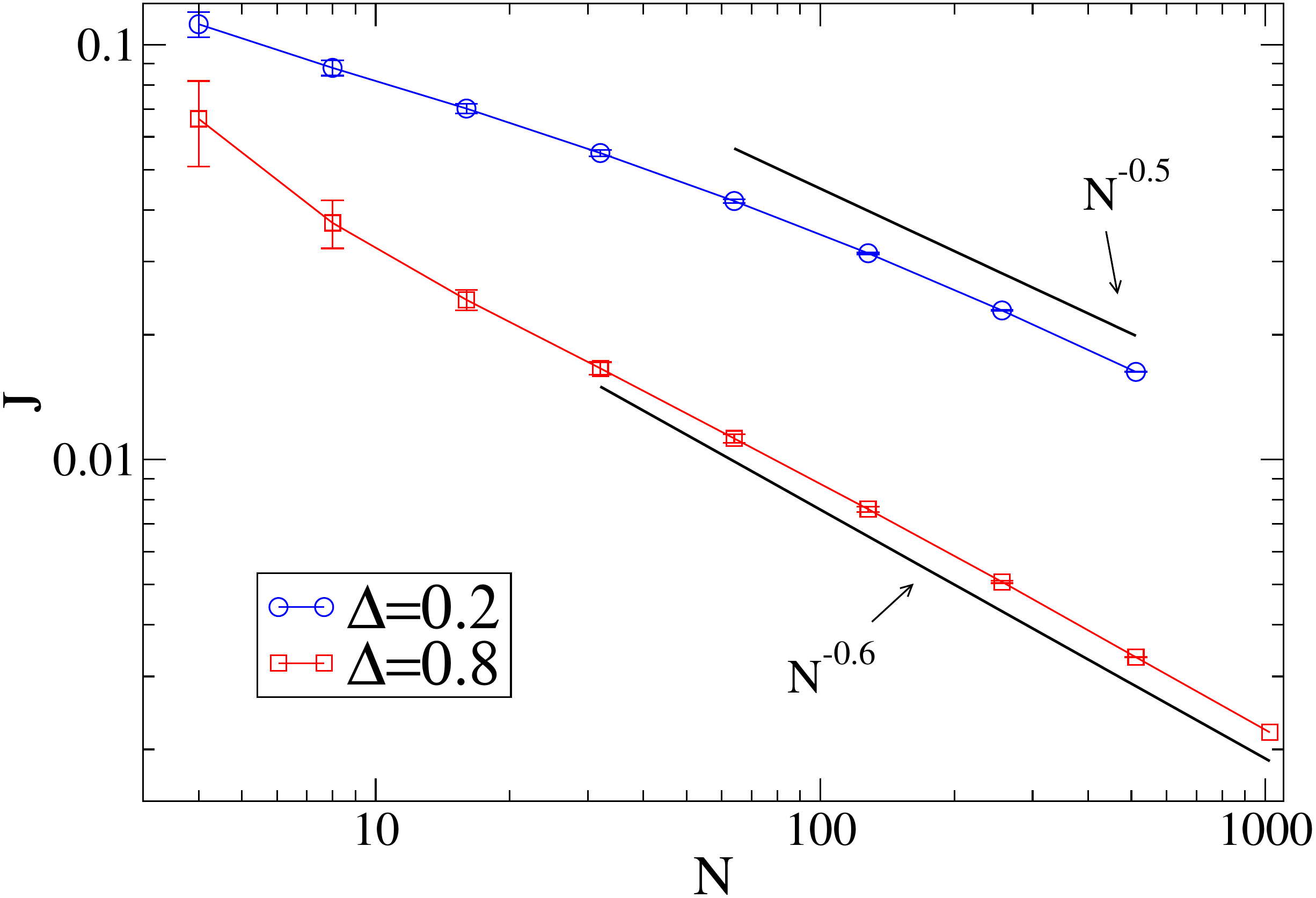}
\caption{(color online) $2D$ unpinned case with free BC. \\
Plot of disorder-averaged
  current $J$ versus system size for two
  different values of $\Delta$. 
 The error-bars  show standard devations due to  sample-to-sample fluctuations. Numerical
  errors are much smaller.}
\label{jvsn2dfree}
\vspace{0.75cm}
\end{figure} 

\begin{figure}
\includegraphics[width=3.2in]{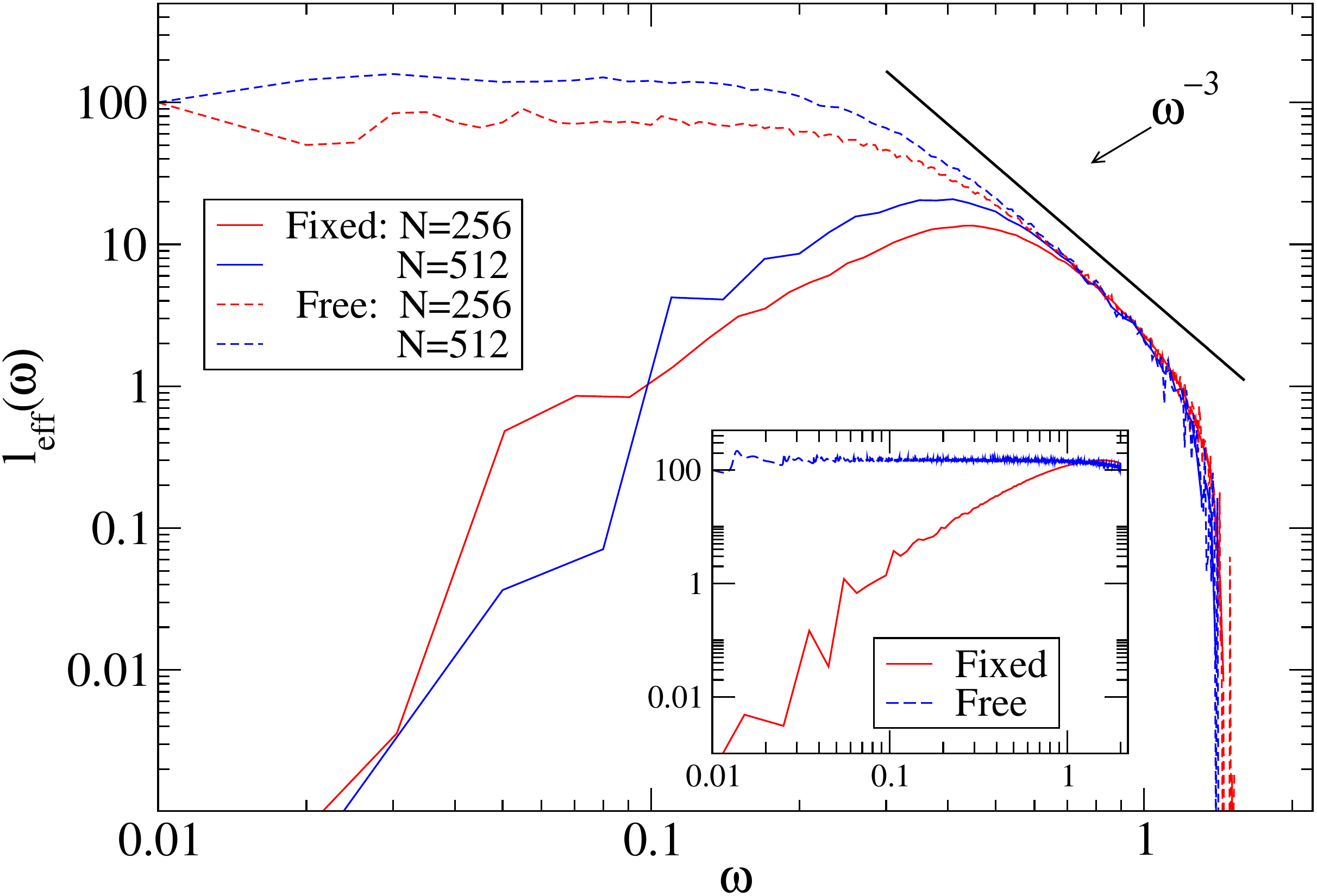}
\caption{
 Plot of the effective mean-free path $l_{\rm
 eff}=NT(\omega)/\om^{d-1}$ in $2D$ 
 with $\Delta=0.8$. The insets show
 $\ell_{\rm eff}$ for the ordered lattices with a single mass. An 
 $\om^{-3}$ behaviour is observed in a small part of the diffusive
 region.The fixed BC data is highly oscillatory and has been smoothed. }  
\label{leff2d}
\end{figure}

\begin{figure}
\includegraphics[width=4in]{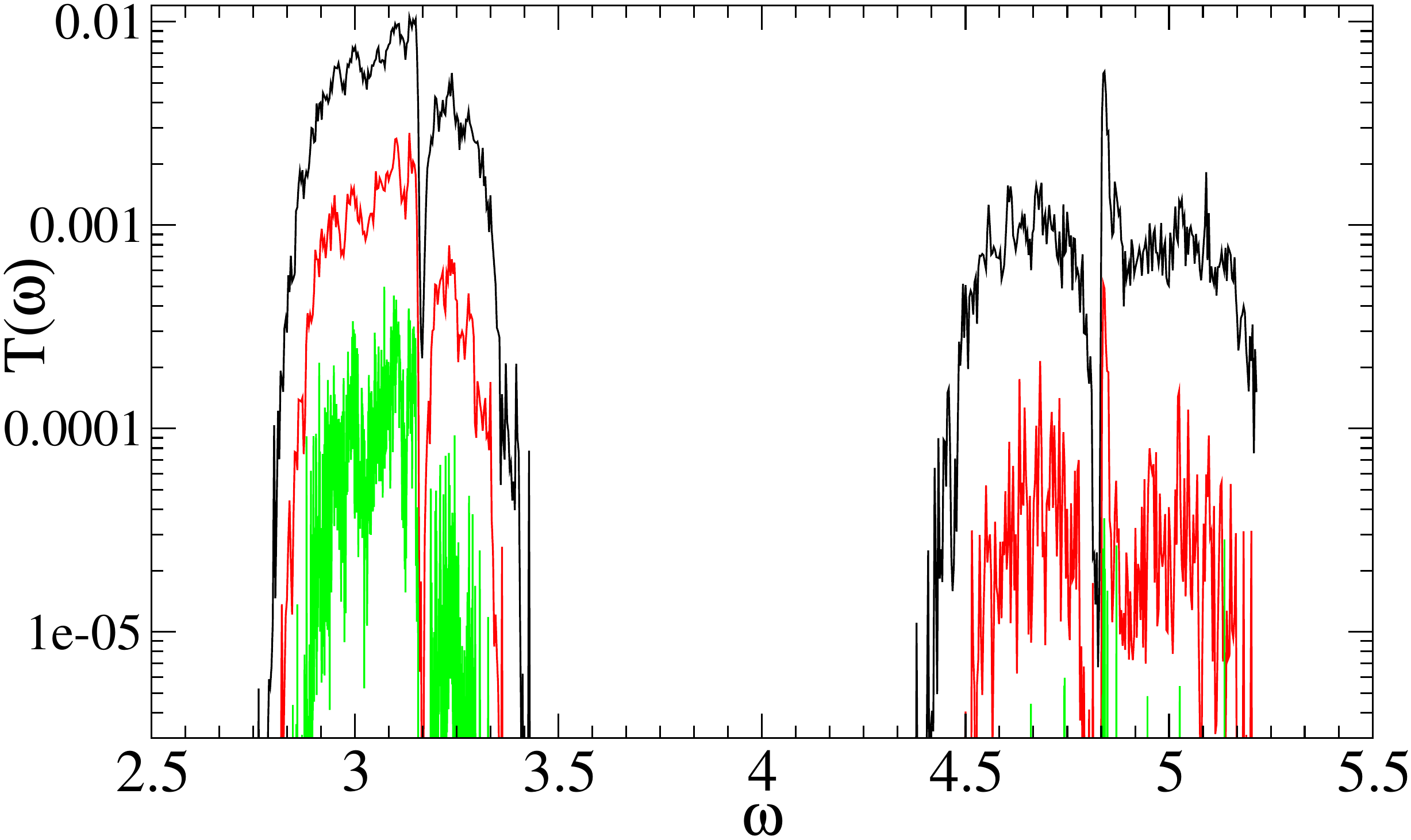}
\includegraphics[width=5in]{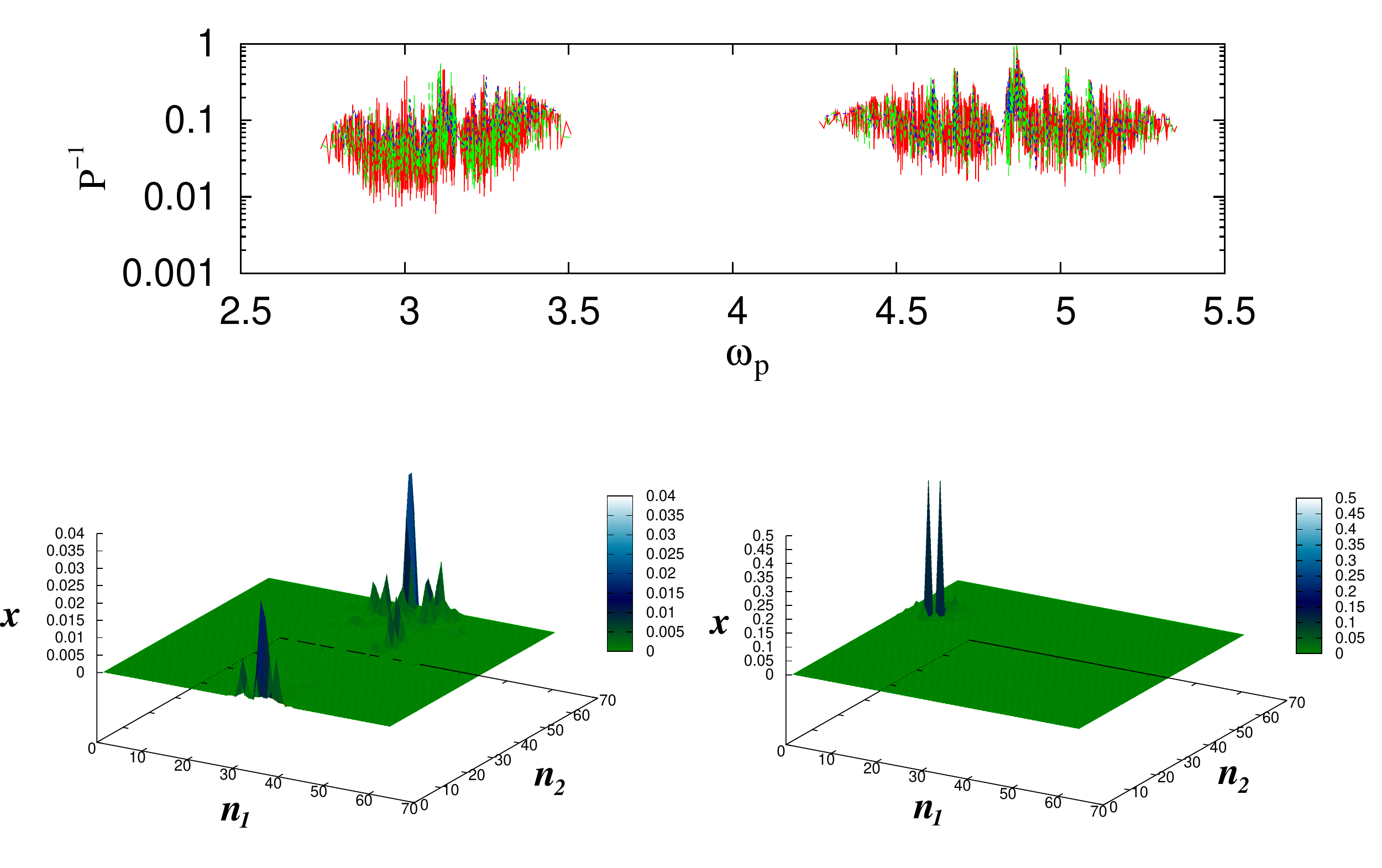}
\caption{(color online) {$2D$
  pinned case for $\Delta=0.4$ and $k_o=10.0$. \\ TOP: Plot of the disorder averaged transmission  
  $T(\om)$ 
  versus $\omega$.  
The various curves (from top to bottom) correspond to 
  lattices with $N=16,32,64 $ respectively. Here we choose 
$\gamma = \sqrt{10}$}. \\
BOTTOM: {Plot of the IPR ($P^{-1}$) as a function of normal mode-frequency 
$\om_p$. The curves are for
  $N=16$ (blue), $32$ (green) and $64$ (red). Also shown are two typical normal modes   for one
  small (left) and one   large value of $P^{-1}$  for $N=64$. }}
\label{tw2dpin10}
\vspace{0.75cm}
\end{figure}

\begin{figure}
\includegraphics[width=4in]{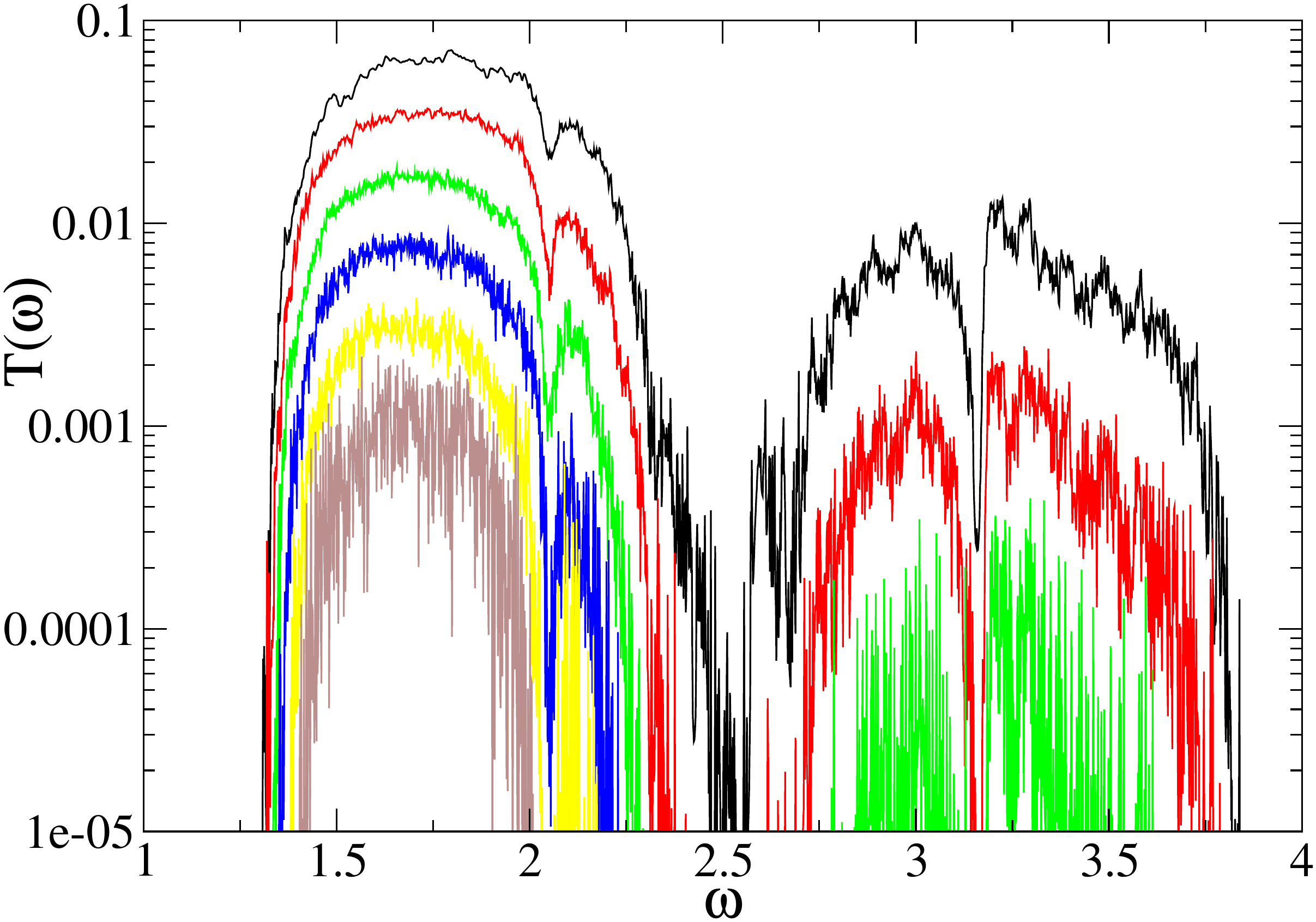}
\caption{(color online) $2D$ pinned case for  $\Delta=0.4$ and $k_o=2.0$. \\
 Plot of the disorder averaged transmission $T(\om)$ versus $\omega$ .  
The various curves (from top to bottom) are for 
$N=16,32,64,128,256,512$ respectively. Here we choose $\gamma = \sqrt{2}$. 
}
\label{tw2dpin02}
\vspace{0.75cm}
\end{figure}

\begin{figure}
\includegraphics[width=4in]{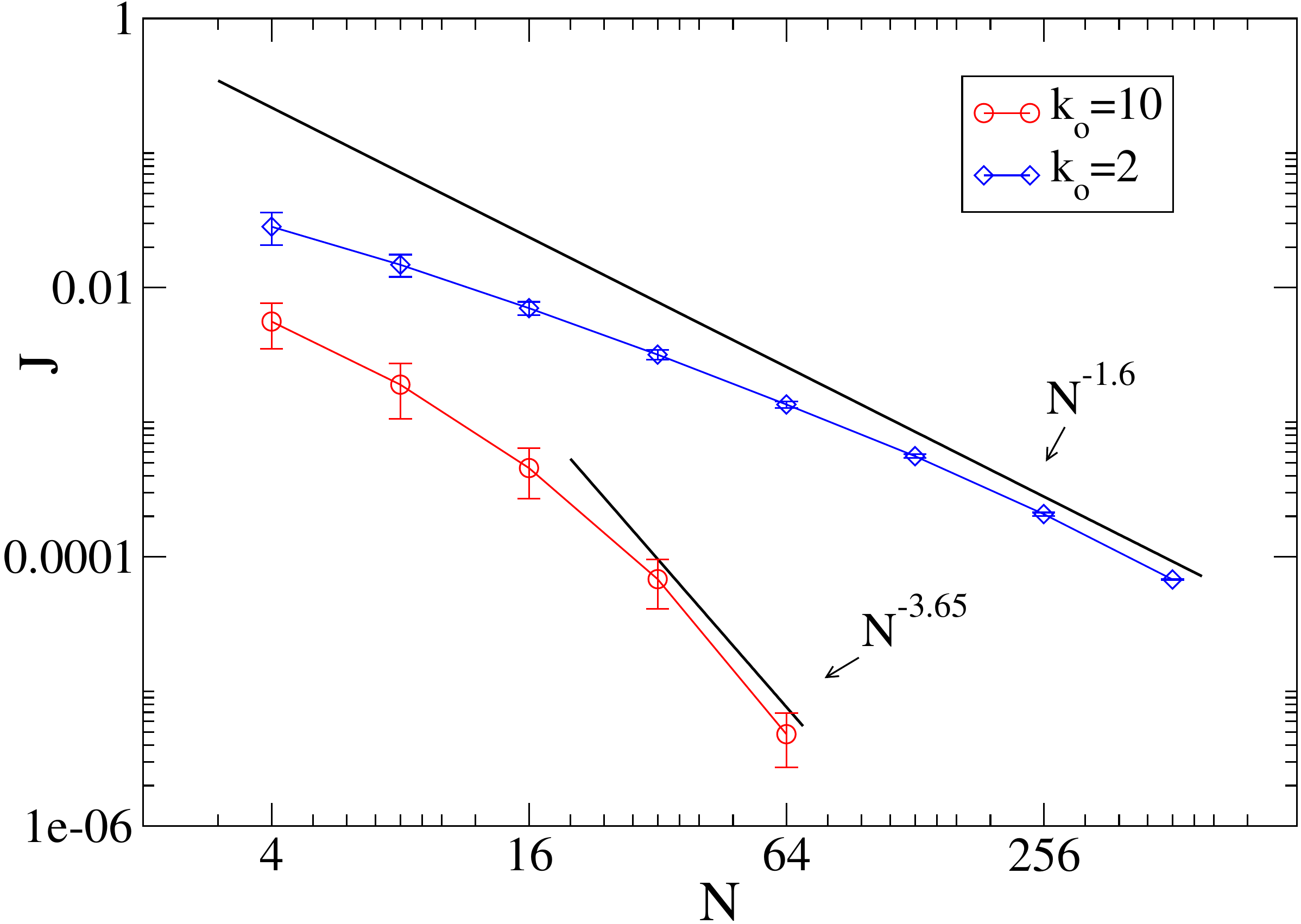}
\caption{ (color online) $2D$ pinned case for $\Delta=0.4$.\\
Plot of disorder-averaged current
  $J$ versus system size  for two different values of $k_o$. Error bars
show standard deviation due to disorder and numerical errors are much 
smaller. Note that the standard deviation do not decrease with system size
for higher $k_o$.
}
\label{jvsn2donpin}
\vspace{0.75cm}
\end{figure}

\begin{figure}
\includegraphics[width=4in]{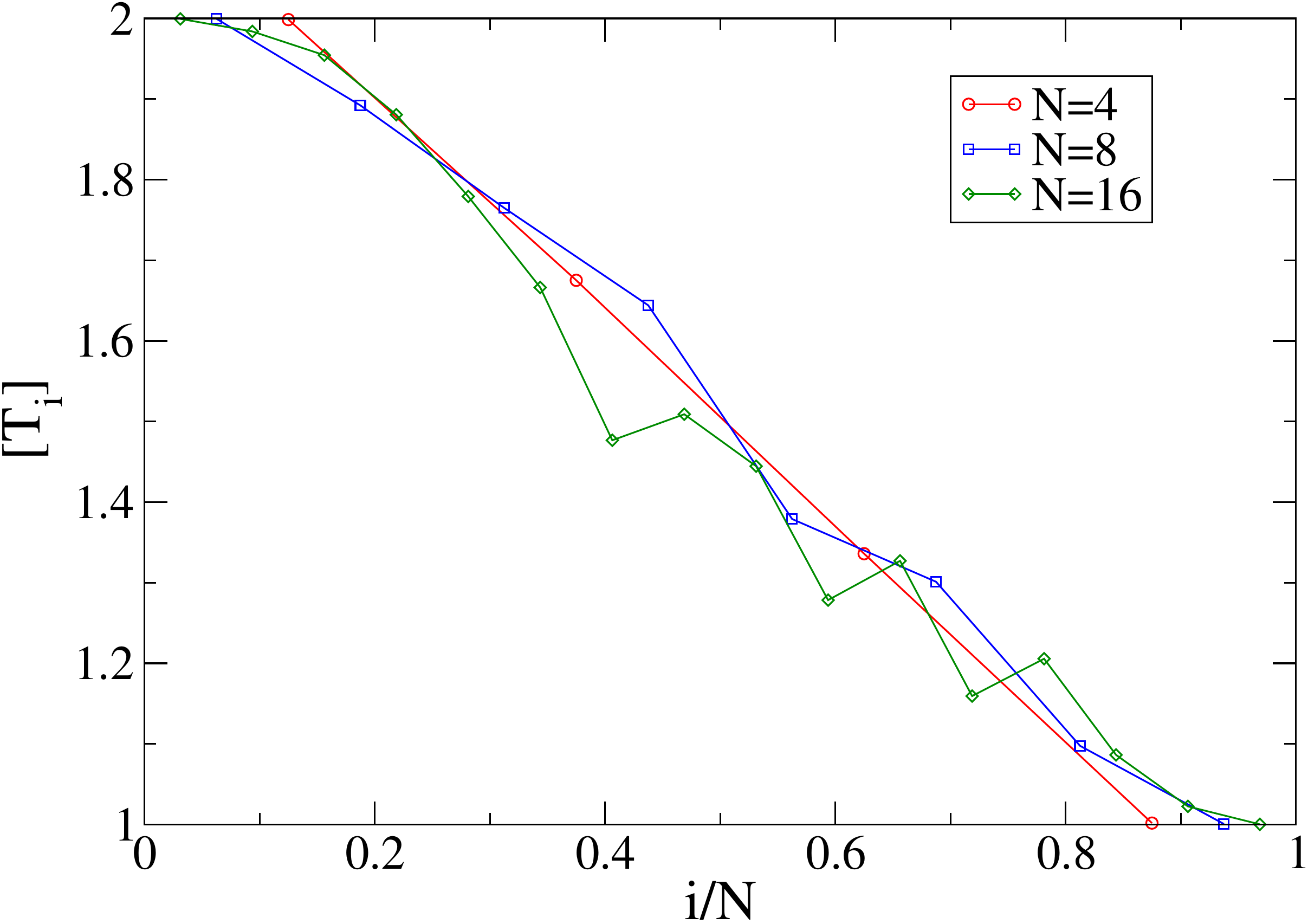}
\caption{ (color online) $2D$ pinned case for $\Delta=0.4$ and $k_o=10.0$. \\
Plot of disorder-averaged temperature
  profile   $[T_i]$  for  different system sizes. The plots are from
  simulations and here we choose $\gamma=\sqrt{10}$. 
}
\label{temp2donpin}
\vspace{0.75cm}
\end{figure}

\begin{figure}
\includegraphics[width=4in]{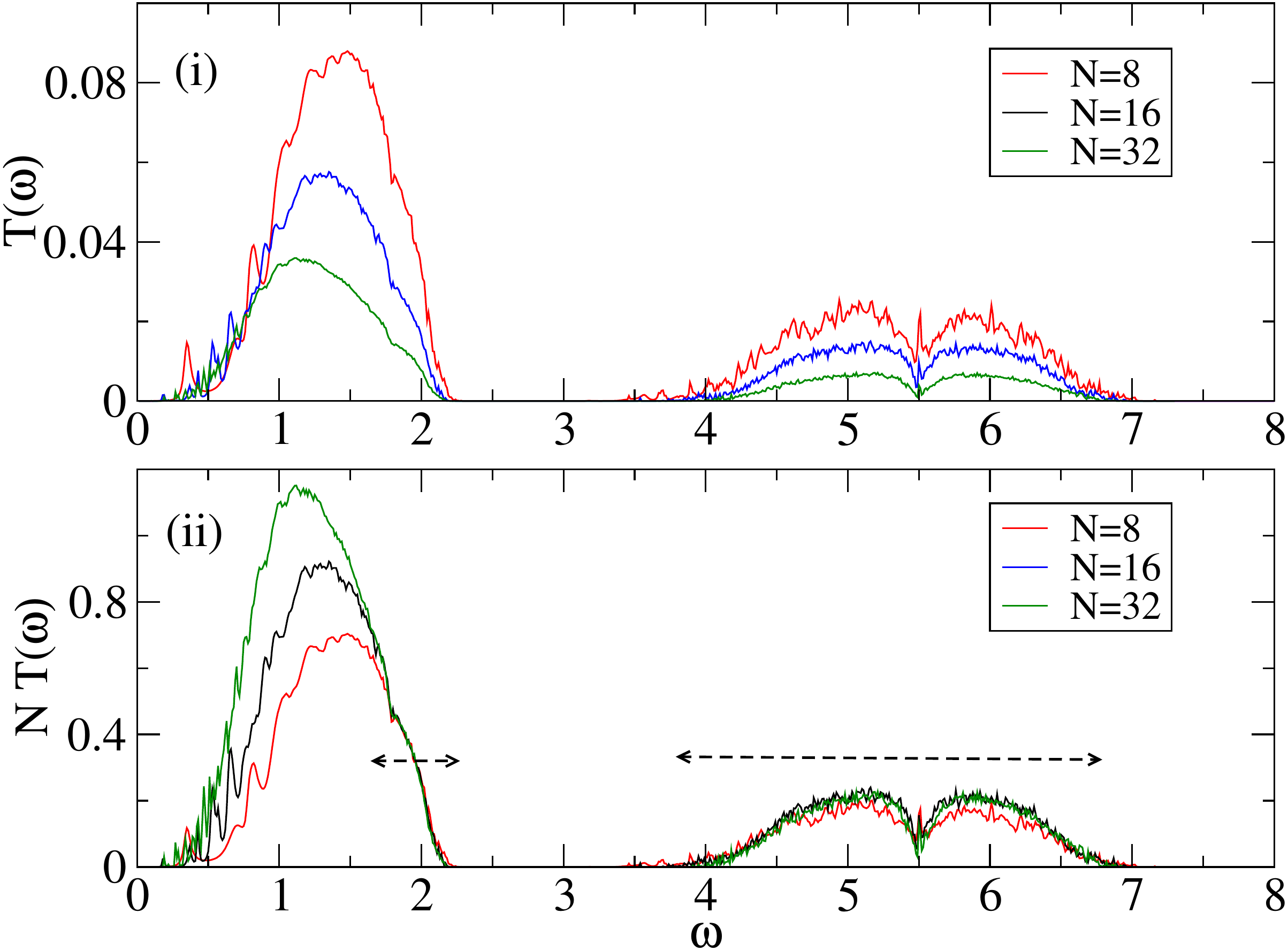}
\includegraphics[width=4in]{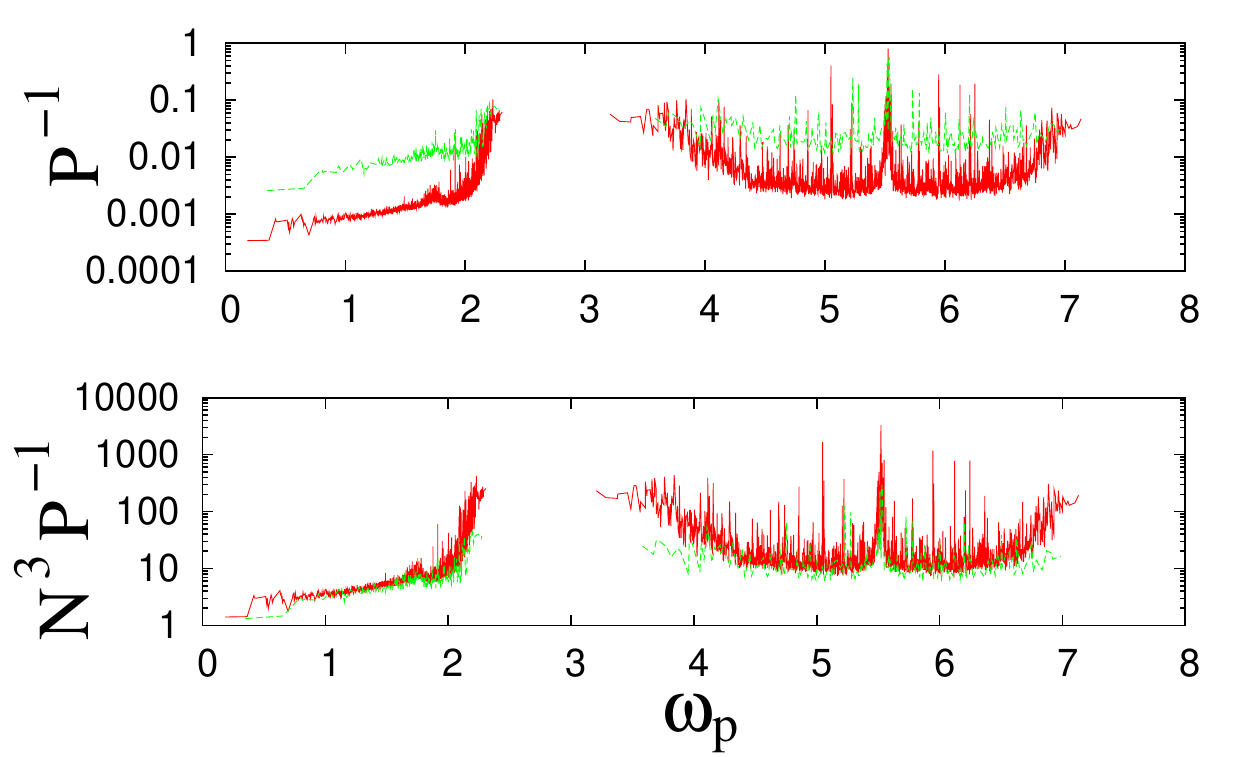}
\caption{(color online) {$3D$ unpinned case with fixed BC for $\Delta=0.8$. 
\\ TOP: Plot of the
  disorder averaged transmission  
  $T(\om)$ versus $\omega$. The inset shows the same data
  multiplied by a factor of $N$. }\\
BOTTOM: {Plot of the IPR ($P^{-1}$) and scaled IPR ($N^3 P^{-1}$) as a
  function of normal mode-frequency $\om_p$ 
  for a fixed disorder-realization.
  The curves  are for $N=8$ (green) and $16$ (red).}}
\label{tw3dfixed0.8}
\vspace{.5cm}
\end{figure}

\begin{figure}
\includegraphics[width=4in]{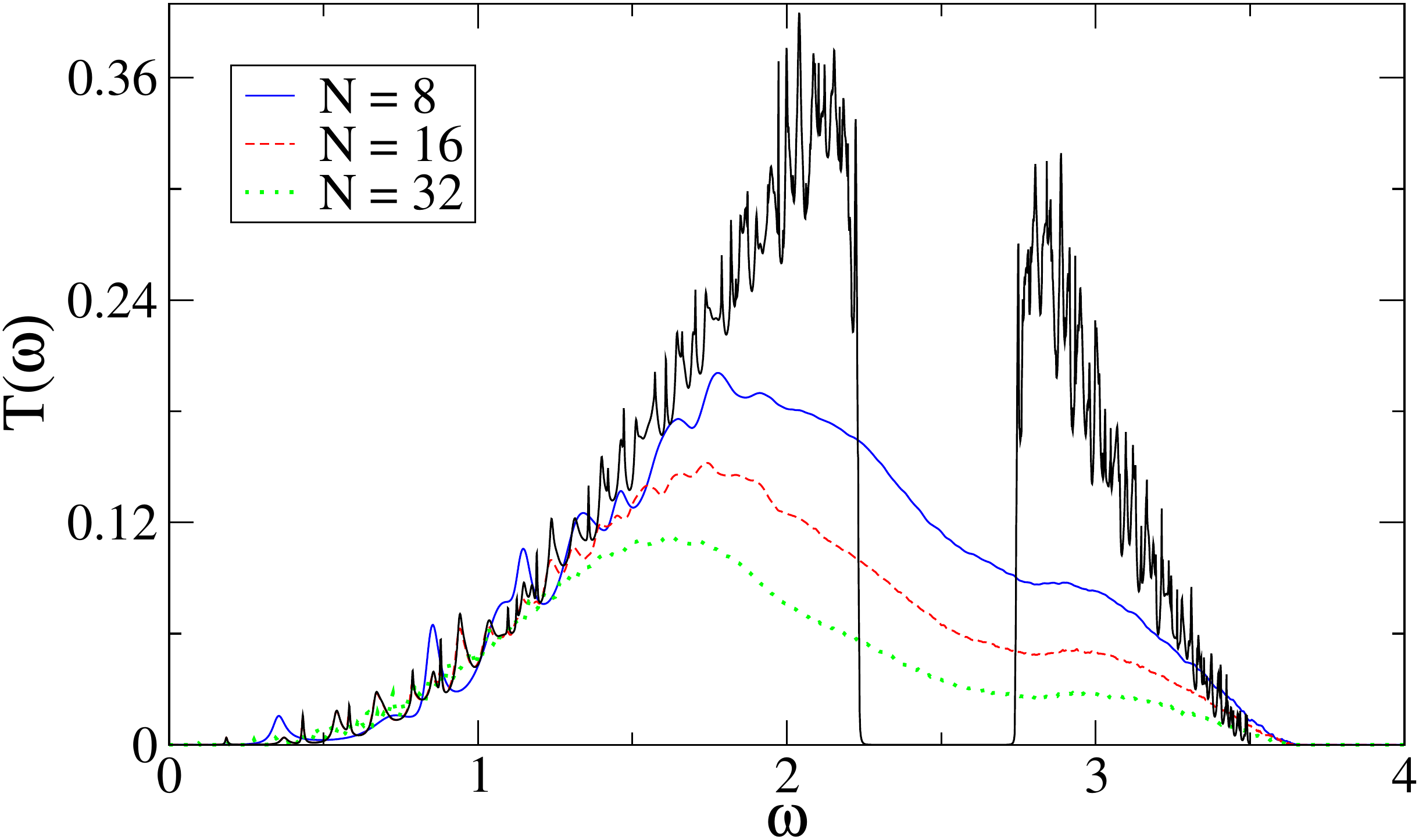} 
\includegraphics[width=4in]{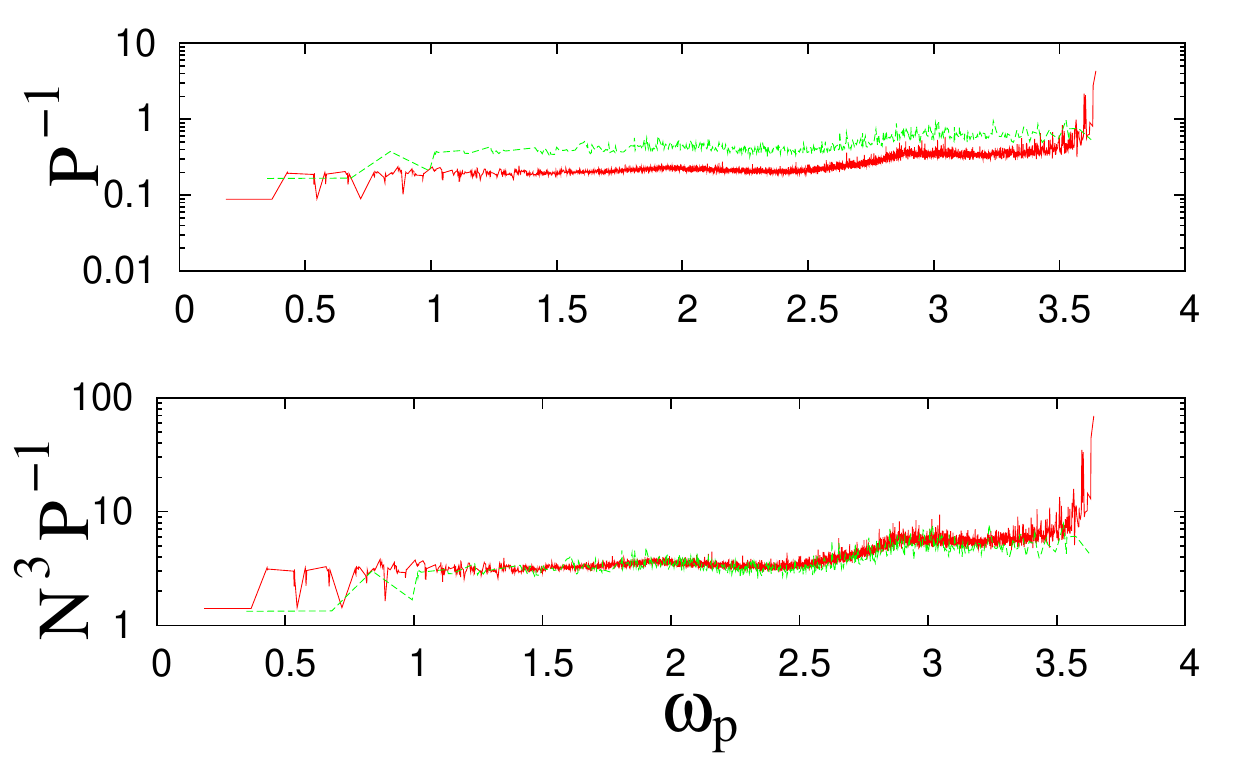}
\caption{(color online) {$3D$ unpinned case with fixed BC for $\Delta=0.2$. \\
TOP: Plot of the disorder averaged transmission 
  $T(\om)$   versus $\omega$. The uppermost curve is
the transmission curve for the binary mass ordered lattice for $N=16$.
  }\\
BOTTOM: {Plot of IPR ($P^{-1}$) and scaled IPR ($N^3 P^{-1}$) as a
  function of normal mode-frequency $\om_p$ for a fixed disorder-realization.
  The curves  are for $N=8$ (green) and $16$ (red)}.} 
\label{tw3dfixed0.2}
\vspace{.5cm}
\end{figure}

\begin{figure}
\includegraphics[width=4in]{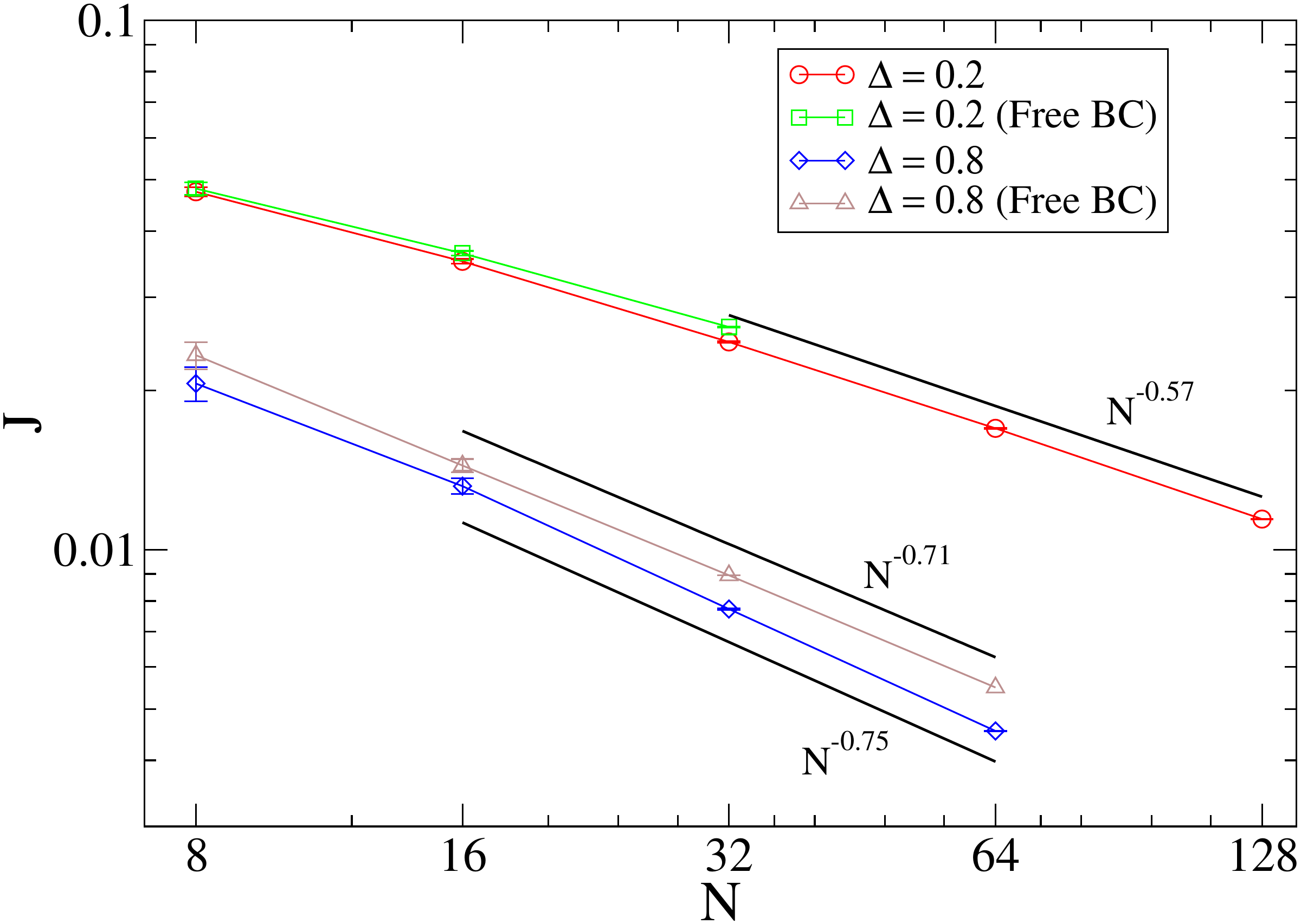}
\caption{ (color online) $3D$ unpinned case with fixed and free BCs. \\
Plot of disorder-averaged
  current $J$ versus system size for  two different values of
  $\Delta$. The data for $\Delta=0.2$  is from simulations. The
  error-bars show standard deviations due to disorder and numerical
  errors are smaller. }
\label{jvsn3dunpin}
\vspace{0.75cm}
\end{figure} 

\begin{figure}
\includegraphics[width=4in]{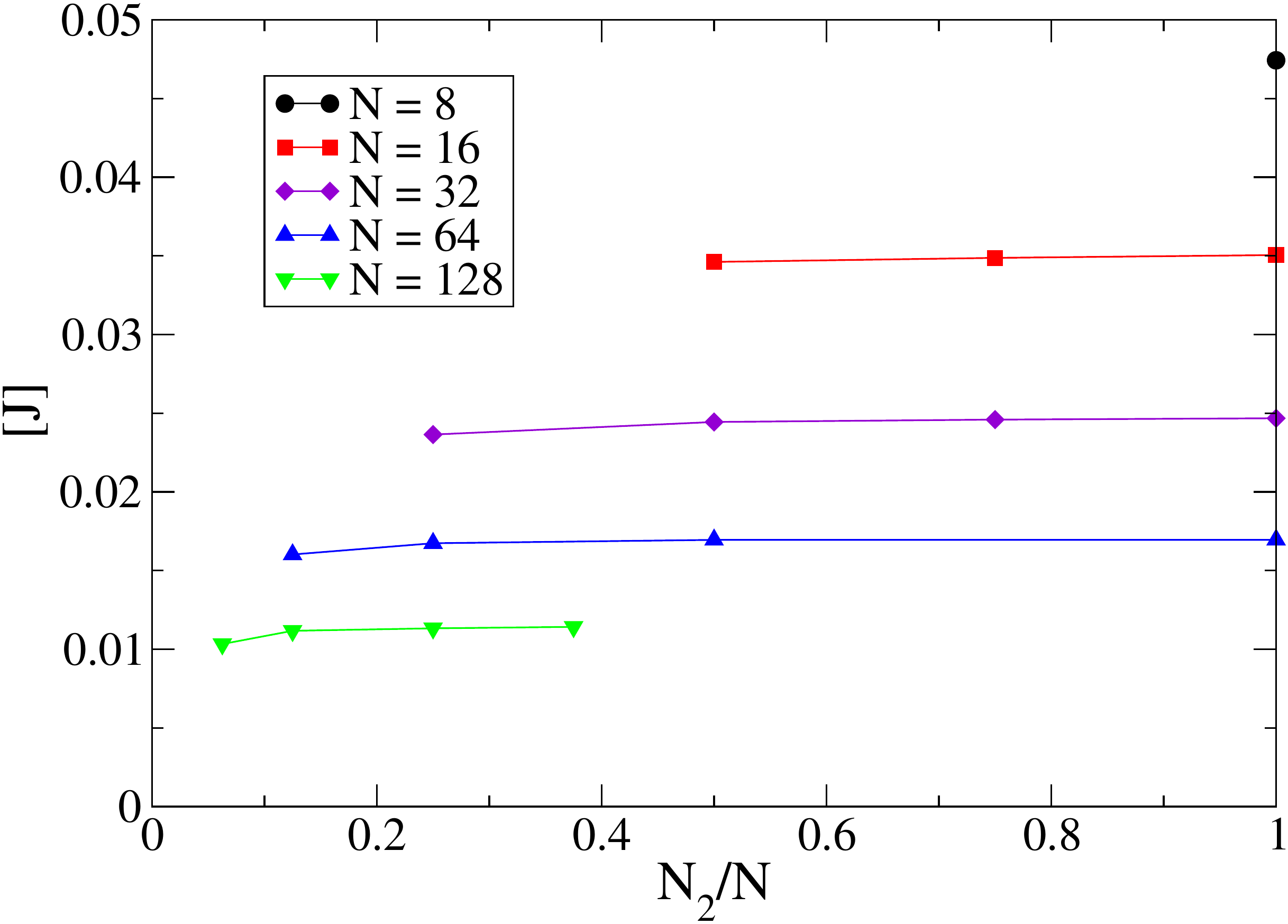}
\caption{ (color online) $3D$ unpinned case with fixed BC for $\Delta = 0.2$. \\
Plot of disorder-averaged
current density $J$ (with the definition $J=I/N_2^2$) versus $N_2/N$
for different fixed values of $N$. We see that the 3D limiting value is 
reached at quite small values of $N_2/N$.
}
\label{scale3d}
\vspace{0.75cm}
\end{figure}

\begin{figure}
\includegraphics[width=4in]{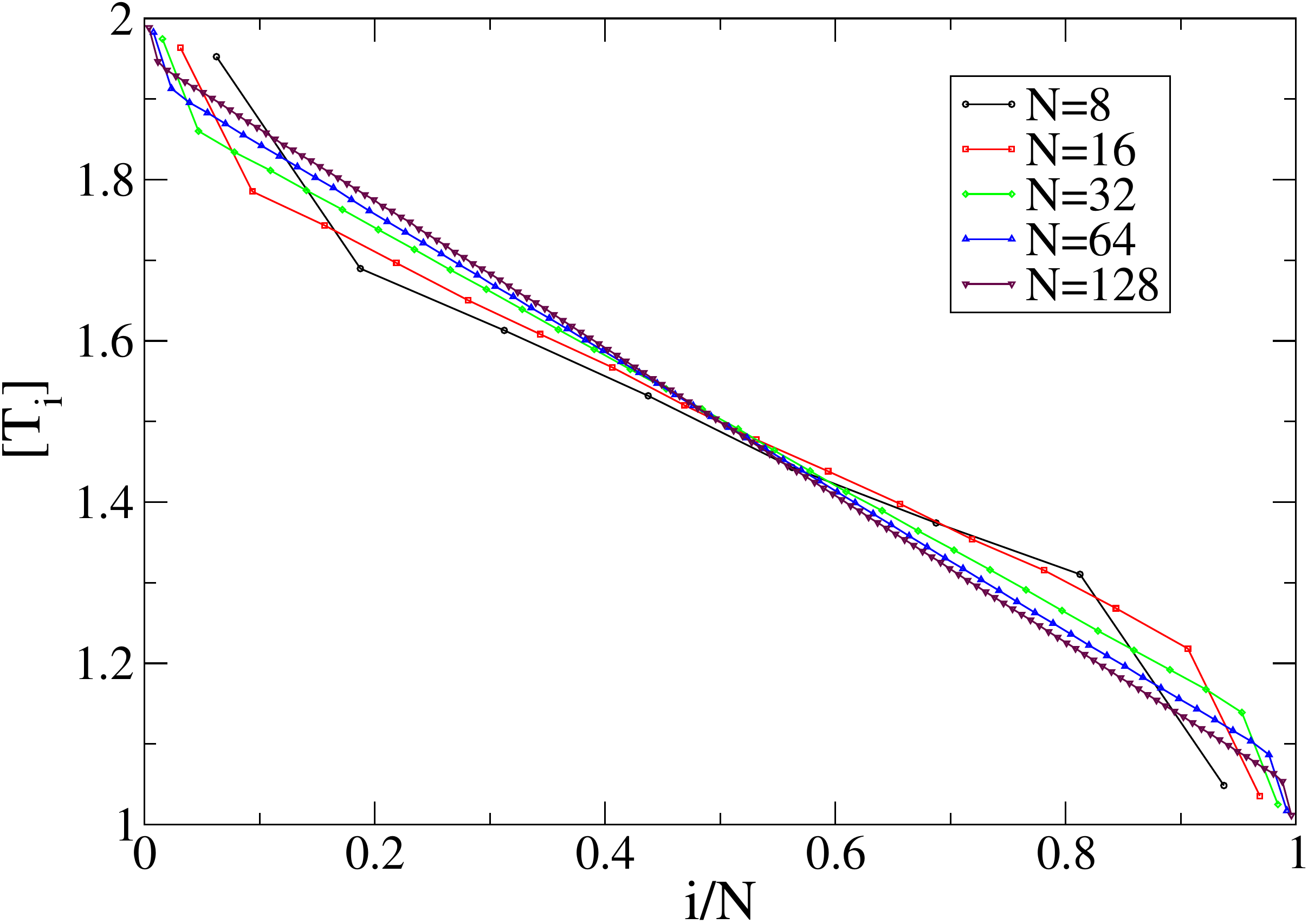}
\caption{ (color online) $3D$ unpinned case with fixed BC for $\Delta = 0.2$.\\
Plot of temperature
  profile   $T_i$  in a single disorder realization for different
  system sizes. The plots are from  simulations.. 
}
\label{temp3dfix}
\vspace{0.75cm}
\end{figure} 

\begin{figure}
\includegraphics[width=4in]{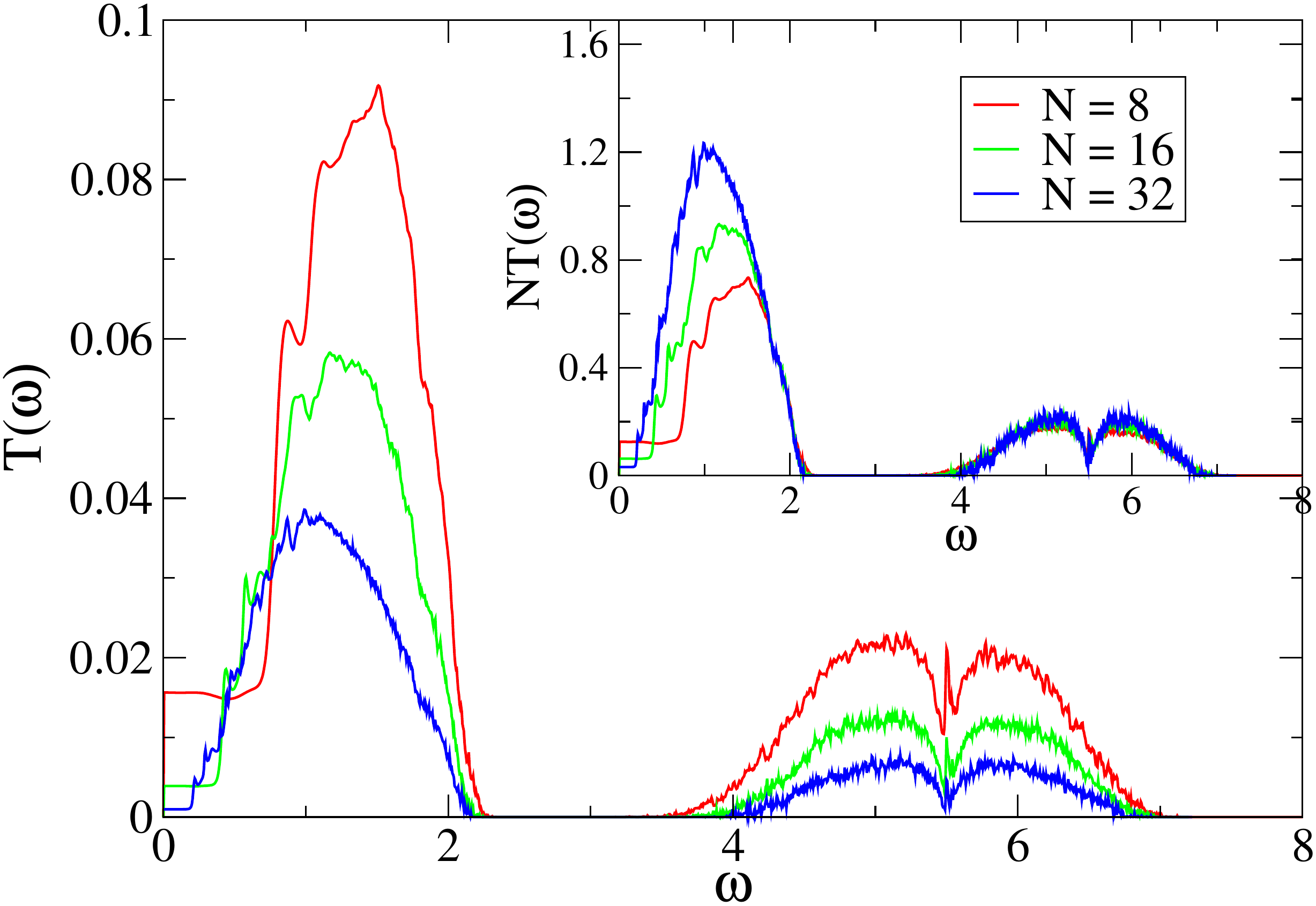}
\caption{(color online) {$3D$ unpinned case with free BC for $\Delta=0.8$. 
\\ TOP: Plot of the
  disorder averaged transmission  
  $T(\om)$ versus $\omega$. The inset shows the same data
  multiplied by a factor of $N$. }}
\label{tw3dfree0.8}
\vspace{.5cm}
\end{figure}

\begin{figure}
\includegraphics[width=3.2in]{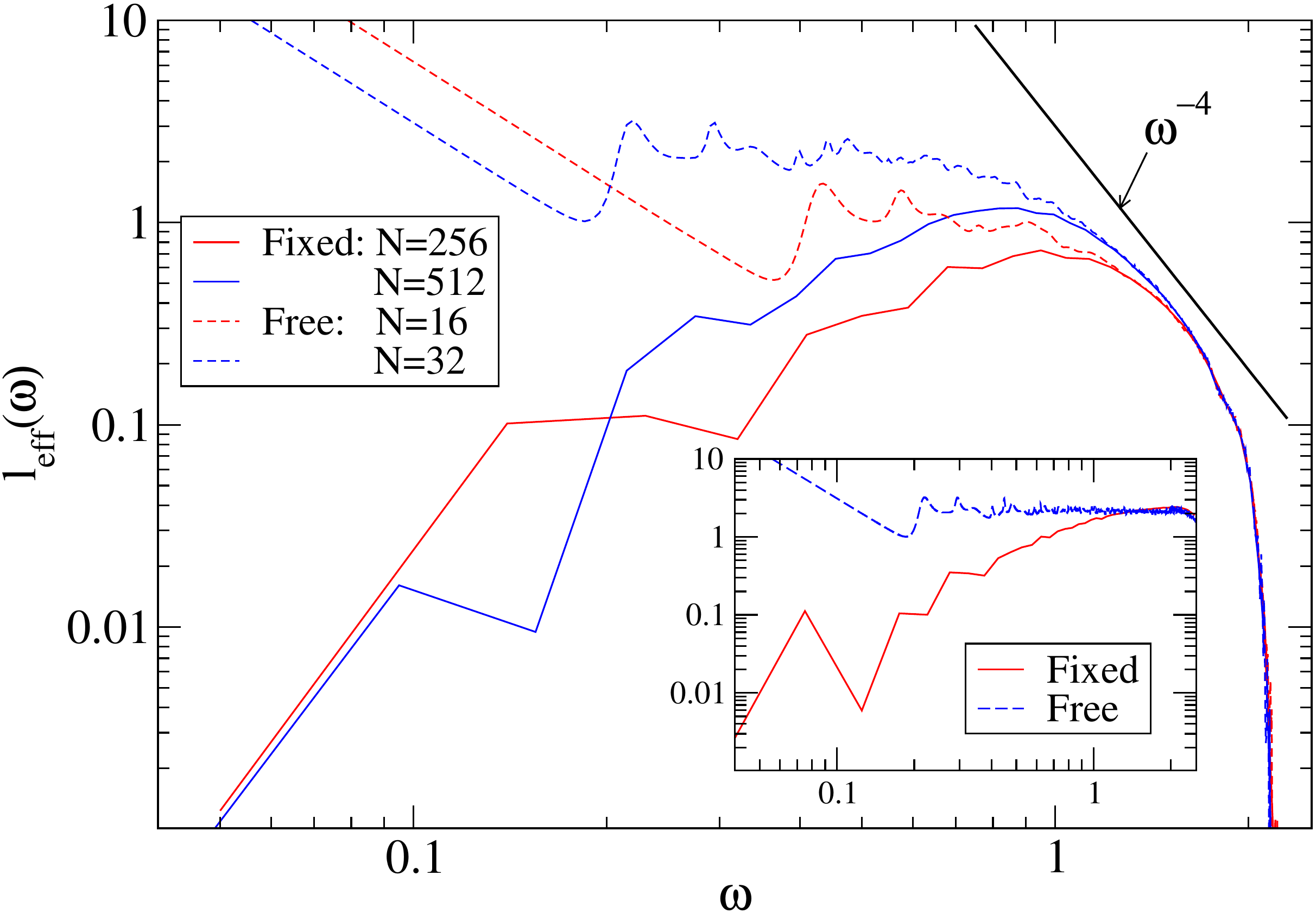}
\caption{
 Plot of the effective mean-free path $l_{\rm
 eff}=NT(\omega)/\om^{d-1}$ in $3D$ 
 with $\Delta=0.8$ for fixed and free BCs. The insets show
 $\ell_{\rm eff}$ for the ordered system with a single mass. An 
 $\om^{-4}$ behaviour is observed in a small part of the diffusive
 region.The fixed BC data is highly oscillatory and has been smoothed. }  
\label{leff3d}
\end{figure}

\begin{figure}
\includegraphics[width=4in]{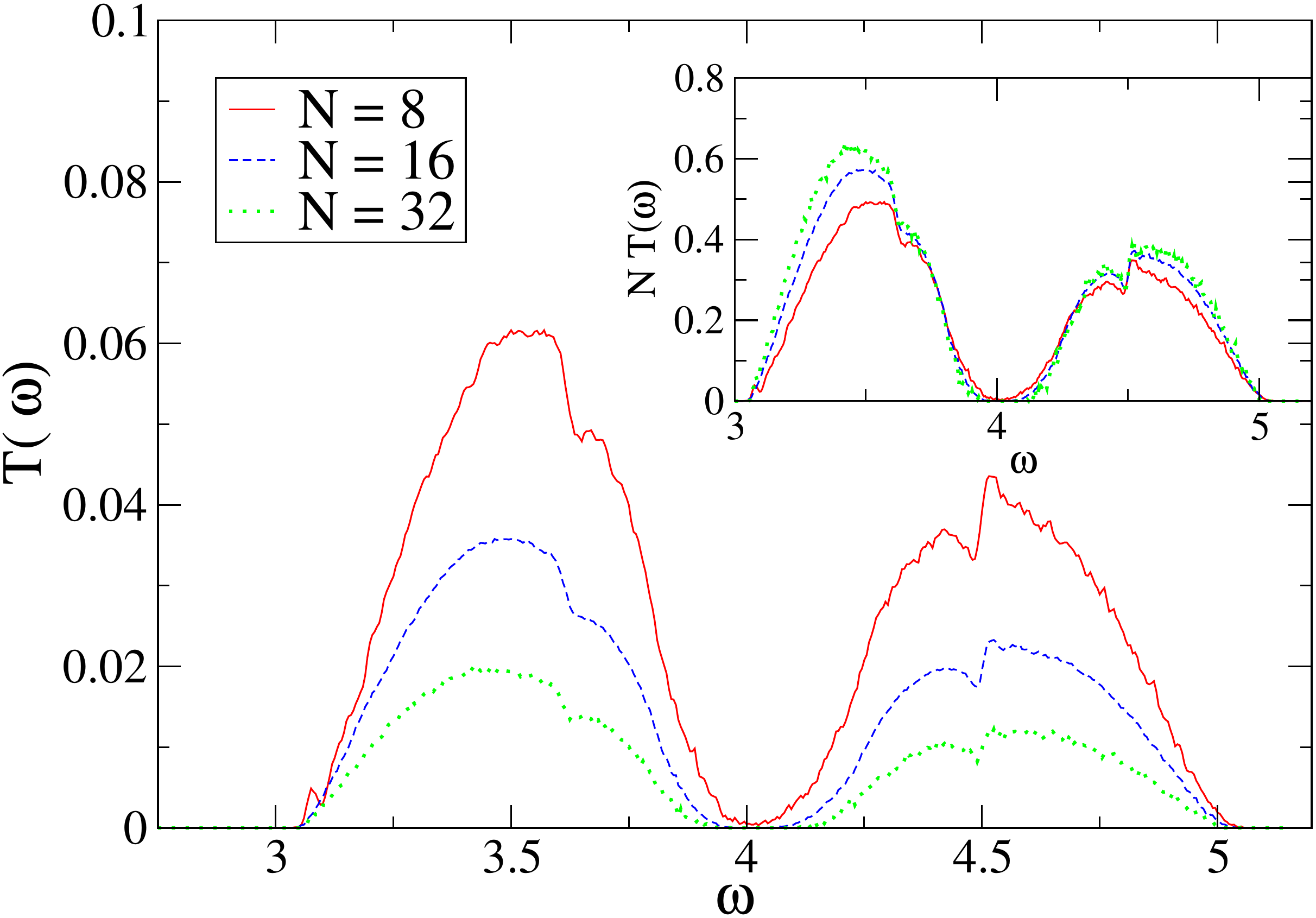}
\includegraphics[width=4in]{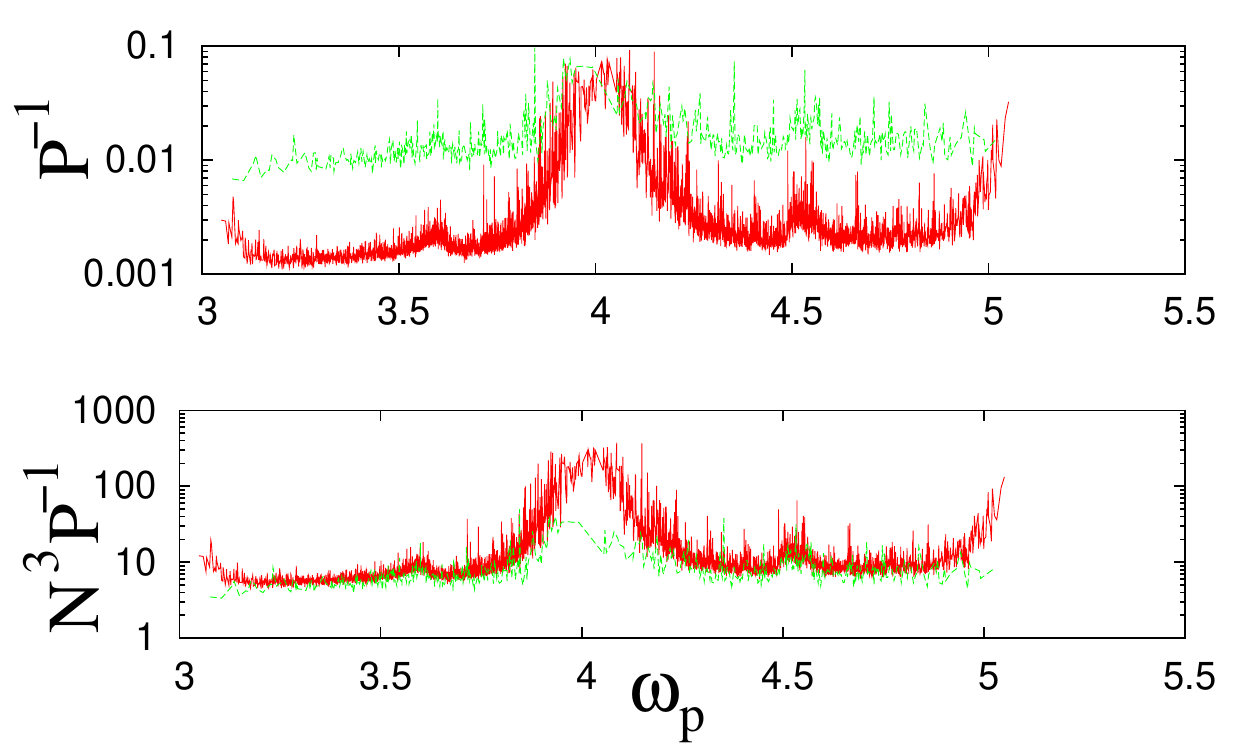}
\caption{(color online) {$3D$ pinned case for $\Delta = 0.2$ and $k_o=10.0$.\\ 
TOP: Plot of the disorder averaged transmission 
  $T(\om)$   versus $\omega$.
}\\
BOTTOM: {Plot of the IPR ($P^{-1}$) and scaled IPR ($N^3 P^{-1}$) as a function of normal mode-frequency 
$\om_p$. The curves are for $N=8$ (green) and $16$ (red).}}
\label{tw3donpin100.2}
\vspace{0.75cm}
\end{figure} 

\begin{figure}
\includegraphics[width=4in]{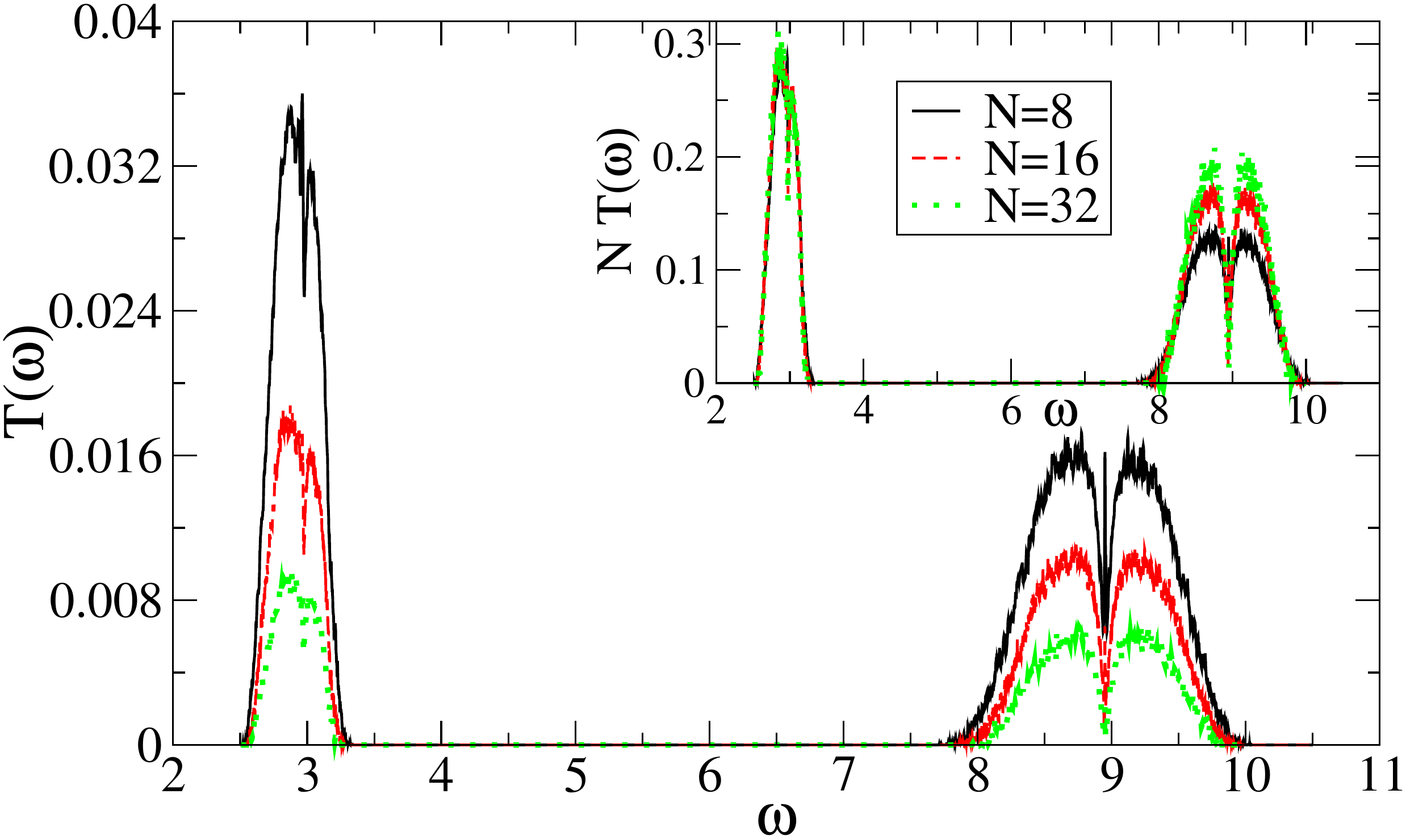}
\includegraphics[width=4in]{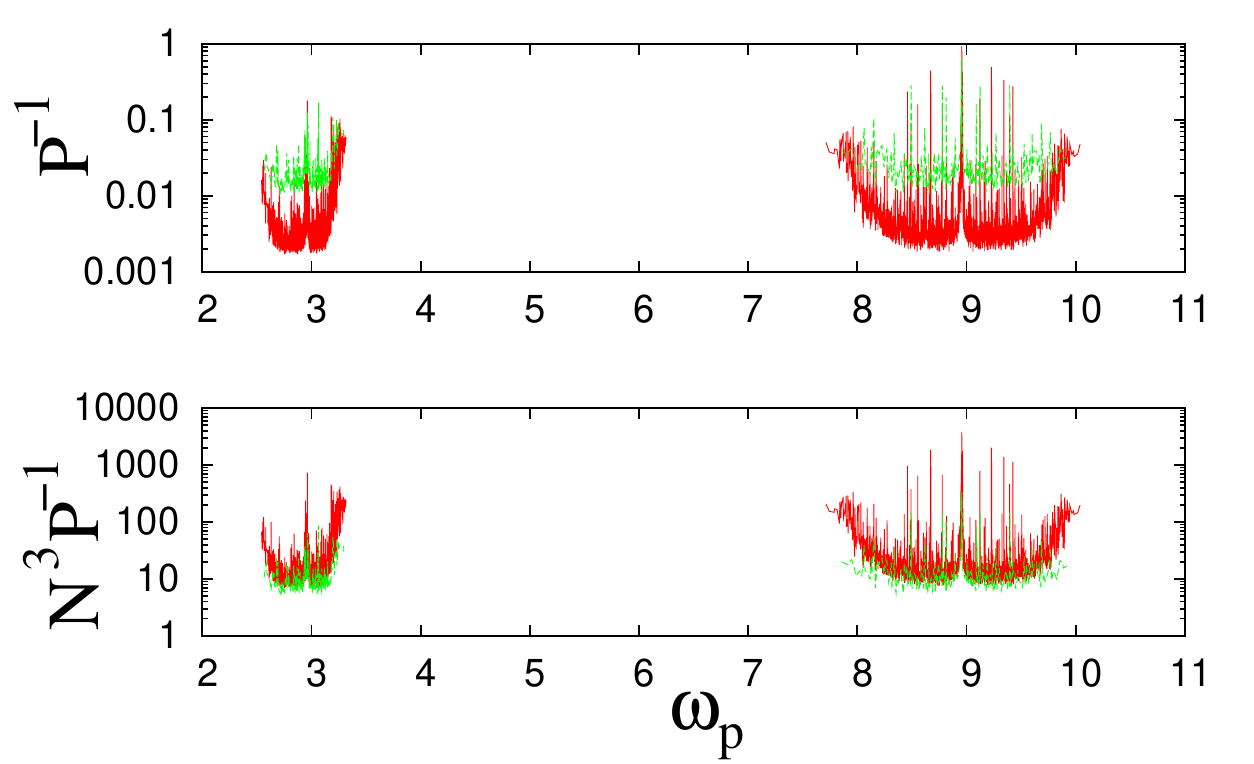}
\caption{ (color online) $3D$ pinned case for  $\Delta=0.8$ and $k_o=10.0$. 
\\ TOP: Plot of the disorder averaged transmission 
  $T(\om)$   versus $\omega$.\\ 
BOTTOM: Plot of the IPR ($P^{-1}$) scaled IPR ($N^3 P^{-1}$) as a 
function of normal mode-frequency $\om_p$.  The curves are for $N=8$ (green) 
and $16$ (red).}
\label{tw3donpin100.8}
\vspace{0.75cm}
\end{figure}

\begin{figure}
\includegraphics[width=4in]{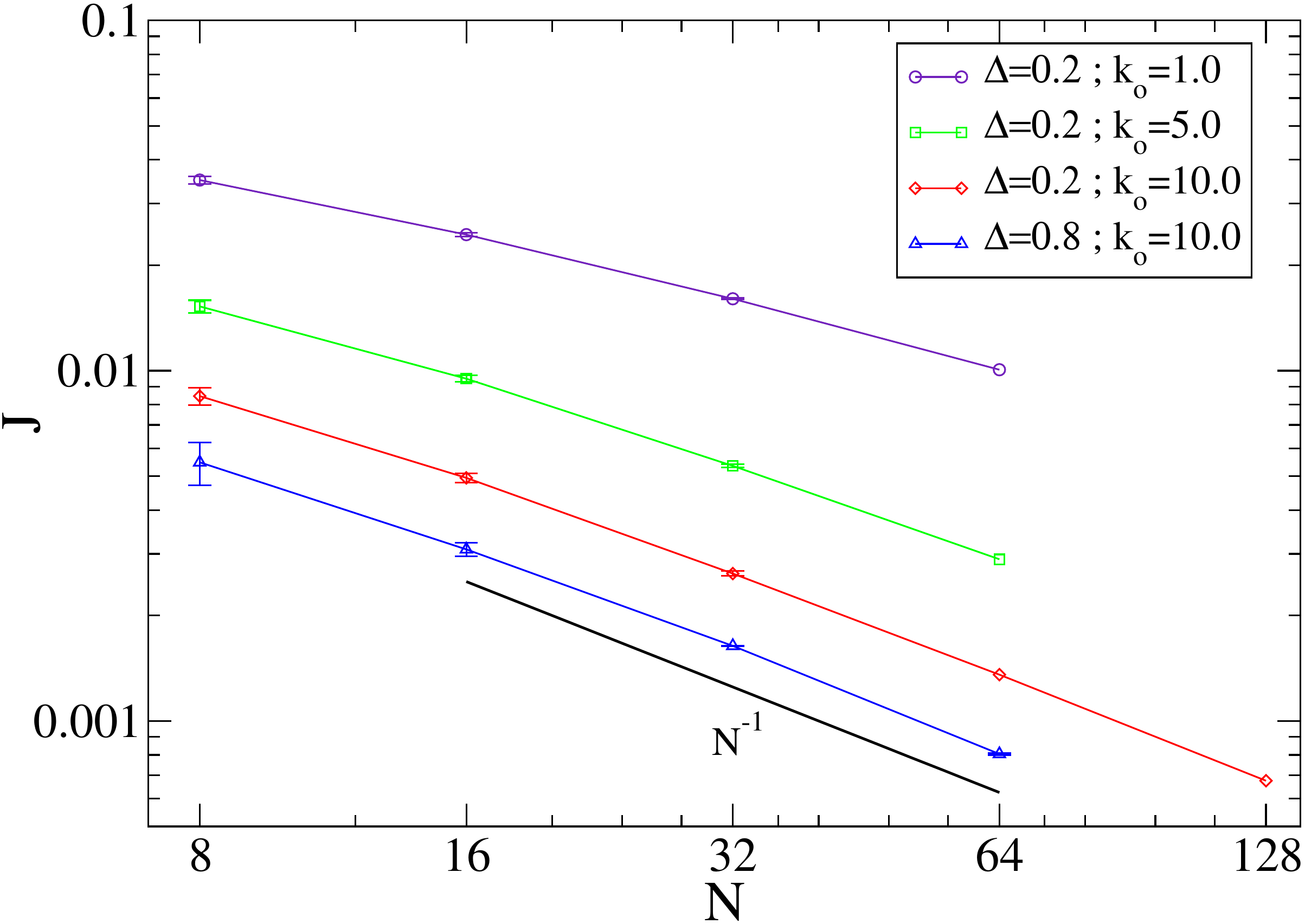}
\caption{ (color online) $3D$ pinned case. \\
Plot of disorder-averaged
  current $J$ versus system size 
 for   different values of $k_o$ and $\Delta$. The data sets for $\Delta=0.2$ for
   different values of $k_o$   are from simulations while the data for
  $\Delta = 0.8$ is from numerics.
\label{jvsn3donpin}
}
\vspace{0.75cm}
\end{figure} 

\begin{figure}
\includegraphics[width=4in]{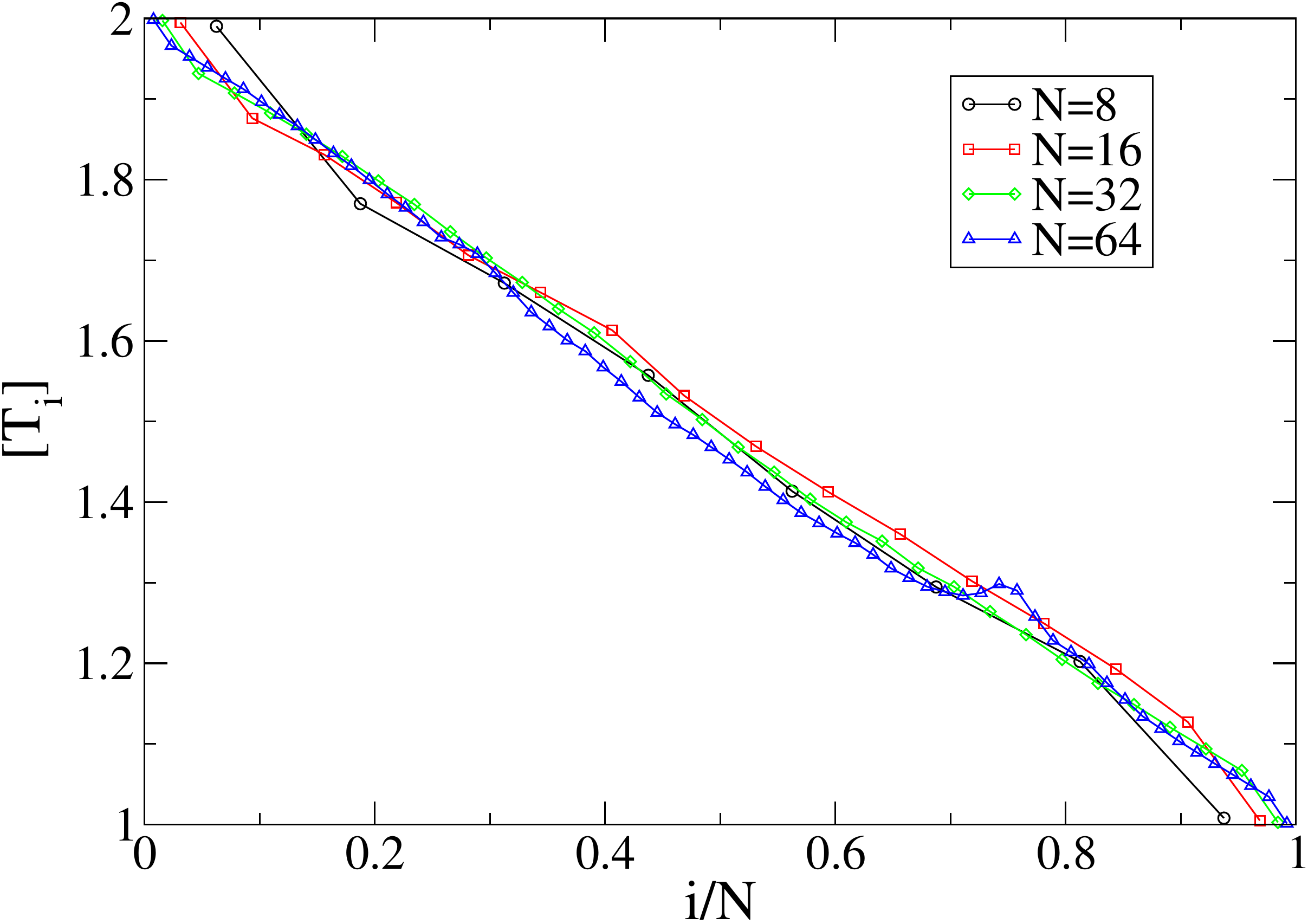}
\caption{ (color online) $3D$  pinned case for $\Delta = 0.2$ and $k_o=10.0$.\\
Plot of temperature
  profile   $T_i$  in a single disorder realization for different system sizes. The plots are from
  simulations. 
}
\label{temp3donpin}
\vspace{0.75cm}
\end{figure}


\end{widetext}

\end{document}